\documentclass[11pt]{article}

\usepackage{amsmath}
\usepackage{graphicx}
\usepackage{indentfirst}
\usepackage{amssymb}
\usepackage{cite}
\usepackage{color}
\usepackage{subfigure}
\usepackage{varwidth}

\begin{document}

\title{ Gravitational lensing and observational features of a dynamic black hole }

\date{}
\maketitle

\begin{center}
\author{Ke-Jian He,}$^{a}$\footnote{E-mail: kjhe94@.163com}
\author{Guo-Ping Li,}$^{b}$\footnote{E-mail: gpliphys@yeah.net}
\author{Li-Fang Li,}$^{c}$\footnote{E-mail: lilifang@imech.ac.cn}
\author{Xiao-Xiong Zeng}$^{d}$\footnote{E-mail: xxzengphysics@163.com (Corresponding author)}
\\

\vskip 0.25in
$^{a}$\it{Department of Mechanics, Chongqing Jiaotong University, Chongqing 400000,  China}\\
$^{b}$\it{School of Physics and Astronomy, China West Normal University, Nanchong 637000,  China}\\
$^{c}$\it{Center for Gravitational Wave Experiment, National Microgravity Laboratory, Institute ofMechanics, Chinese Academy of Sciences, Beijing 100190, China}\\
$^{d}$\it{College of Physics and Electronic Engineering, Chongqing Normal University, Chongqing 401331, China}\\
\end{center}
\vskip 0.6in
{\abstract
{In this work, we investigate the gravitational lensing effects and  the dynamic evolution of the shadow of Vaidya black holes by employing backward ray-tracing techniques. Within the celestial sphere framework, the black hole shadow exhibits a complete evolutionary sequence, transitioning from an initial stable configuration through continuous expansion to a final static state. Notably, during and after the active accretion phase, a distinct lensing ring emerges outside the shadow. Extending this analysis to the thin accretion disk model reveals richer observational signatures. A bright ring, formed by the superposition of the photon ring and lensing ring, appears outside the shadow but persists only during the initial and final stages of accretion, vanishing entirely when accretion becomes active. Interestingly, as the accretion process progresses, an additional ring-like structure, which is caused by the dynamical redshift effect, emerges in the image. This ring-like structure   not only contracts inward but also brightens continuously as accretion proceeds. Under varying observational inclinations, the Doppler effect and the dynamical redshift effect jointly modulate the brightness distribution of the image, resulting in significant asymmetry in the inner shadow, bright ring, and additional ring. Our findings uncover dynamical redshift as a novel observable phenomenon intrinsic to evolving spacetimes, offering a potential discriminant for identifying accreting black holes and providing observational access to the imprints of temporal spacetime evolution on black hole images.}}

\thispagestyle{empty}
\newpage
\setcounter{page}{1}

\section{Introduction}
\label{sec:intro}
As a fundamental prediction of general relativity, the physical existence of black holes has been conclusively verified through multiple independent observational means. These encompass the detection of binary black hole mergers by the Laser Interferometer Gravitational Wave Observatory (LIGO) \cite{LIGOScientific:2016vlm}, as well as the direct imaging of the supermassive black hole at the core of the M87 galaxy \cite{EventHorizonTelescope:2019dse} and of Sgr A* at our Galactic Center \cite{EventHorizonTelescope:2022wkp} by the Event Horizon Telescope (EHT). Crucially, the characteristic structure revealed by EHT images, i.e., a central dark region encircled by a luminous ring, has been unambiguously interpreted as the black hole shadow, a direct imprint of photon trajectories within the extreme gravitational field near the event horizon \cite{Virbhadra:1999nm, Claudel:2000yi, Virbhadra:2007kw, Falcke:1999pj,Tsukamoto:2018prd,Tsukamoto:2014tja}. This interpretation has established the black hole shadow as a powerful probe, enabling investigations into black hole properties that were previously inaccessible. In this context, comprehensive research on the black hole shadow has been conducted, and several remarkable results have been obtained. These involve constraining the deviation parameters of rotating black holes via angular radius statistics \cite{Vagnozzi:2022moj}; applying polarization signatures to verify the local validity of the equivalence principle in strong field scenarios \cite{Yan:2019hxx}; serving as a standard scale to quantify and test various cosmological phenomena \cite{Tsupko:2019pzg,Chen:2022nbb}; exploring fundamental physical issues such as dark matter and quasinormal modes \cite{Konoplya:2019xmn,Hou:2018bar,Konoplya:2019sns,Haroon:2018ryd,Luo:2024avl}; distinguishing black holes from other compact objects \cite{He:2025qmq,
Li:2025awg,Zeng:2025fjg}; and testing various gravitational theories  \cite{He:2024bll,He:2024yeg,Abdujabbarov:2016hnw,Zeng:2021dlj,Wei:2015dua,Perlick:2015vta,Atamurotov:2021cgh,Wang:2022kvg,Chen:2022scf,
He:2024qka,Guo:2020zmf,Atamurotov:2013sca,Shaikh:2018lcc,Qu:2023hsy,Meng:2023wgi,Zhang:2023bzv,Yang:2025jheap}.

Current theoretical investigations of black hole shadows predominantly rely on static spacetime backgrounds. While static models provide a fundamental framework for understanding shadow formation, realistic astrophysical black hole systems are inherently dynamic, involving processes such as radiation \cite{Hawking:1975vcx}, accretion \cite{Frank}, and binary mergers \cite{LIGOScientific:2016vlm}. Consequently, conventional static models with time-invariant mass and spin parameters are insufficient for capturing phenomena like fluctuating accretion rates or horizon phase transitions. A paradigmatic model for describing black hole evaporation and accretion dynamics is the Vaidya solution, formulated using ingoing null coordinates \cite{Vaidya:1951zz}. As an exact dynamical solution to the Einstein field equations, it incorporates a time-dependent mass function, thereby capturing distinct evolutionary timescales. However, the classical Vaidya solution, which includes only a pure radiation field, does not account for more complex matter fields present in realistic environments. To address this, Wang and Wu constructed a generalized Vaidya solution by linearly combining type-I and type-II energy-momentum tensors, enabling the description of scenarios involving both radiation and other matter fields like dust or dark energy \cite{Wang:1998qx}. This approach has led to various generalized models, including the Bonnor-Vaidya \cite{Bonnor:1970zz} and Vaidya-de Sitter solutions \cite{Mallett:1985lsn}. Furthermore, for rotating dynamical black holes, Dahal et al. extended the Vaidya solution using the Newman-Janis algorithm, proposing the Kerr-Vaidya metric \cite{Dahal:2020hsj}. These solutions have facilitated further studies on gravitational collapse, horizon evolution, energy conditions, and naked singularities \cite{Vertogradov:2018ora, Vertogradov:2022zuo, Vertogradov:2024fbd}.

Unlike in static spacetimes, the energy of particles moving along geodesics in a dynamical black hole spacetime is not conserved, rendering the shadow analysis more challenging. Nevertheless, valuable information can still be obtained by directly solving the second order geodesic equations \cite{Heydarzade:2017nfp} or by  exploiting  additional symmetries to reduce the order of the equations \cite{ Ojako:2019gwc,Koh:2020hta}. Building on these techniques, significant progress has been made in understanding the shadow and photon ring structures in Vaidya spacetimes. For instance, Mishra et al. systematically investigated the photon sphere structure of Vaidya black holes with several mass functions, uncovering its evolutionary behavior within future characteristic time regions \cite{Mishra:2019trb}.  Solanki et al. analytically solved the  null geodesic equation in the Vaidya spacetime, and conducted a comparative analysis of the evolution patterns under the two scenarios of accretion  and radiation \cite{Solanki:2022glc}.  In the Bonnor-Vaidya spacetime, Heydarzade et al. found that charge reduces the photon sphere radius and that a naked singularity might cast a shadow \cite{Heydarzade:2023gmd}.  In addition, Vertogradov et al. took into account the scenario in which the accretion rate of the Vaidya black hole is relatively low, and analyzed the dynamic evolution of the Vaidya black hole's shadow by integrating the induced force, centrifugal force and relativistic force \cite{Vertogradov:2024eim}. However, these studies have primarily focused on the geometric properties of the shadow itself (e.g., its size and shape) or on the behavior of photon rings. The critical question of how a dynamic black hole would actually look when illuminated by a practical light source, such as an accretion disk, remains largely uninvestigated. Against the backdrop of static black  hole spacetime, a comprehensive system has been developed for investigating black hole shadows, observational characteristics, and photon  ring structures via various accretion  models, including the spherically symmetric model, thin accretion disk and thick accretion disk \cite{Luminet:1979nyg,Narayan:2019imo,Gralla:2019xty,Zeng:2020dco,Peng:2020wun,He:2022yse,Zeng:2021mok,Li:2021riw,Gao:2023mjb,Wang:2022yvi,Hou:2022eev,He:2024amh,Zeng:2022fdm}. In view of this, we intend to investigate the evolutionary behavior of shadows and the observational characteristics of dynamic black holes surrounded by a thin accretion disk. In this work, we study the evolutionary behavior of the shadow and the observational characteristics of a Vaidya black hole surrounded by two idealized accretion models: a celestial sphere and a geometrically thin accretion disk. Our primary objective is to ascertain whether the temporal evolution of the shadow features can function as a diagnostic tool for the underlying dynamic spacetime structure, thus differentiating it from a static one.

The structure of this paper is organized as follows: Section \ref{sec1} briefly introduces the Vaidya black hole, presents a smooth mass function for accretion, and analyzes its geodesic equations. In Section \ref{sec2}, we examine the gravitational lensing effects and the dynamic evolutionary behavior of the shadow using the celestial sphere model. Section \ref{sec3} presents the observational features of the Vaidya black hole surrounded by a thin accretion disk, detailing the disk parameters and imaging method. Finally, a concise conclusion and discussion are provided in Section \ref{conclusion}.

\section{Vaidya black hole and geodesic motion }
\label{sec1}
As the first exact solution within the framework of general relativity that depicts a non-static spherically symmetric gravitational field, the Vaidya metric serves as a significant theoretical instrument for the study of dynamic black holes. Unlike the Schwarzschild metric with a constant mass, the Vaidya metric can describe the continuous change in the mass of a black hole during the accretion or evaporation process by introducing a time-evolving mass function, thus being more in line with the real astrophysical environment. The Vaidya metric can be expressed in  terms of the Eddington-Finkelstein coordinates, which is \cite{Wang:1998qx}
\begin{align}
ds^{2} =-f(v,r) dv^{2} \pm 2dv\,dr + r^{2}d\Omega^{2}, \label{metric1}
\end{align}
and
\begin{align}
f(v,r) = 1 - \frac{2M(v)}{r}. \label{metric2}
\end{align}
Here, $v$ is Eddington's advanced time coordinate, $M(v)$ is the mass function, and $d\Omega^2 = d\theta^2 + \sin^2\theta d\phi^2$ is the solid angle line element on the unit two-sphere. This form of the line element not only maintains spherical symmetry but also clearly demonstrates the dynamic nature of spacetime. When $M(v)$ is a constant, the spacetime (\ref{metric1}) degenerates to the Schwarzschild metric. When $M(v)$ is non-constant, the metric characterizes the gravitational field surrounding a central object with a mass that either increases or decreases, where the mass is dependent on the coordinate $v$.
Since the Einstein tensor depends on the derivatives of the metric, the energy-momentum tensor associated with the Vaidya spacetime corresponding to equation (\ref{metric1}) has the following form \cite{Mishra:2019trb}
\begin{align}
T^{\mu\nu} = \pm \frac{1}{4\pi r^2}\frac{dM(v)}{dv} K^\mu K^\nu, \quad K^\mu \partial_\mu=\mp  \partial_r,  \label{EM1}
\end{align}
which is typically interpreted as that of a null dust moving in the direction of decreasing $r$. It is worth noting that the vector field $K^\mu \partial_\mu$ consistently points towards the future. It is assumed that the time orientation has been chosen in such a way that the coordinate $v$ increases with time towards the future. In addition, the selection of the upper and lower symbols in the aforementioned equation must consistently maintain coherence. If the upper sign is selected, it denotes the ingoing Eddington-Finkelstein coordinates. In this situation, the null dust moves in the direction of decreasing $r$. Moreover, for the energy-momentum tensor to satisfy the null energy condition, the constraint $\frac{d M(v)}{d v} > 0$ must be satisfied. This indicates that the central object accretes by absorbing the falling matter with positive energy density.  Conversely, a radiating (evaporating) black hole would be described by the outgoing (retarded) coordinates with $\frac{d M(v)}{d v}<0$.

In this work, we primarily concentrate on accreting black holes. The core of the Vaidya metric lies in the selection of the mass function $M(v)$, which completely determines the evolution behavior of the black hole. To model a black hole that grows smoothly from an initial state to a final, asymptotically static configuration, one can adopt the following form \cite{Solanki:2022glc}
\begin{align}
ds^{2} = -\left(1 - \frac{2M(v)}{r}\right)dv^{2} + 2dv\,dr + r^{2}d\Omega^{2},  \label{metric3}
\end{align}
and
\begin{align}
M(v) = \frac{M_{0}}{2}\left[1 + \tanh(v)\right], \label{Mass}
\end{align}
where \(v \in (0, \infty)\) describes a smooth transition from a low-mass state in the past to a final mass \(M_0\) in the future. This function tends to a constant \(M_0\) in the asymptotic future (\(v \rightarrow \infty\)). And, the crucial feature allows us to impose well-defined future boundary conditions for dynamical quantities, such as the photon orbit radius \(r_{ph}\), which will approach its Schwarzschild limit \(r_{ph} \rightarrow 3M_0\) and \(\dot{r}_{ph} \rightarrow 0\).
The apparent horizon, defined locally by the vanishing expansion of outgoing null geodesics, is located at \(r_{\text{AH}} = 2M(v)\). The event horizon, being a global concept, is more complex. Its location \(r_{\text{EH}}(v)\) can be found by solving the ordinary differential equation for outgoing radial null geodesics that just manage to escape to future null infinity, that is\cite{Nielsen:2010gm}
\begin{align}
\frac{dr}{dv} = \frac{1}{2}\left(1 - \frac{2M(v)}{r}\right),  \label{Event horizon}
\end{align}
with the boundary condition that \(r_{\text{EH}}(v) \rightarrow 2M_0\) as \(v \rightarrow \infty\).
Unlike the event horizon, which is a global concept and requires knowledge of the entire future evolution of the spacetime to be located, the apparent horizon is locally defined. This makes it a more practical tool for studying dynamic black holes.
After establishing the basic framework of the Vaidya spacetime, we next analyze the motion behavior of photons within it. Since the Vaidya metric maintains spherical symmetry, this guarantees the existence of a conserved angular momentum for null geodesics. By confining the analysis to the equatorial plane ($\theta = \pi/2$), the Lagrangian for the photon can be expressed as\cite{Mishra:2019trb}
\begin{align}
\mathcal{L} = \frac{1}{2}\left[-f(r,v)\left(\frac{dv}{d\lambda}\right)^2 +2\left(\frac{dv}{d\lambda}\right)\left(\frac{dr}{d\lambda}\right) + r^2\left(\frac{d\theta}{d\lambda}\right)^2 +r^2\sin^2\theta \left(\frac{d\phi}{d\lambda}\right)^2\right], \label{Lagrangian}
\end{align}
where $\lambda$ is the affine parameter.
Since the metric is independent of the azimuthal angle $\phi$, its conjugate momentum gives a strict constant of motion
\begin{align}
p_\phi = \frac{\partial \mathcal{L}}{\partial \dot{\phi}} = r^2 \dot{\phi} \equiv L = \text{constant}. \label{ang_mom_cn2}
\end{align}
However, the situation for energy is completely different. The metric explicitly depends on the coordinate $v$, $\partial_v$ is not a Killing vector field, and thus the energy of the photon is no longer conserved. Nevertheless, one can still define a time-dependent specific energy $E(v)$ from the Lagrangian, that is \cite{Mishra:2019trb, Vertogradov:2024eim}
\begin{align}
E(v) = -\frac{\partial \mathcal{L}}{\partial \dot{v}} = f(v,r)\dot{v} - \dot{r}. \label{energy_def_cn2}
\end{align}
The evolution of $E(v)$ along the geodesic is governed by the equation $dE/d\lambda = \partial \mathcal{L}/\partial v$, which directly couples the photon's energy with the dynamics of spacetime. Specifically, its evolution equation can be written as
\begin{align}
\frac{dE}{d\lambda} = -\frac{\dot{M}(v)}{r^2} \left(\frac{dv}{d\lambda}\right)^2,
\end{align}
in which  $\dot{M}(v) = dM/dv$. This relationship clearly indicates that the rate of change of photon energy is proportional to the accretion rate of the black hole mass, $\dot{M}(v)$.

At this point, we have obtained the fundamental equation governing the motion of photons in the Vaidya spacetime. Compared with static spacetimes (such as the Schwarzschild metric), a fundamental difficulty has emerged, i.e., the energy $ E $ is no longer a conserved quantity. In static spacetimes, one  can define the impact parameter $ b \equiv L/E $ and rewrite the null geodesic condition as a first-order radial equation involving an effective potential, thereby directly and analytically analyzing the trajectory of light and the boundary of the shadow. However, within the dynamic Vaidya spacetime, the quantity \( E \) exhibits variation with respect to the time coordinate $v$, rendering the radial motion equation non-separable. That is to say, it is impossible to obtain the critical impact parameter that determines the shadow boundary by solving a simple algebraic equation, such as the extremum condition of the effective potential at $r$, as is the case in the static scenario. Therefore, it is necessary to resort to numerical integration methods to trace a large number of light rays in order to determine the real-time shape and size of the shadow that an observer at infinity (or at a certain finite distance) can see during the dynamic evolution process. This is precisely the work that will be carried out in the subsequent parts of this article.

\section{Dynamic gravitational lensing effects and shadows}
\label{sec2}
In this section, we intend to employ the celestial  sphere \cite{Bohn:2014xxa} model to visually illustrate the gravitational lensing effect induced by black holes and the dynamic evolutionary characteristics of the black hole shadow. Among them, the celestial sphere is partitioned into four quadrants and colored with four distinct colors, and the intersection points of the four quadrants are colored white. In addition, the grid composed of both longitude and latitude lines is depicted by brown lines, with a spacing of $10^{\circ}$ between each line. Owing to the symmetry of spacetime, the observer can be positioned at an eccentric location within the celestial sphere, where the distance between the observer and the center of the boson star or black hole is $r_{obs} = 500 M_0$.

\begin{figure}[!t]
\centering 
\subfigure[$v=0$]{\includegraphics[width=.2\textwidth]{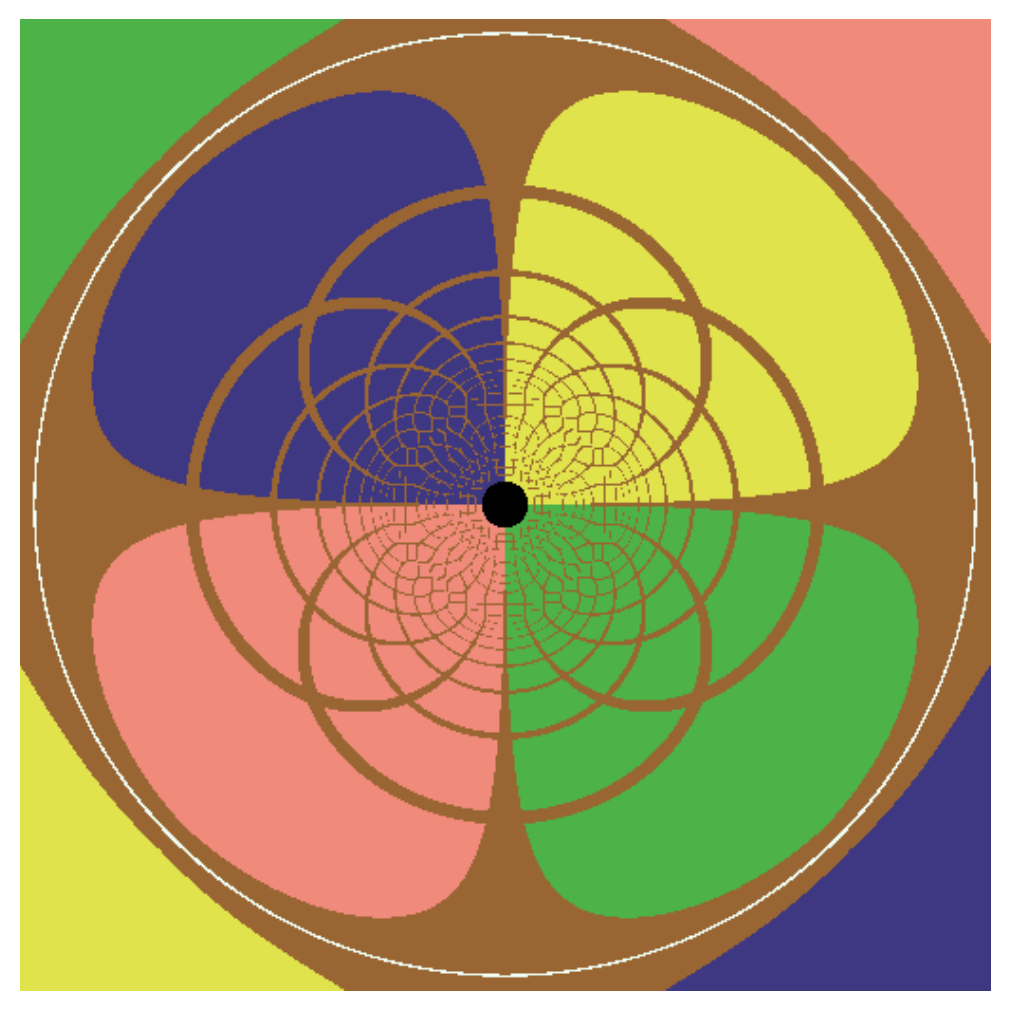}}
\subfigure[$v=0.1$]{\includegraphics[width=.2\textwidth]{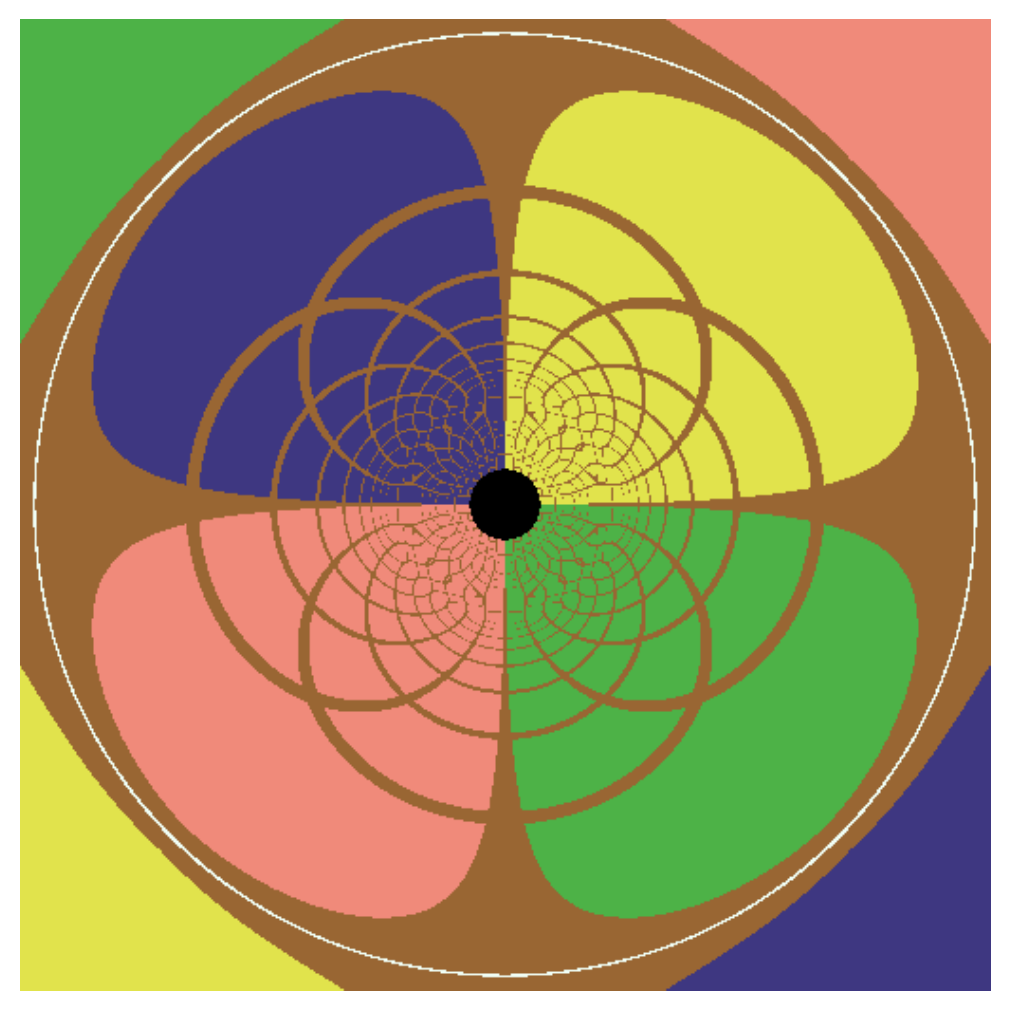}}
\subfigure[$v=0.3$]{\includegraphics[width=.2\textwidth]{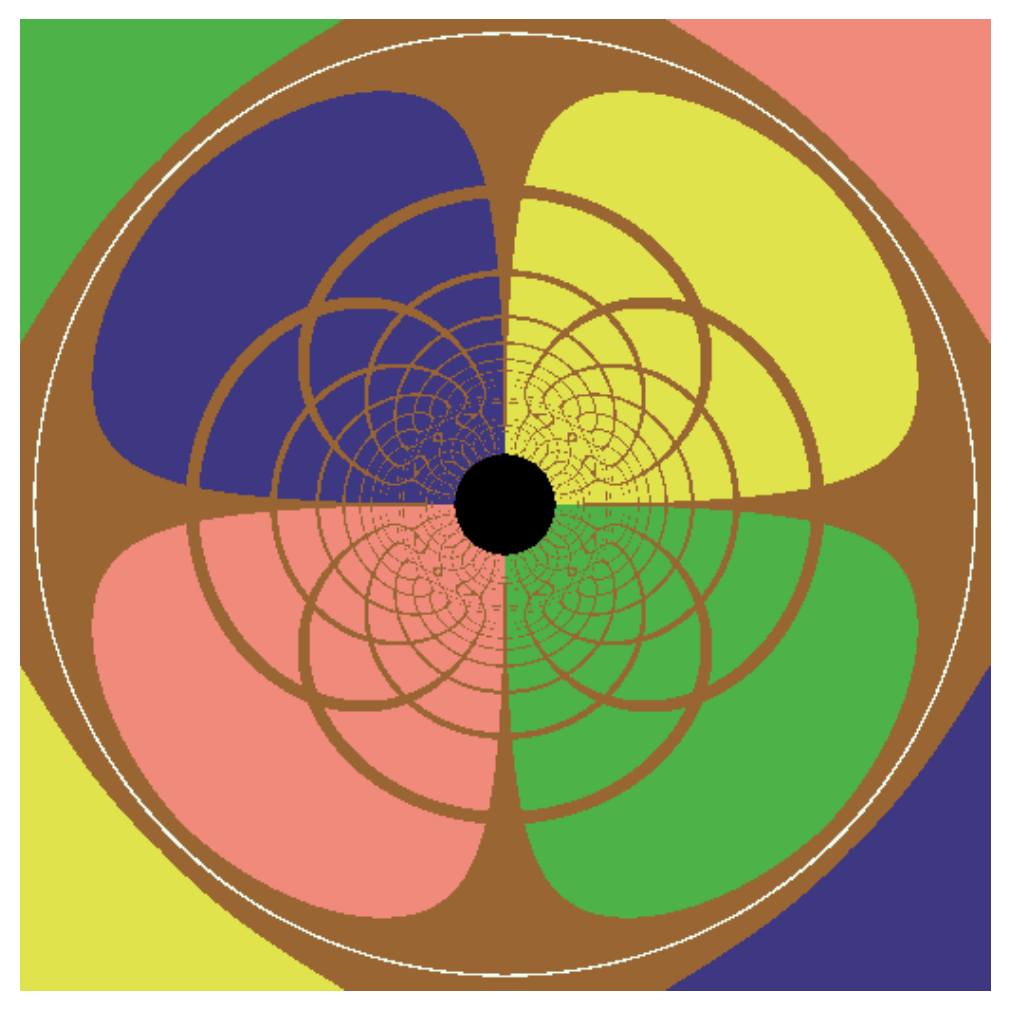}}
\subfigure[$v=0.5$]{\includegraphics[width=.2\textwidth]{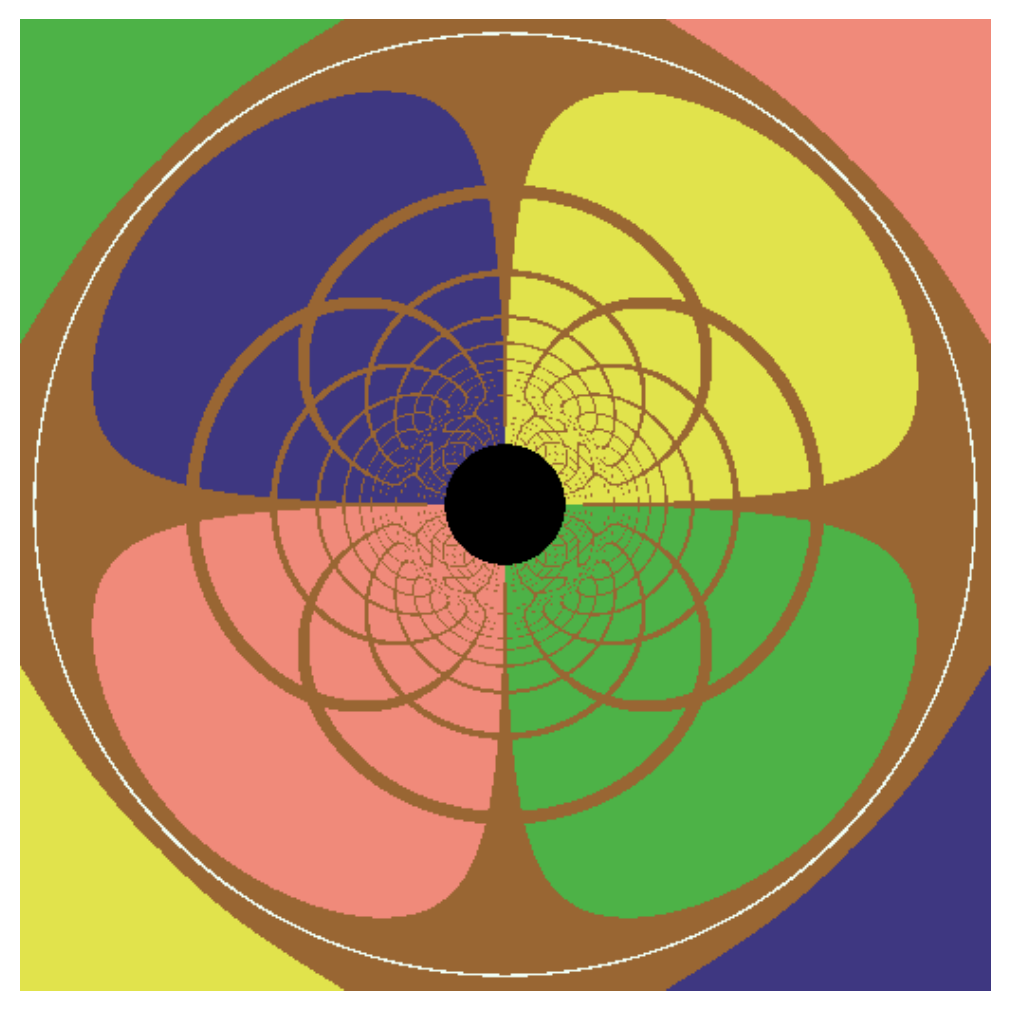}}
\subfigure[$v=0.8$]{\includegraphics[width=.2\textwidth]{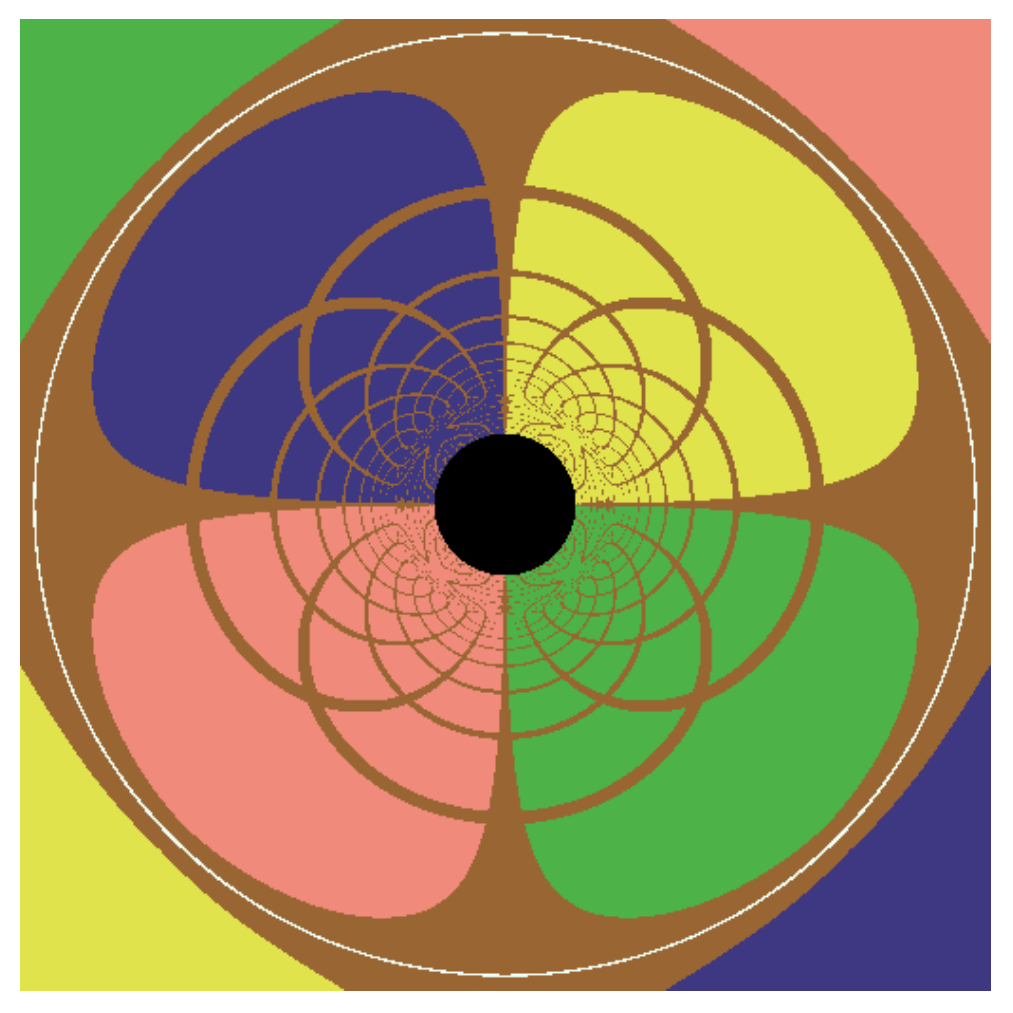}}
\subfigure[$v=1.1$]{\includegraphics[width=.2\textwidth]{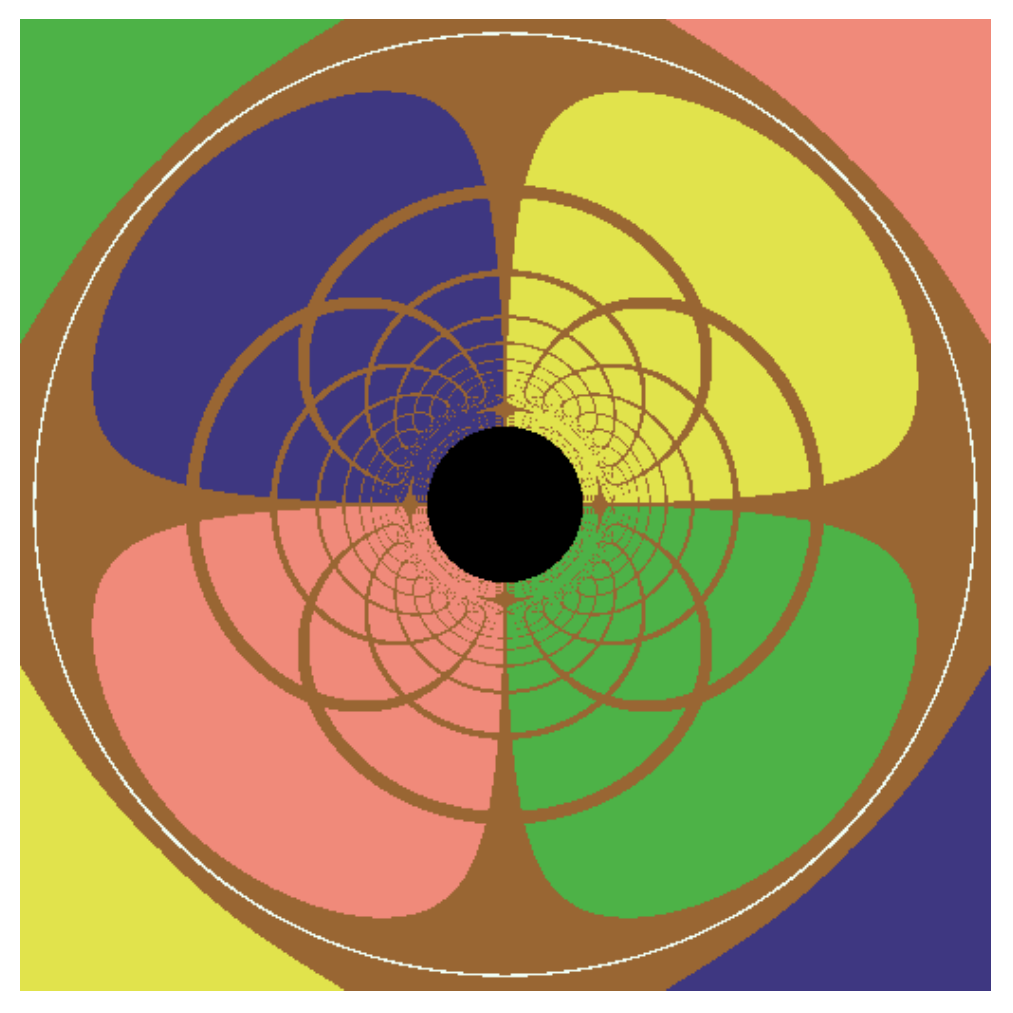}}
\subfigure[$v=1.4$]{\includegraphics[width=.2\textwidth]{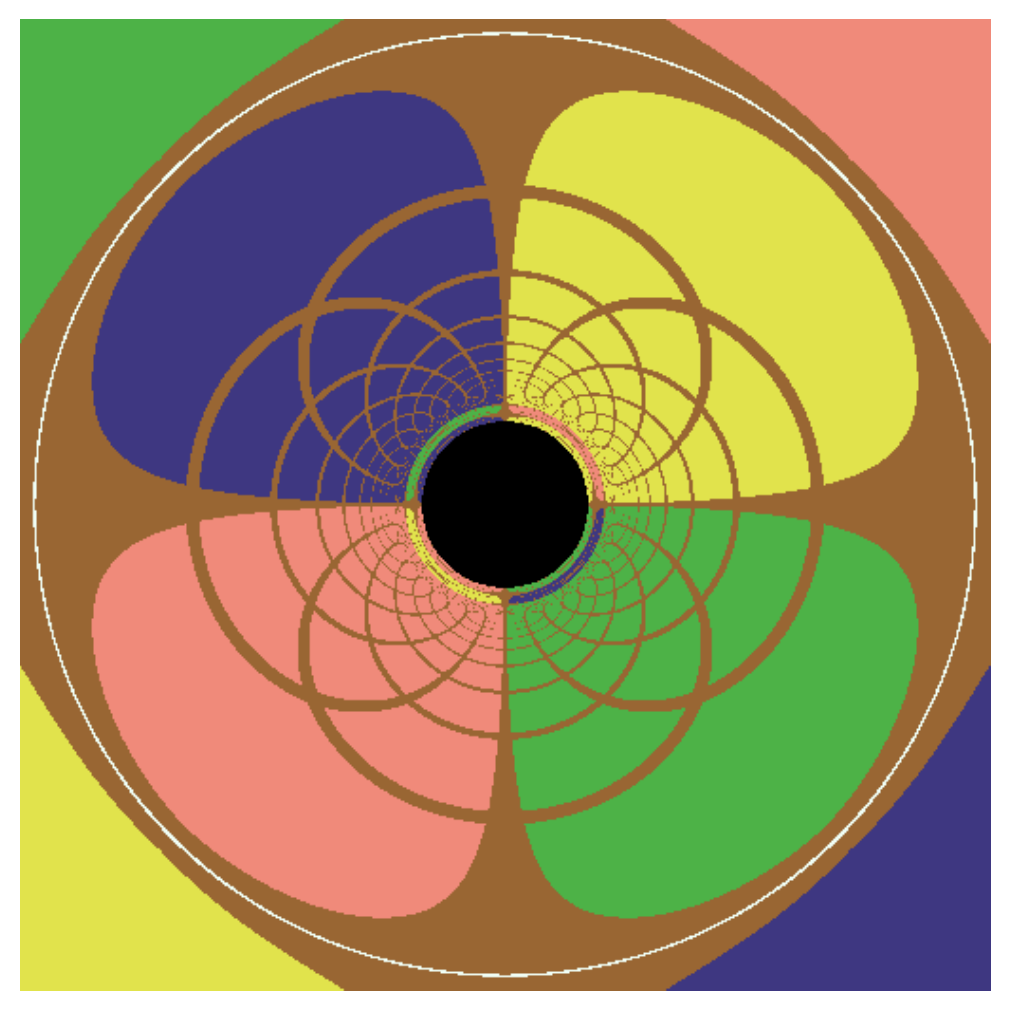}}
\subfigure[$v=1.7$]{\includegraphics[width=.2\textwidth]{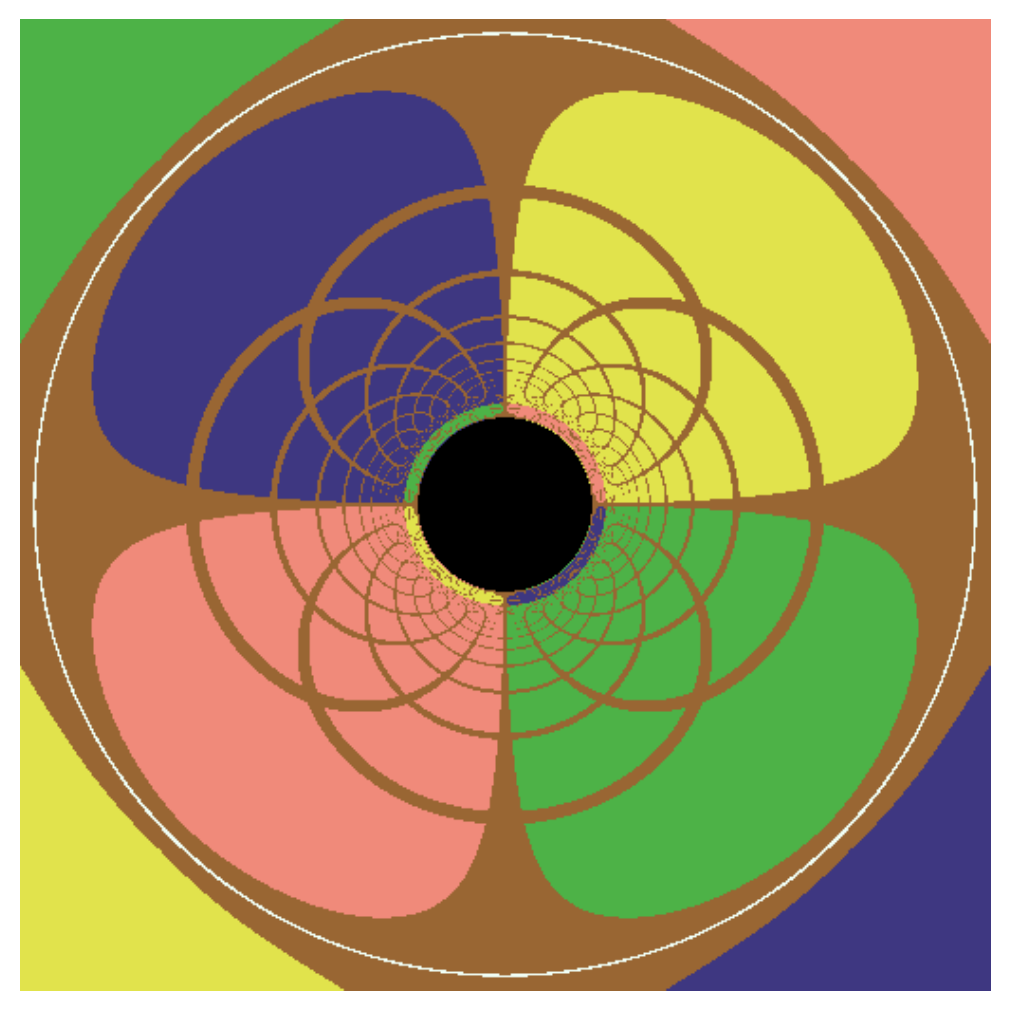}}
\subfigure[$v=2.2$]{\includegraphics[width=.2\textwidth]{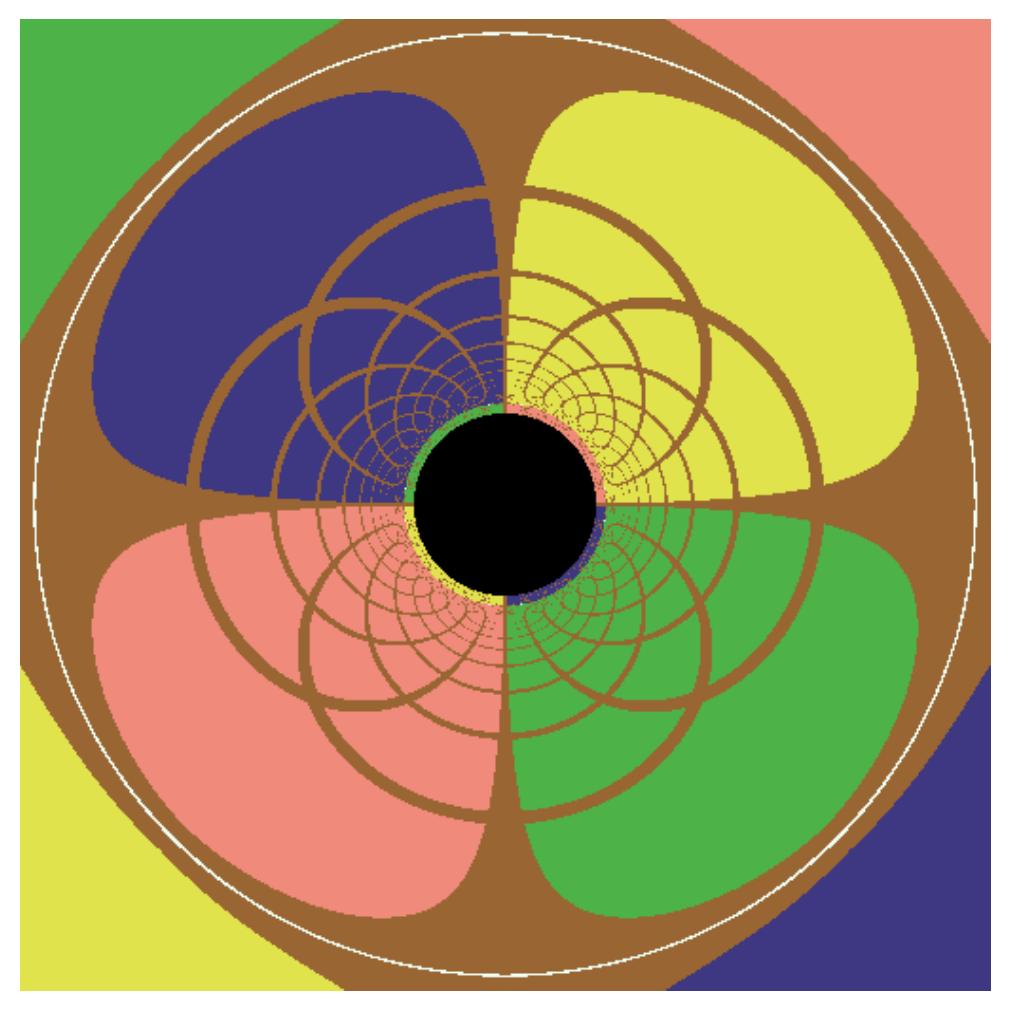}}
\subfigure[$v=2.7$]{\includegraphics[width=.2\textwidth]{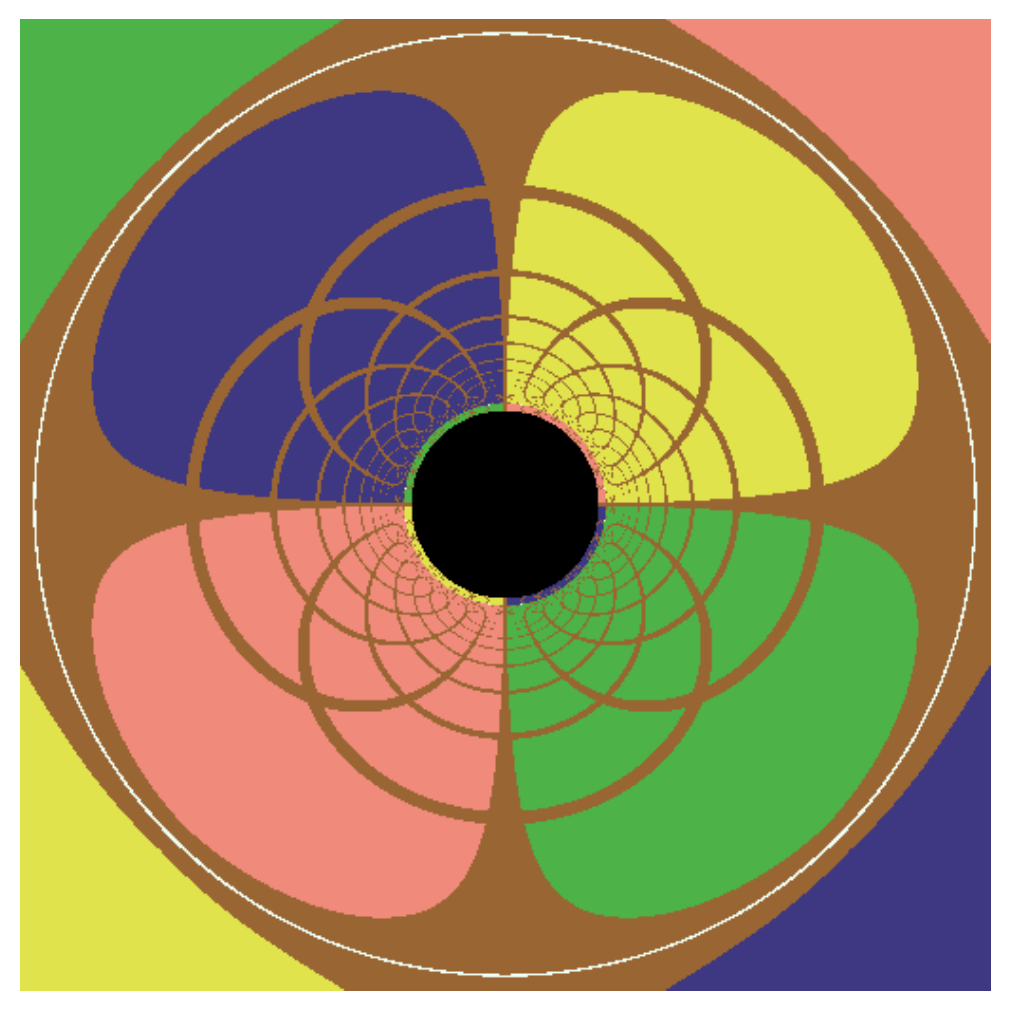}}
\subfigure[$v=3.2$]{\includegraphics[width=.2\textwidth]{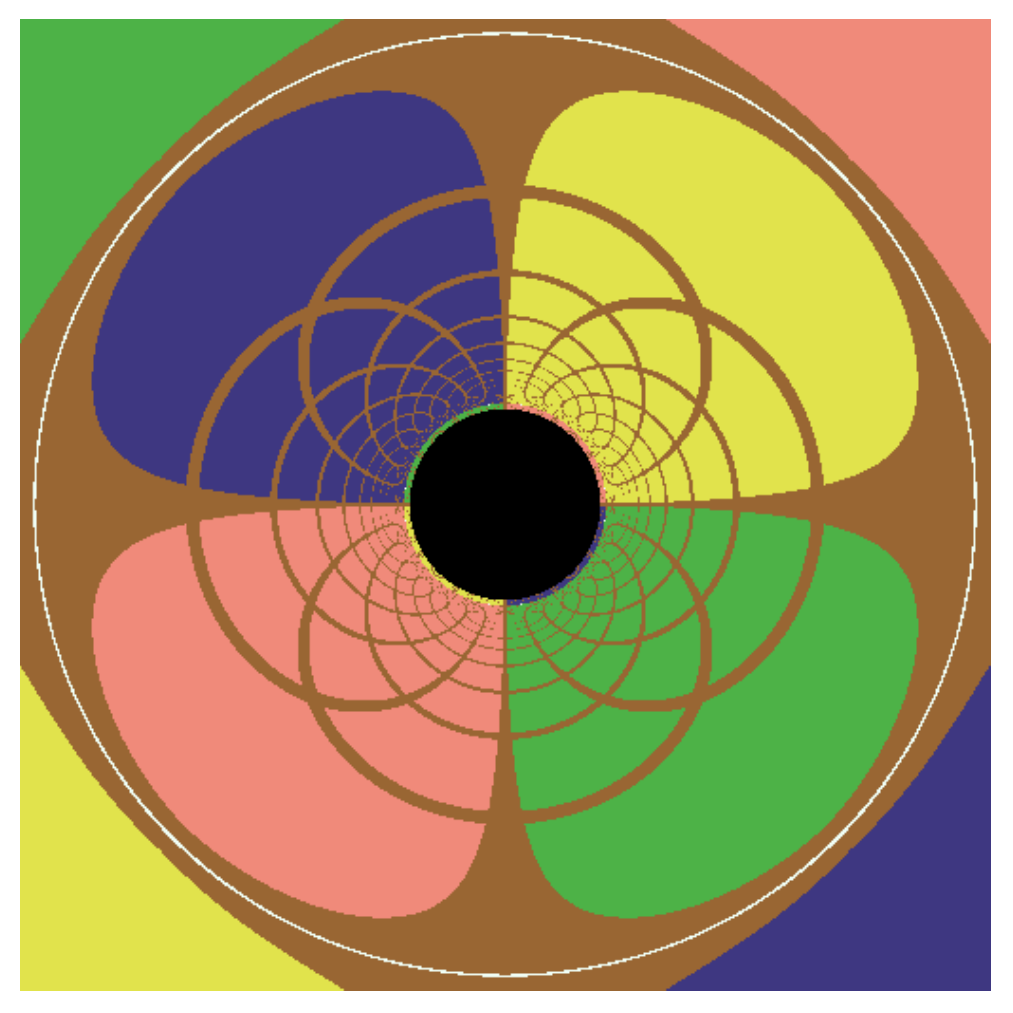}}
\subfigure[$v=4.0$]{\includegraphics[width=.2\textwidth]{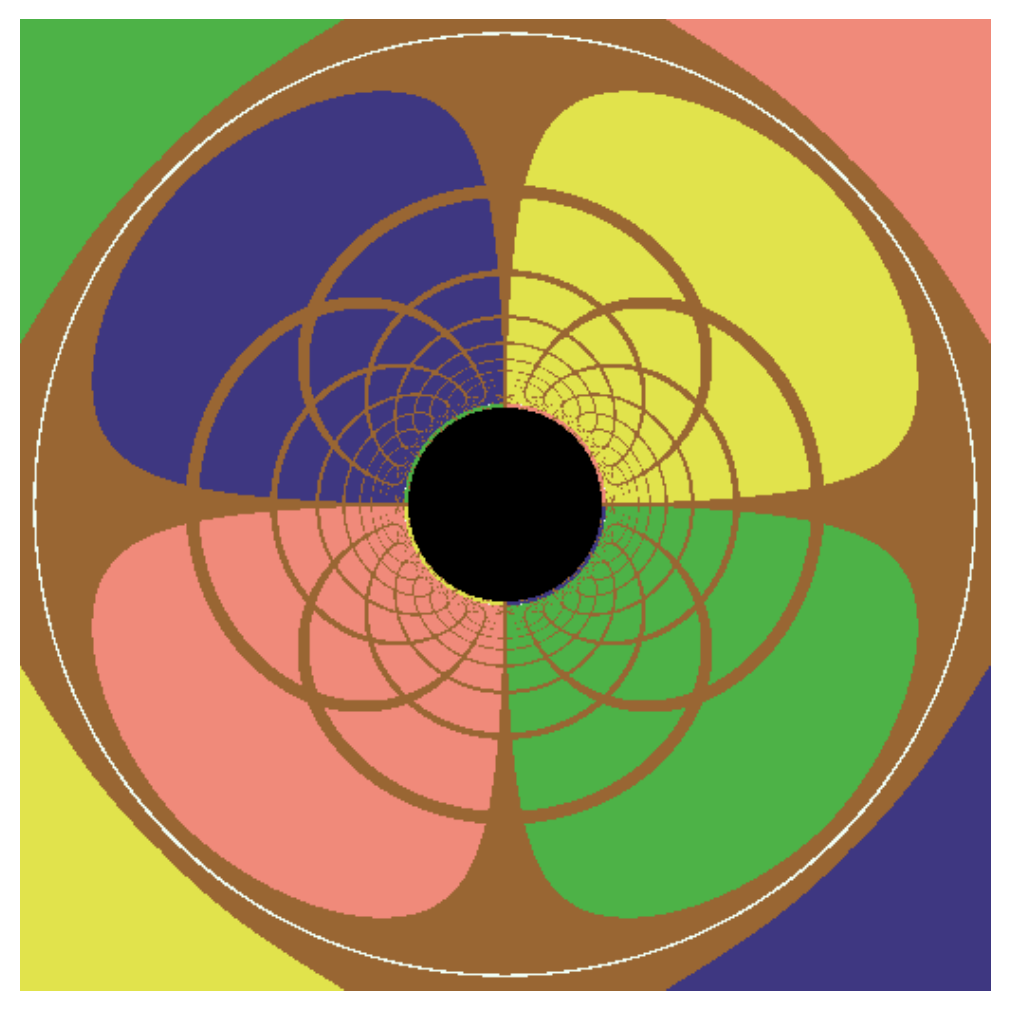}}
\caption{\label{fig1}  The gravitational lensing effect and the dynamic evolution characteristics of the shadow of the Vaidya black hole in the celestial sphere model. }
\end{figure}

During the actual numerical simulation process, two distinct operation schemes are present. The first approach considers each spacetime slice as a fixed background and directly computes the trajectory of the space time geodesic to investigate the lensing effect and shadow \cite{Liang:2025bbn}. Although the light trajectory obtained in this manner fails to capture the dynamic changes of spacetime, this assumption can still conduct a qualitative analysis of the changes in the gravitational lensing effect and shadow under different density states throughout the process. The second approach takes into account the time delay of light and performs the time-reversed geodesic evolution within the dynamic spacetime\cite{Zhang:2025xnl,Zhang:2023qxf}. In this approach, null geodesics are integrated backwards in time from the observer, with the metric components interpolated from the pre-computed numerical grid at each integration step. This allows us to track photon trajectories through the dynamically evolving spacetime. Unlike the first scheme, which can only provide a ``snapshot'' of the lensing structure at a fixed coordinate time, this dynamic ray-tracing method captures the continuous formation process of the shadow as it would be seen by a real distant observer. As demonstrated in the work \cite{Zhang:2025xnl}, such an approach is essential for revealing physically realistic features, such as the gradual growth of the shadow from a minute central dot, which is a key signature of time delay in a collapsing spacetime.

Given that spacetime is continuously evolving as light propagates in a dynamic spacetime, we will implement the second scheme. We will track the trajectories of light rays emitted from fixed spatial coordinates at different times and trace them backward along the time axis. The results are shown in Figure \ref{fig1}. In Figure \ref{fig1}, the white circles denote Einstein rings, which are a direct consequence of the gravitational lensing effect. The black disk signifies the shadow region of the black hole. It can be noted that as $v$ continues to increase, the size of the Einstein ring and the grid line structure of the gravitational lensing effect in the adjacent regions remain nearly unchanged. Nevertheless, the distribution of the pixel points near the shadow area and the grid  line structure have changed substantially. Since the black hole evolves smoothly from the initial state to the final asymptotically static configuration, the spacetime structures of both the initial and final black holes at infinity asymptotically approach the Schwarzschild spacetime. For an observer at infinity, both the increase in mass and the radius of the apparent horizon of a black hole resulting from the accretion process are extremely minute. Therefore, in the regions distant from the center of the black hole, the external gravitational lensing structure remains nearly unchanged.

Interestingly, as depicted in Figure \ref{fig1} (a), a small shadow region emerges at the center of the image during the early stage of accretion (or prior to the onset of accretion). This phenomenon differs from the gradually expanding shadow originating from a tiny central point, as described in work \cite{Zhang:2025xnl}.  In our model, a compact object (a low-mass black hole) exists from the outset, whereas in a collapse scenario, a black hole forms only at the end of the dynamical evolution. Consequently, the shadow is present throughout the entire accretion process, with its size evolving as the mass grows. Such a contrastive outcome can effectively determine whether the spacetime is undergoing a collapse or an accretion process. However, at this point, the black hole has not undergone significant matter accretion; thus, the distribution of pixel points near the shadow region and the grid line structure will not change much.  Another notable feature is that a new lensing ring emerges outside the shadow region during and after the active accretion period, a characteristic that is absent in the early stage of accretion, see Figure \ref{fig1} (g). During the accretion process, the mass density of the central black hole continuously increases, which strengthens its gravitational effect, enabling light to have the chance to orbit the center once. It can be observed that as the accretion process advances, the radius of the lens ring remains nearly constant, while its width exhibits an increasing trend. In the later stage of accretion, the width of the lens ring starts to decline, presenting a compressed effect, until it attains a stable state. When a black hole is in an asymptotically static configuration (either in the late stage of accretion or when accretion ceases), the distribution of image pixels and the lens structure within the image no longer experience substantial alterations, thereby presenting the static structure of a Schwarzschild black hole.

\section{Image of the  black hole with thin disk}
\label{sec3}
It is well established that the characteristics of a black hole's shadow depend not only on the underlying spacetime geometry but also critically on the nature of the surrounding accretion flow and the observational frequency band. Accordingly, a variety of accretion disk models have been developed, ranging from idealized configurations, such as celestial spheres and thin accretion disks, to more realistic descriptions like geometrically thick accretion flows, and general relativistic magnetohydrodynamic (GRMHD) simulations \cite{EventHorizonTelescope:2019pcy,McKinney:2012vh,Hou:2023bep}.

The central aim of this work is to probe the evolutionary signatures of dynamical spacetimes through their shadow evolution. Given this objective, it is methodologically advantageous to disentangle the pure gravitational effects from the intricate details of radiative transfer and magnetohydrodynamic processes inherent in more complex models. To isolate the key imprints of spacetime dynamics while retaining the essential physical content of an emitting source, we strategically adopt a simplified yet highly effective light source: a geometrically and optically thin accretion disk. This model serves as a clean probe, allowing us to systematically investigate how the shadow of a dynamic black hole evolves, free from the degeneracies introduced by more complicated accretion physics.

\subsection{Accretion disk and imaging method}
\begin{figure}[!h]
\makeatletter
\renewcommand{\@thesubfigure}{\hskip\subfiglabelskip}
\makeatother
\centering 
\subfigure[]{
\setcounter{subfigure}{0}\subfigure[]{\includegraphics[width=0.55\textwidth]{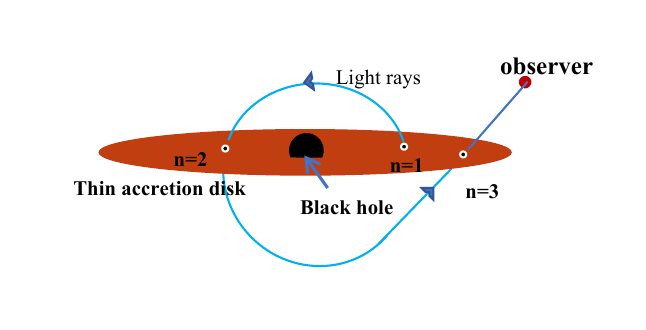}}}
\caption{\label{figAC1} Imaging the black hole with a thin accretion disks, where the black sphere represents the black hole, the red elliptical disk represents the thin accretion disk, and the blue curve indicates the complete path of the light that the observer can receive.}
\end{figure}

For the sake of simplicity, we postulate that the accretion disk is both geometrically and optically thin and resides precisely on the equatorial plane, as illustrated in Fig.~\ref{figAC1}. The inner edge of the thin disk extends all the way to the apparent event horizon of the black hole and expands in tandem with the dynamic evolution of the horizon. A crucial simplification in our model is that we do not distinguish between regions inside and outside the innermost stable circular orbit (ISCO). Given the dynamical nature of the Vaidya spacetime, the concept of a fixed ISCO becomes ambiguous. As a first step to capture the dominant effects of the time-varying geometry on photon propagation, we assume that all fluid elements in the disk, from the horizon outward, move on stable circular orbits. This approximation allows us to focus on the novel temporal features of the image while deferring a more detailed treatment of accretion dynamics to future work.

We employ the backward ray-tracing method to construct the image. Starting from a distant observer, light rays are traced back into the spacetime. As depicted in Figure~\ref{figAC1}, these rays may intersect the equatorial accretion disk once ($n=1$), twice ($n=2$), or more times ($n>2$). Herein, the first, second, and third intersections correspond to the direct image, the lensed image, and the higher order image, respectively. In fact, there exist distinct definitions of the concept of the photon ring. One definition delineates the photon ring as encompassing all images situated outside the direct image \cite{Gralla:2019drh,Hou:2022gge}. The other specifically employs the term photon ring to denote those images located outside both the direct image and the lens image\cite{Gralla:2019xty,He:2024amh}, i.e., the higher-order images. In our work, we adhere to the latter definition. The radial coordinate of the $n$-th intersection, $r_n$, defines the transfer function. Each crossing contributes to the total observed intensity, which is accumulated as
\begin{align} \label{IO1}
\mathcal{I}_{\text{obs}} = \sum_{n=1}^{N} f_n \, g_n^3 \, \mathcal{J}(r_n),
\end{align}
where $f_n$ is the fudge factor which is fixed to $f_n=1$. Taking into account that the black hole image captured by the EHT is observed at a wavelength of 1.3 mm (230 GHz), one can select the emissivity of the thin disk to be a  second order polynomial in log-space, which is
\begin{align} \label{EM1}
\mathcal{J}(r)=\exp \left[-\frac{1}{2}\mathcal{Z}^2-2 \mathcal{Z}\right], \qquad \mathcal{Z}=\log\frac{r}{r_{AH}}.
\end{align}
The redshift factor $g_n$ for a photon emitted from a circularly orbiting fluid element at radius $r_n$ is defined as the ratio of the observed frequency $\nu_{\text{obs}}$ to the emitted frequency $\nu_n$. In the above equation, the third power of the redshift shadow $g_n$ is employed, as it corresponds to the intensity at the specific frequency of 230 GHz \cite{Wang:2023fge}.  In the work of Gralla\cite{Gralla:2020srx}, the application of the fourth power form of the redshift factor $g_n$ was also elaborately introduced. The redshift factor can be calculated from the four-momentum of the photon and the four-velocity of the emitter, that is
\begin{align} \label{RF_new}
g_n = \frac{\nu_{\text{obs}}}{\nu_n} .
\end{align}
For a distant observer in an asymptotically flat spacetime, the observed photon energy can be normalized, i.e., $(p_{\mu} u^{\mu})|_{\text{obs}} = 1$. The four-velocity of the emitter on a circular orbit in the equatorial plane is
\begin{align} \label{FV}
u^{\mu} = u^t (1,0,0,\Omega),
\end{align}
where $\Omega = d\phi/dt$ is the angular velocity. Consedering the condition $g_{\mu\nu}u^{\mu}u^{\nu} = -1$, one can obtain that
\begin{align} \label{FV2}
u^t = 1/\sqrt{-g_{tt} - 2g_{t\phi}\Omega - g_{\phi\phi}\Omega^2}.
\end{align}
For the spherically symmetric spacetime ($g_{t\phi}=0$), the angular velocity for a circular geodesic simplifies to
\begin{align} \label{CG}
\Omega(r,v) = \sqrt{-\partial_r g_{tt} / \partial_r g_{\phi\phi}},
\end{align}
The above equation  is a function of both the radial coordinate $r$ and the advanced time $v$ due to the spacetime's dynamics. Consequently, the redshift factor $g_n$ for a given image point carries an implicit dependence on the time at which the light ray intersected the disk, thereby encoding the temporal evolution of the black hole.  At this stage, selecting a suitable imaging model enables the numerical simulation of the visual appearance of a dynamic black hole on a display screen. Regarding the imaging model, we will adopt the imaging method elaborated in \cite{Hou:2022eev}, which is commonly known as the fisheye lens camera model. Through the application of stereographic projection technology, the photons detected on the image plane correspond to the optical perspective of the observer, who can be likened to the camera.

\subsection{The observational appearance of dynamic black holes}
\begin{figure}[!t]
\centering 
\subfigure[$v=0$]{\includegraphics[width=.2\textwidth]{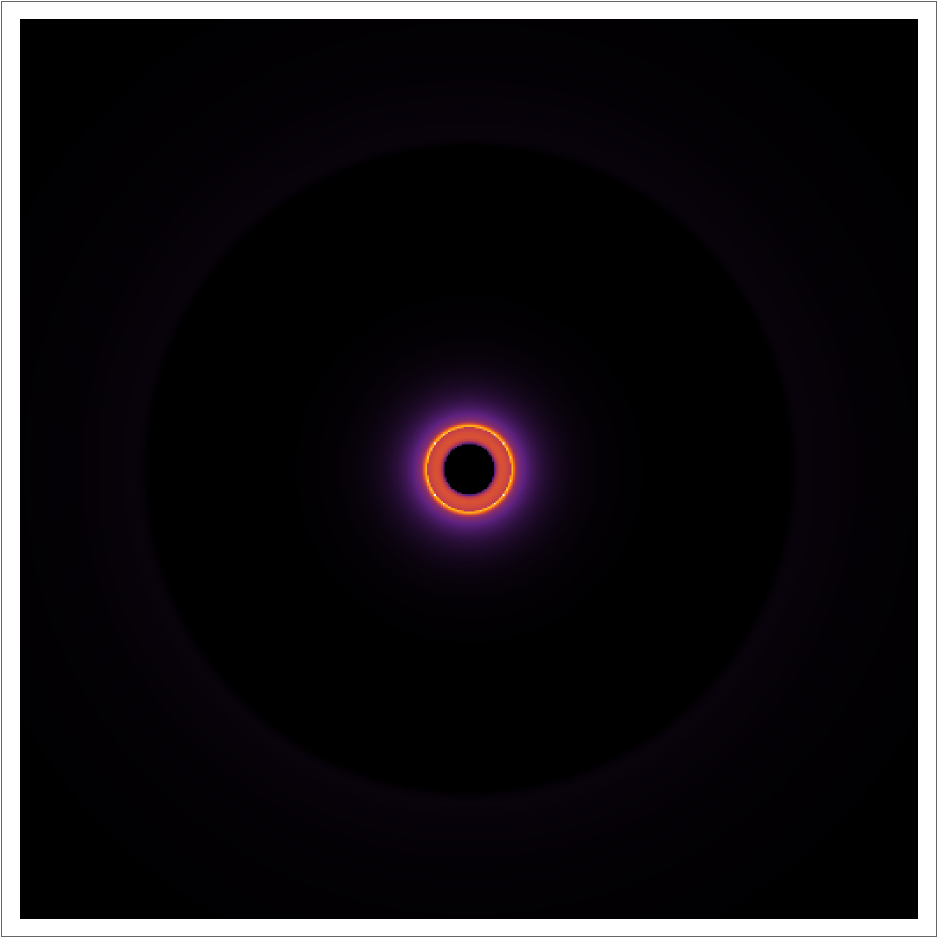}}
\subfigure[$v=0.1$]{\includegraphics[width=.2\textwidth]{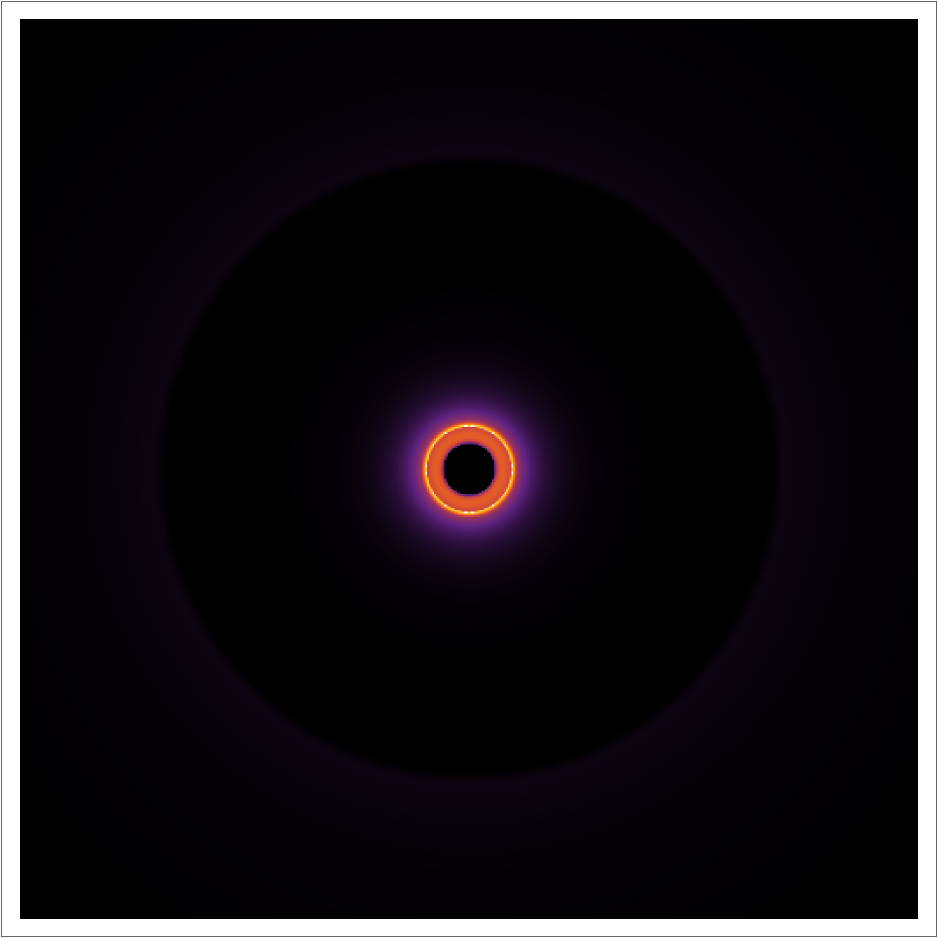}}
\subfigure[$v=0.3$]{\includegraphics[width=.2\textwidth]{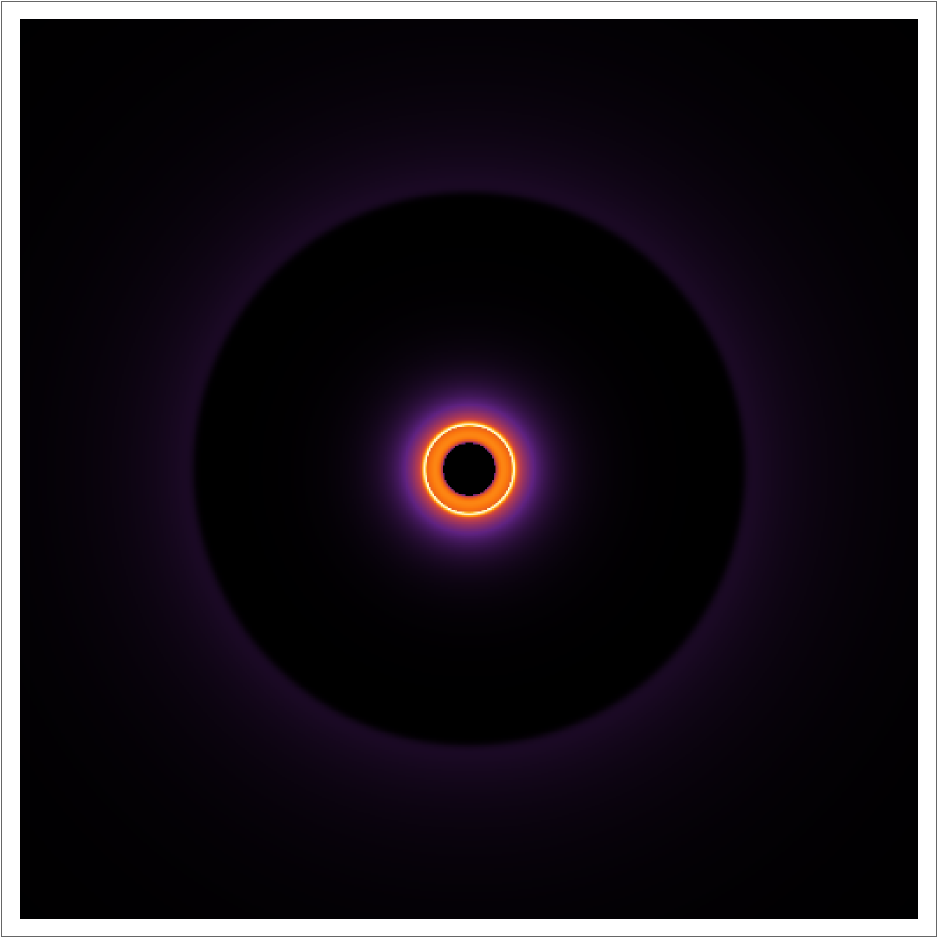}}
\subfigure[$v=0.5$]{\includegraphics[width=.2\textwidth]{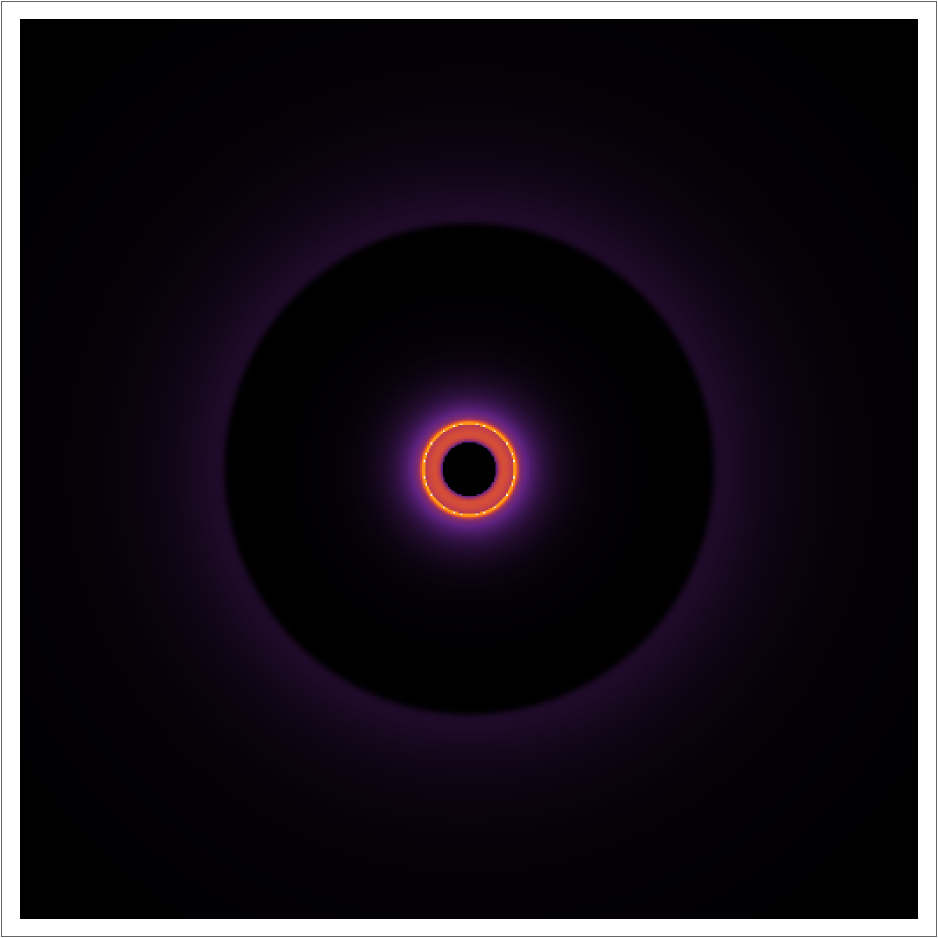}}
\subfigure[$v=0.8$]{\includegraphics[width=.2\textwidth]{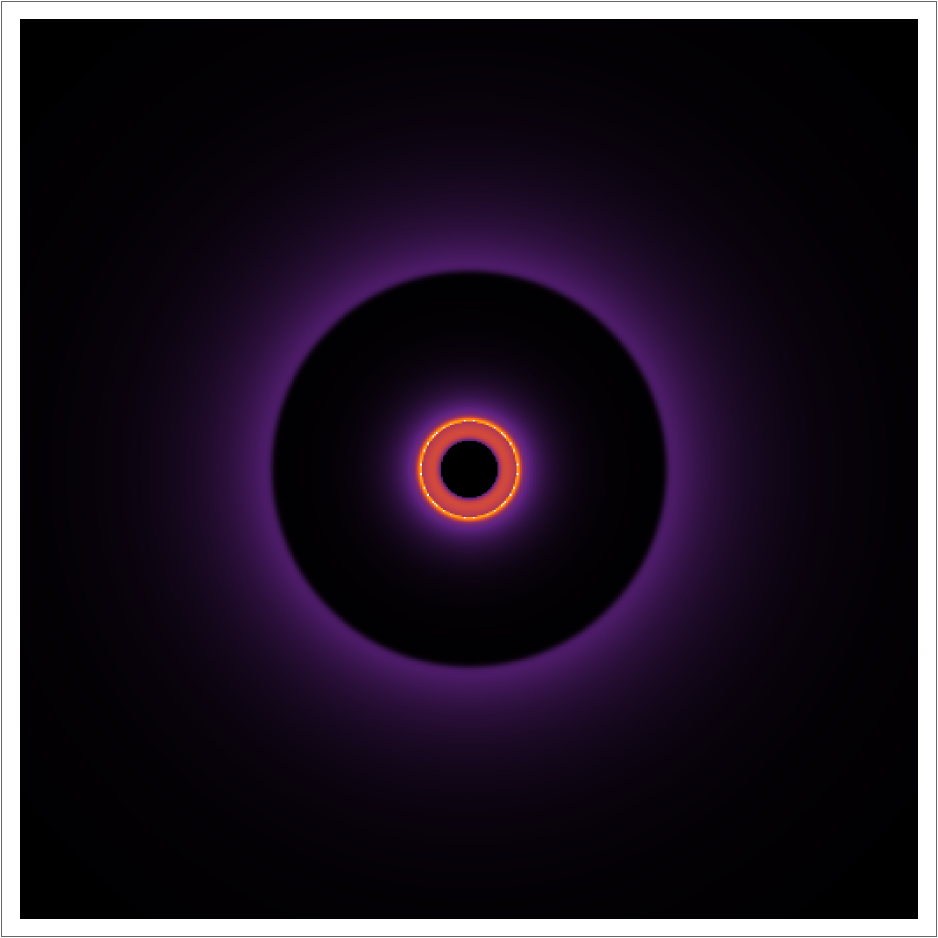}}
\subfigure[$v=1.1$]{\includegraphics[width=.2\textwidth]{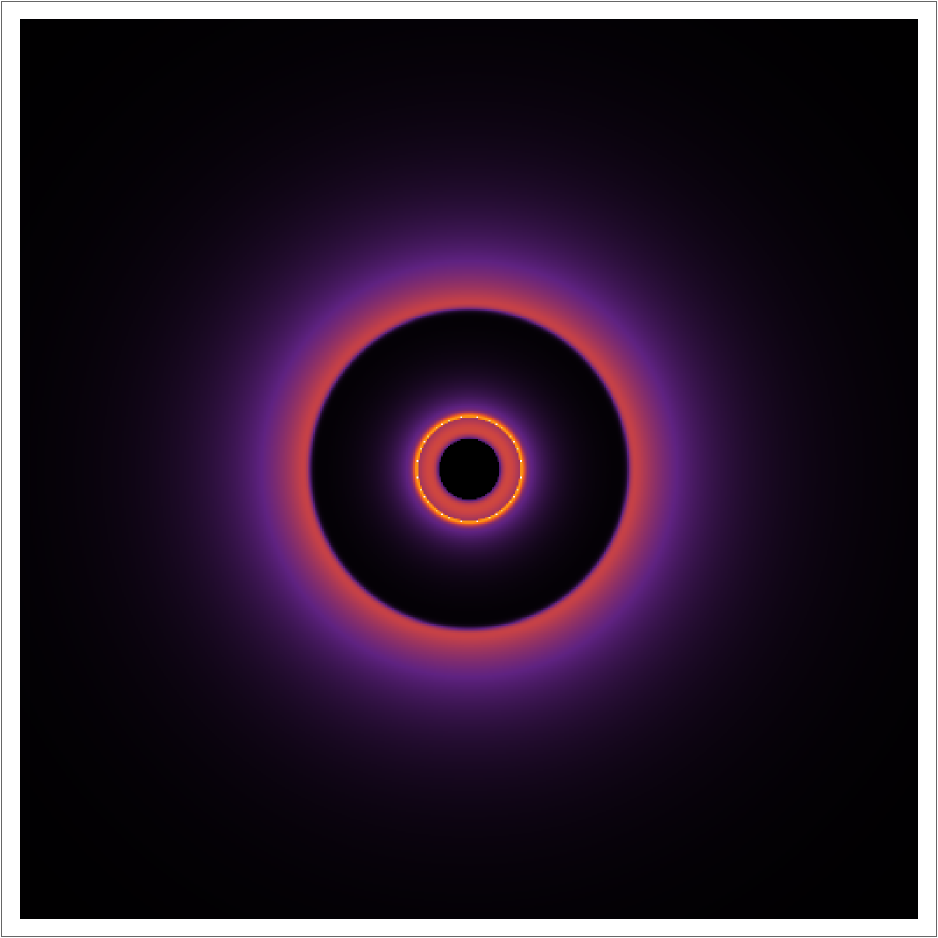}}
\subfigure[$v=1.5$]{\includegraphics[width=.2\textwidth]{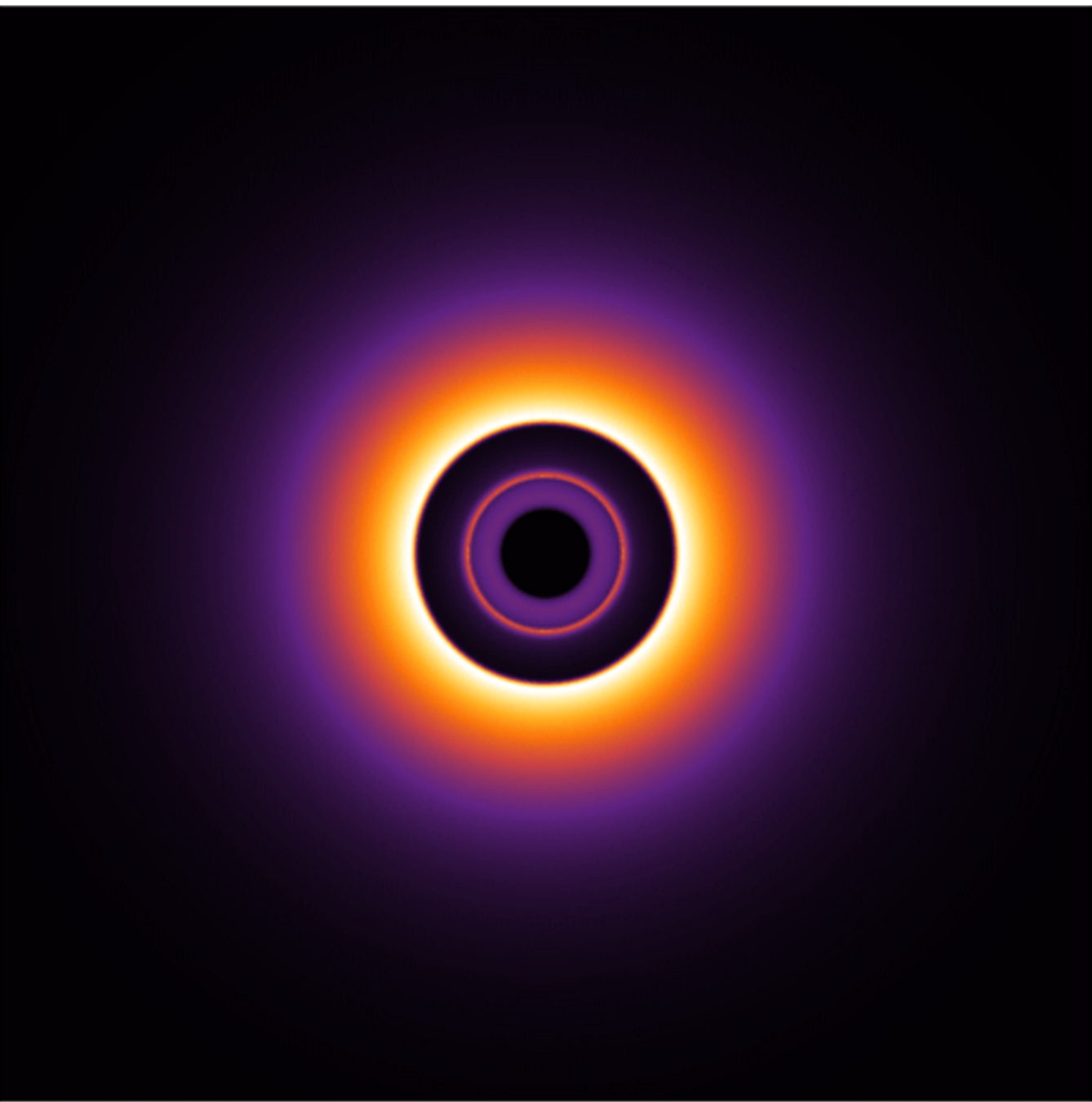}}
\subfigure[$v=2.0$]{\includegraphics[width=.2\textwidth]{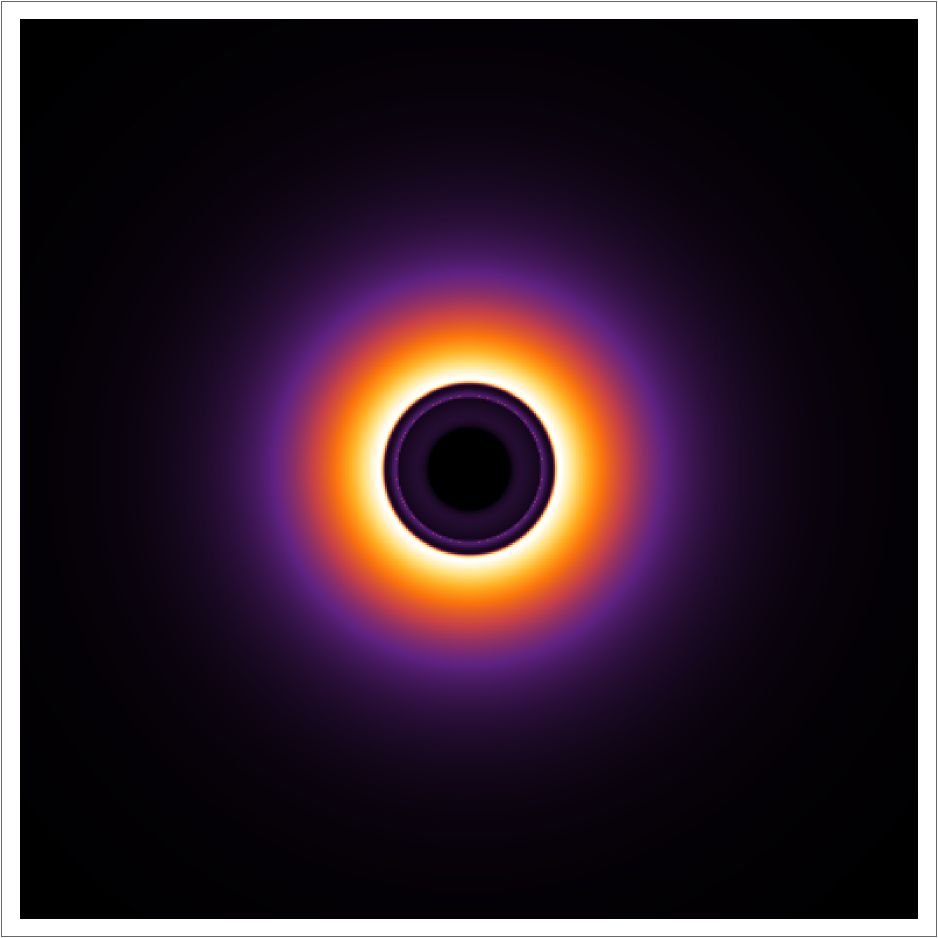}}
\subfigure[$v=2.3$]{\includegraphics[width=.2\textwidth]{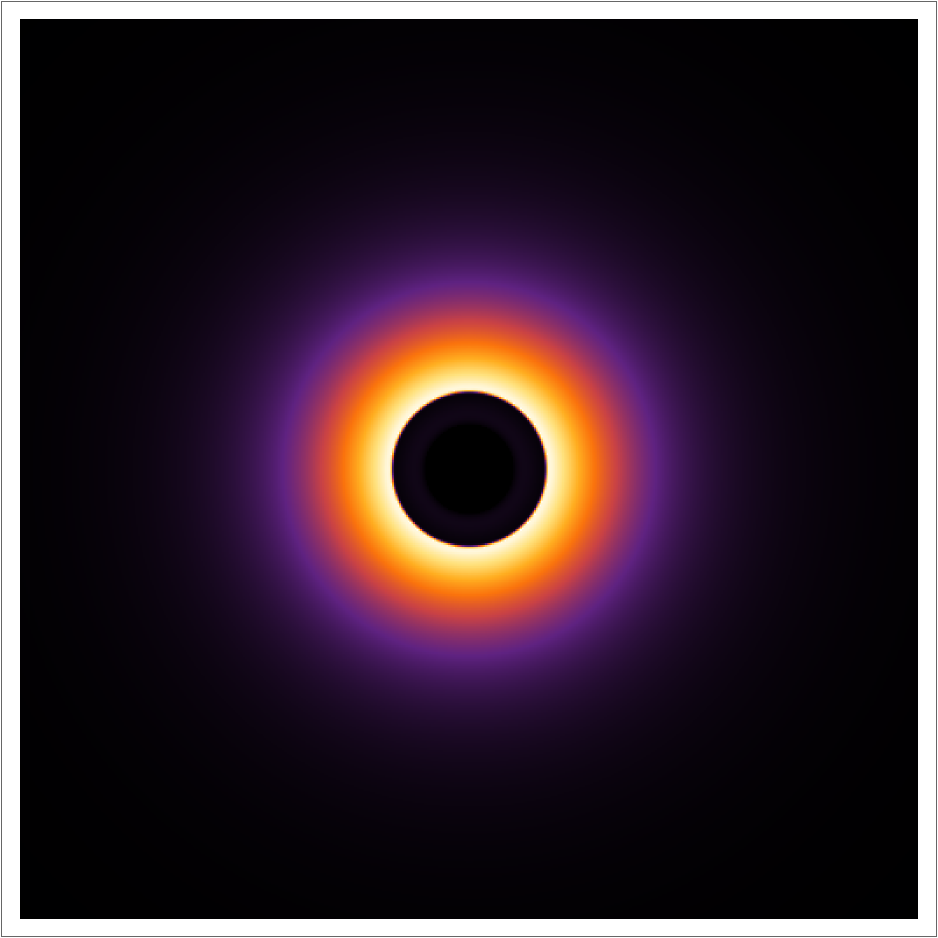}}
\subfigure[$v=2.6$]{\includegraphics[width=.2\textwidth]{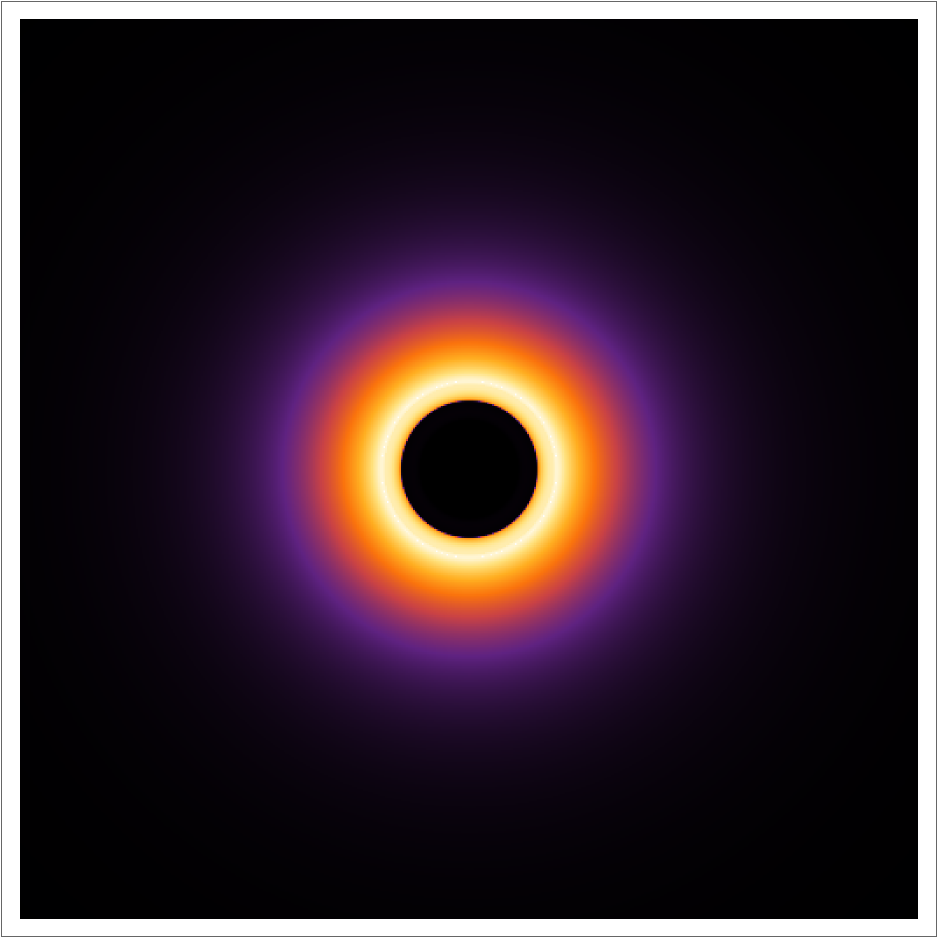}}
\subfigure[$v=3.0$]{\includegraphics[width=.2\textwidth]{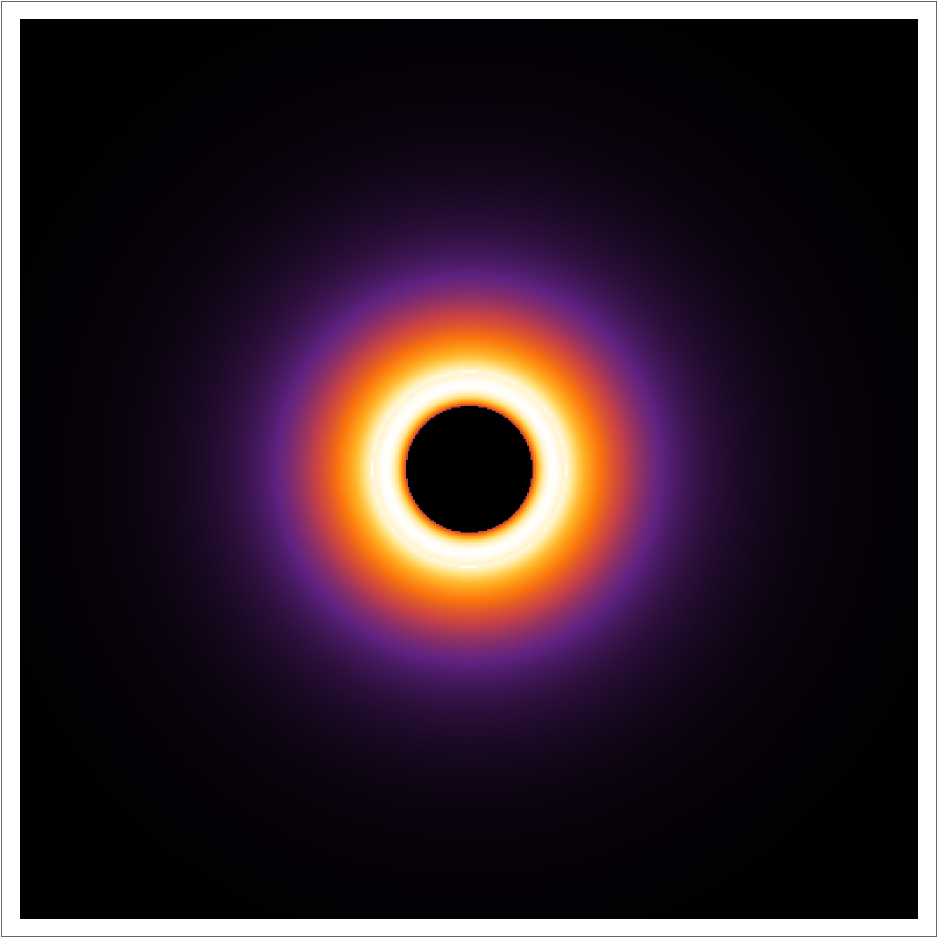}}
\subfigure[$v=3.5$]{\includegraphics[width=.2\textwidth]{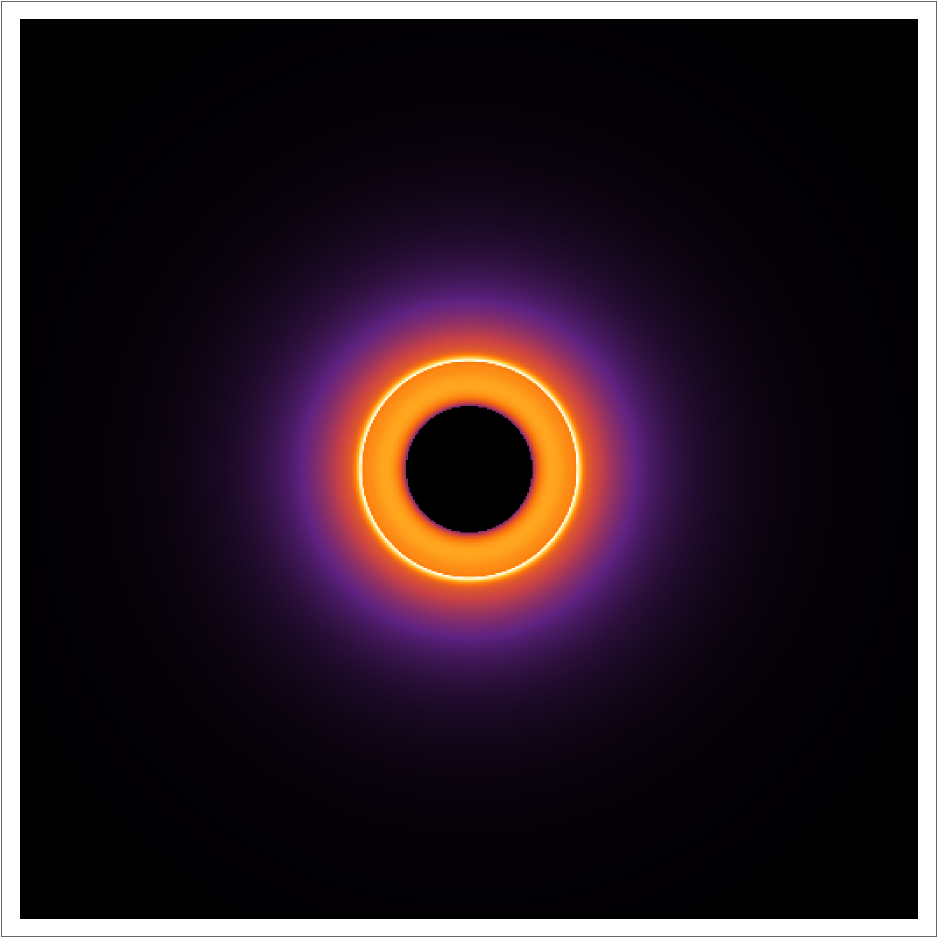}}
\caption{\label{figTD0}  In the thin accretion disk model, the evolutionary characteristics of the optical observational appearance of the Vaidya black hole, with the observational inclination angle being $\theta_{obs}= 0^{\circ}$ . }
\end{figure}

In the context of thin disk accretion, Figure \ref{figTD0} shows the evolution of the optical observational appearance of the Vaidya black hole. Here, the observer's position is set at $r_{obs} = 500M_0$, and is located at the pole with $\theta_{obs} = 0^{\circ}$. As depicted in Figure \ref{figTD0}, there is consistently a black, axially symmetric circular area at the center of the screen, which represents the inner shadow region of the black hole. Moreover, as the accretion process progresses, the inner shadow region exhibits continuous expansion until it attains a stable state. In the initial state of a black hole (before the accretion process begins), there exists a bright ring around the inner shadow of the black hole, which is formed due to the overlap between the lens ring and the photon ring, see Figure \ref{figTD0} (a). This result is consistent with the previous one obtained when a static spherically symmetric black hole is encircled by a thin accretion disk. As the accretion process commences, the bright ring outside the inner shadow region endures and demonstrates a growth tendency; however, its brightness displays a weakening tendency, as depicted in Figure \ref{figTD0} (b)-(f).

On the other hand, an additional, less prominent annular structure appears in the more distant region of the bright ring, as shown in Figure \ref{figTD0} (c). Interestingly, as the accretion process advances, this additional ring not only becomes progressively brighter but also exhibits an evolutionary characteristic of contracting towards the inner shadow. When tracing the light path in the reverse direction of time, the photon energy $E(v)$ is not conserved along the null geodesic due to the time-dependence of the metric. It varies with the advanced time $v$ at which the photon propagates through different regions of the dynamical spacetime. This variation in photon energy, analogous to cosmological redshift but arising from the mass evolution, induces an additional frequency shift, which we term dynamical redshift$\footnote{This effect is induced by the dynamic evolution of the background spacetime and is analogous to cosmological redshift. To prevent confusion, it is referred to as dynamical redshift.}$.  This additional ring-shaped structure cannot be attributed to the standard classification of images (direct, lensed, or photon ring) based on the number of disk intersections. Instead, it is a distinctive feature arising from the dynamical redshift effect, encoding the temporal evolution of the spacetime.

During the active accretion period of a black hole, the bright ring outside the inner shadow gradually fades away, and the direct image predominantly appears on the screen, as demonstrated by  Figure \ref{figTD0} (g)-(j). Due to the significant dynamic alterations in spacetime, there exists no stable photon circular orbit beyond the black hole. In other words, a photon ring structure cannot be formed. Consequently, no distinct bright ring structure emerges in the adjacent region outside the inner shadow. In addition, the extra ring has contracted to the outer edge of the direct image. It is worthy of note that during the later stage of black hole accretion, as the black hole approaches a static configuration, the bright ring structure outside the inner shadow reappears and remains in a stable state, see  Figure \ref{figTD0} (l).

\begin{figure}[!t]
\centering 
\subfigure[$v=0$]{\includegraphics[width=.2\textwidth]{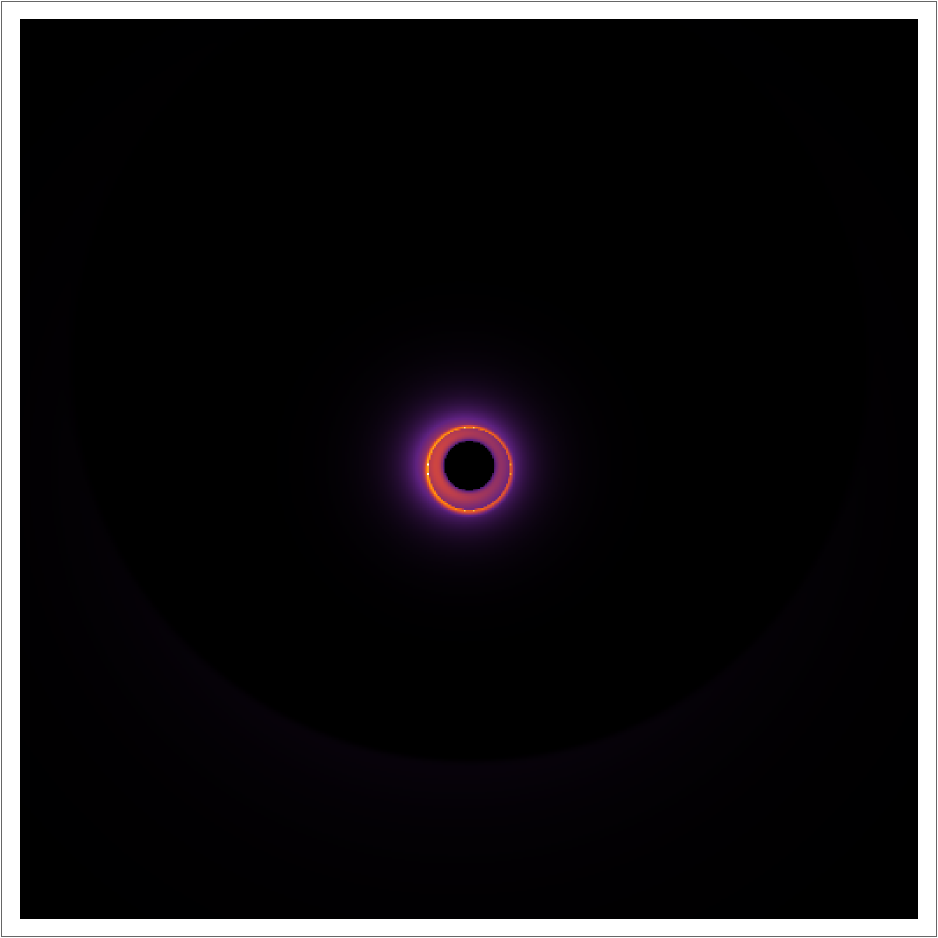}}
\subfigure[$v=0.1$]{\includegraphics[width=.2\textwidth]{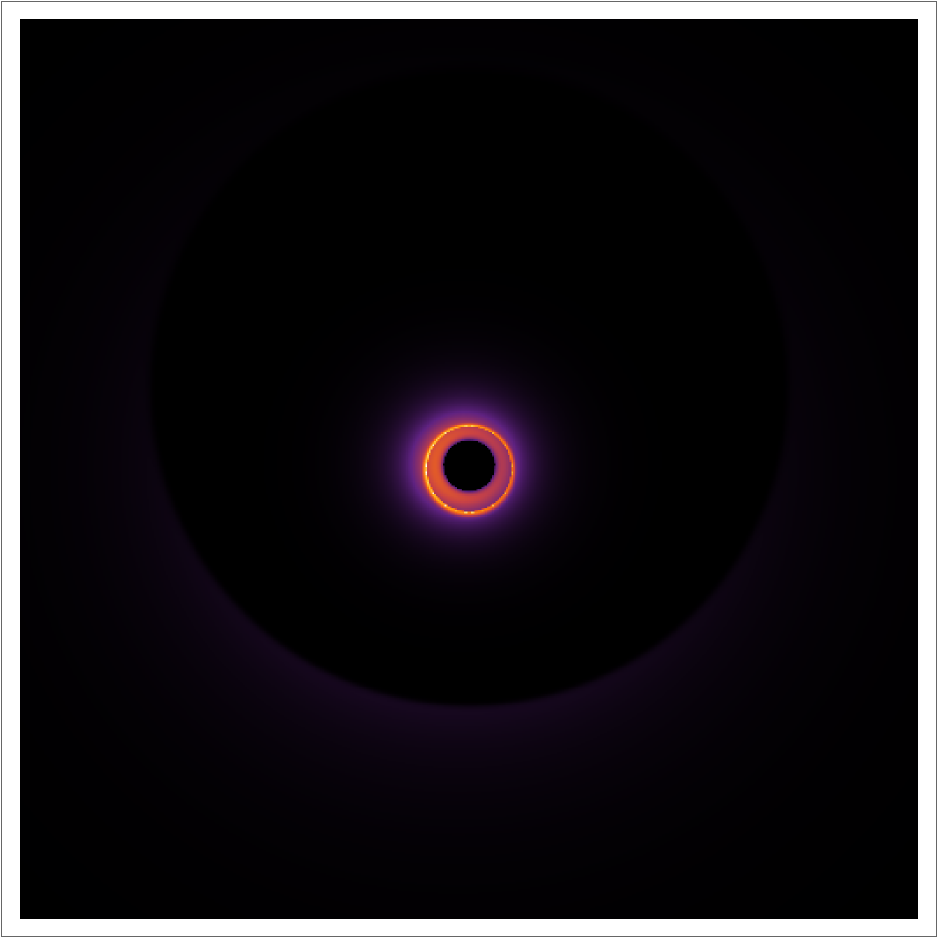}}
\subfigure[$v=0.3$]{\includegraphics[width=.2\textwidth]{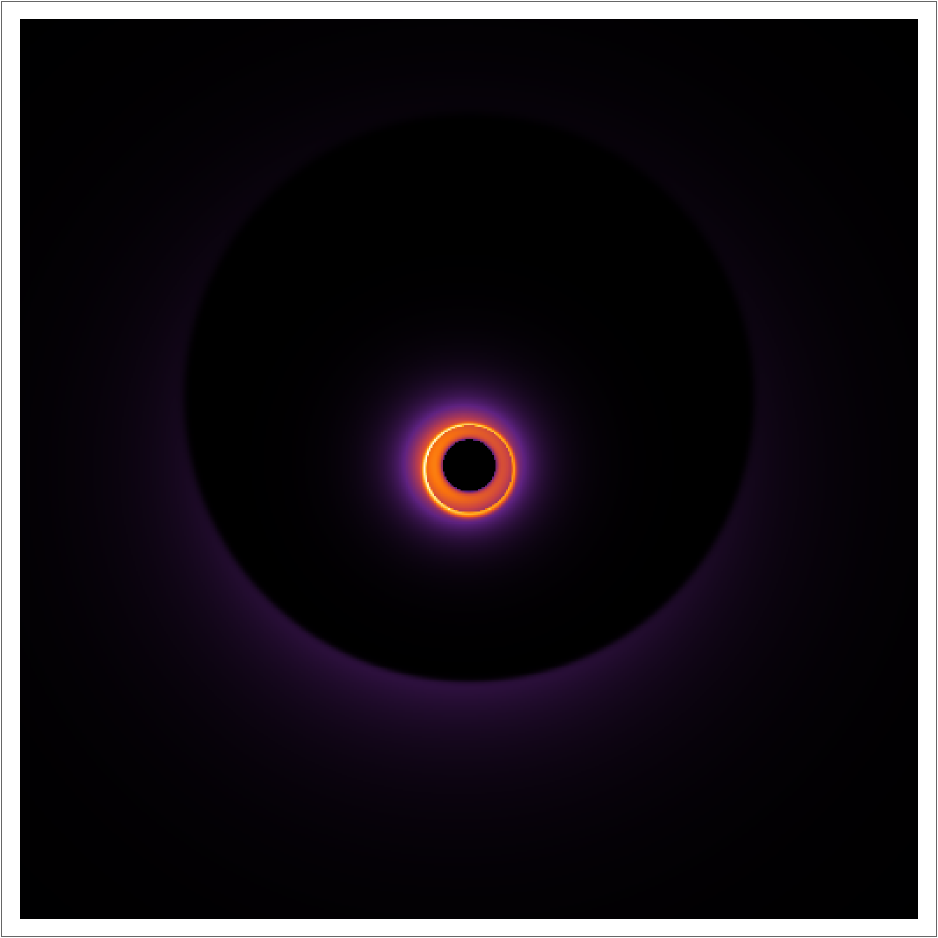}}
\subfigure[$v=0.5$]{\includegraphics[width=.2\textwidth]{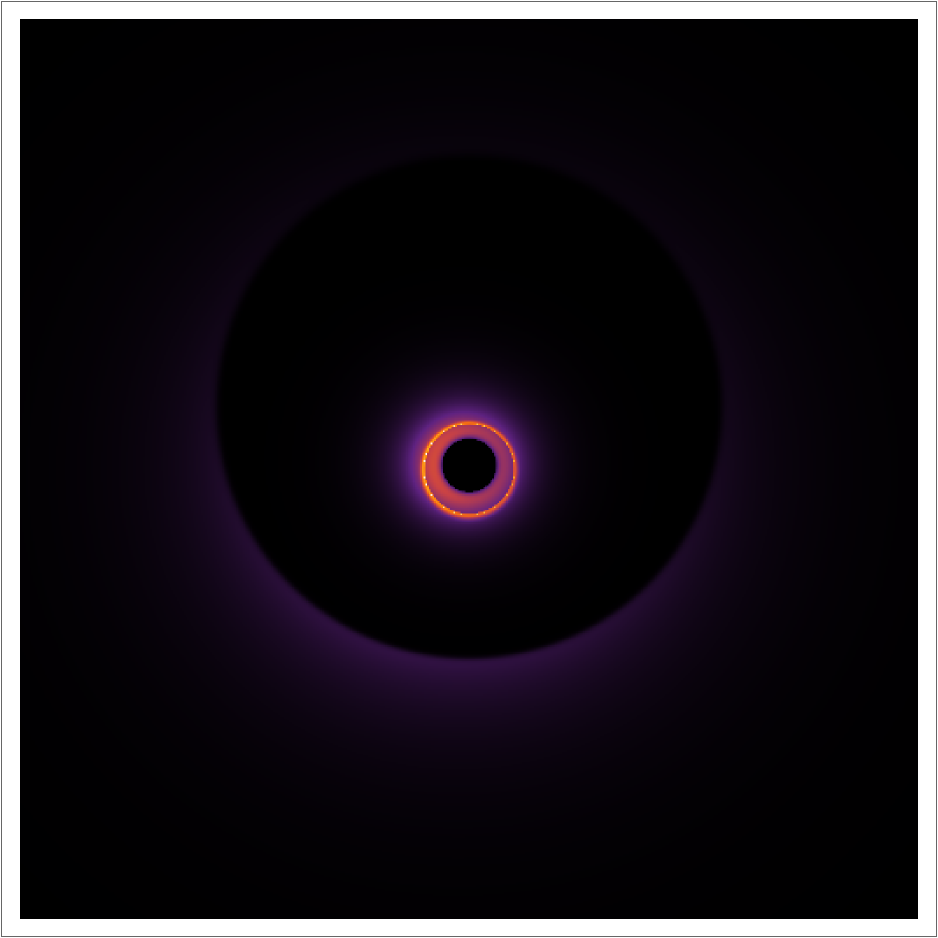}}
\subfigure[$v=0.8$]{\includegraphics[width=.2\textwidth]{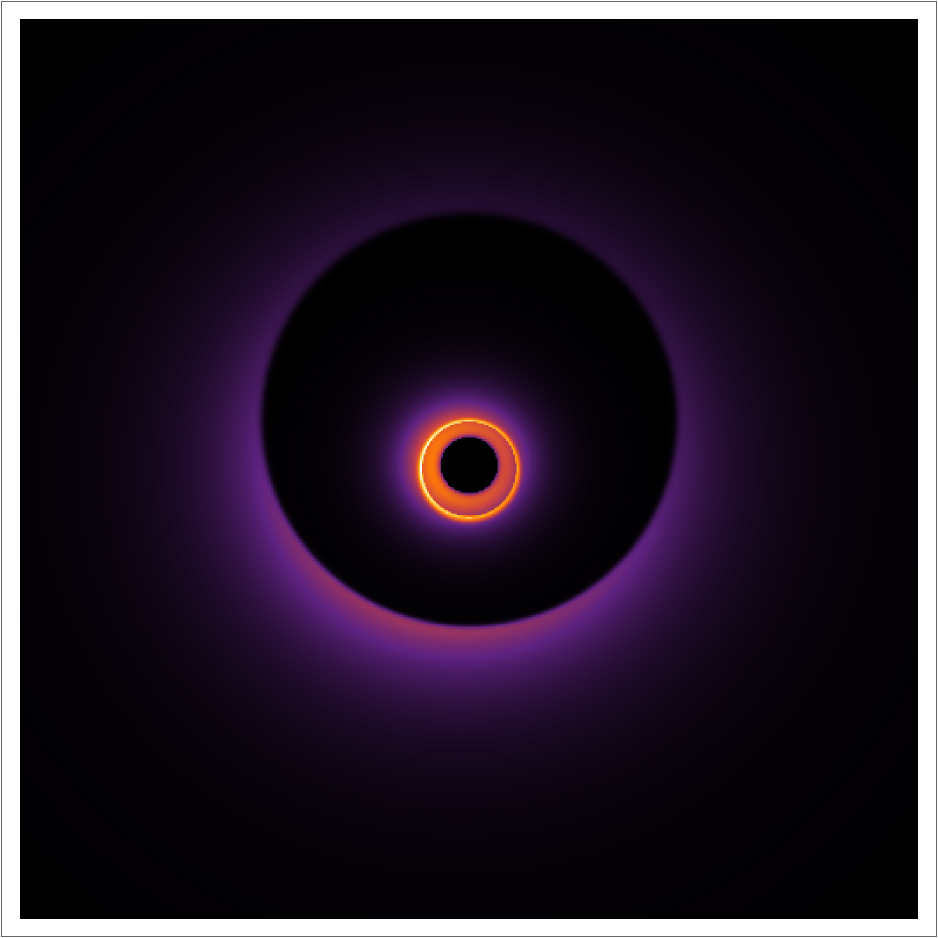}}
\subfigure[$v=1.1$]{\includegraphics[width=.2\textwidth]{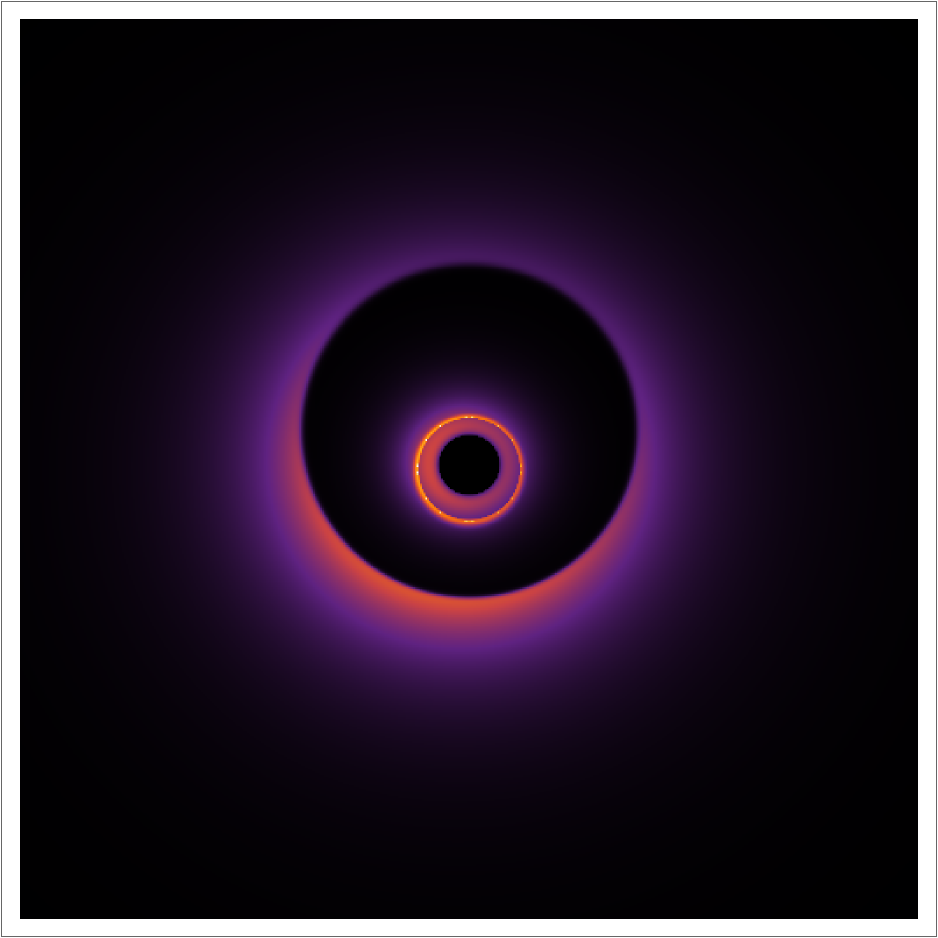}}
\subfigure[$v=1.5$]{\includegraphics[width=.2\textwidth]{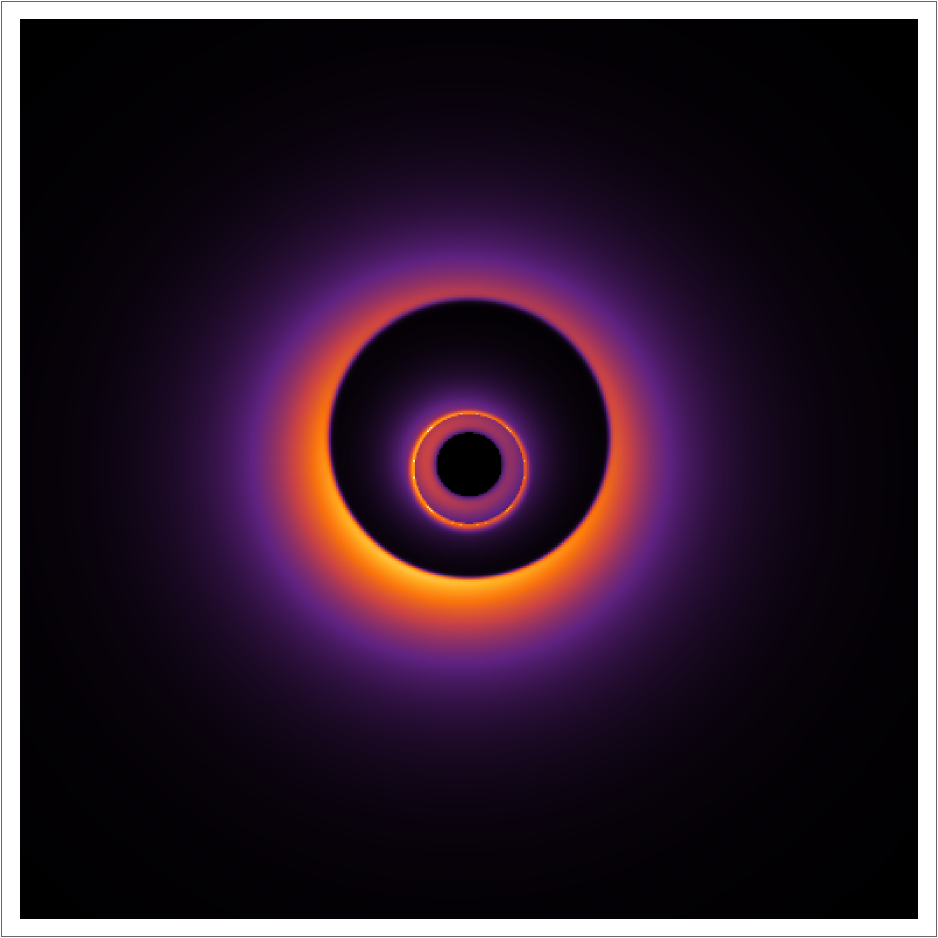}}
\subfigure[$v=2.0$]{\includegraphics[width=.2\textwidth]{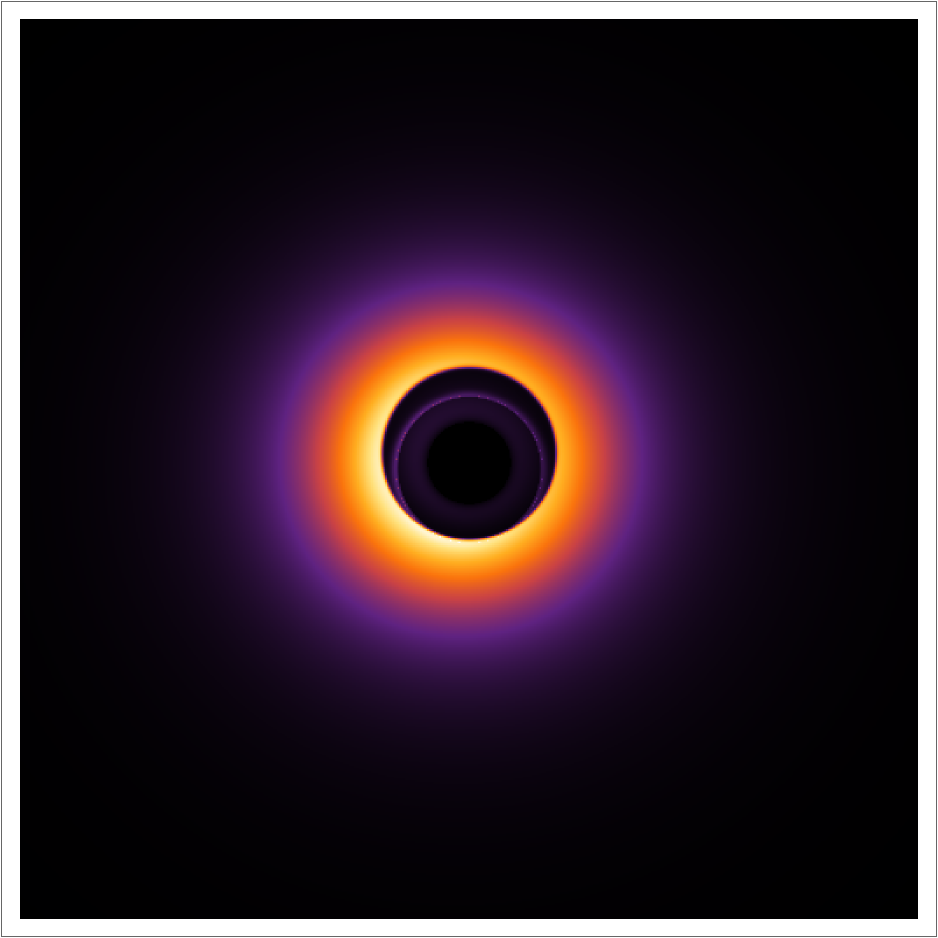}}
\subfigure[$v=2.3$]{\includegraphics[width=.2\textwidth]{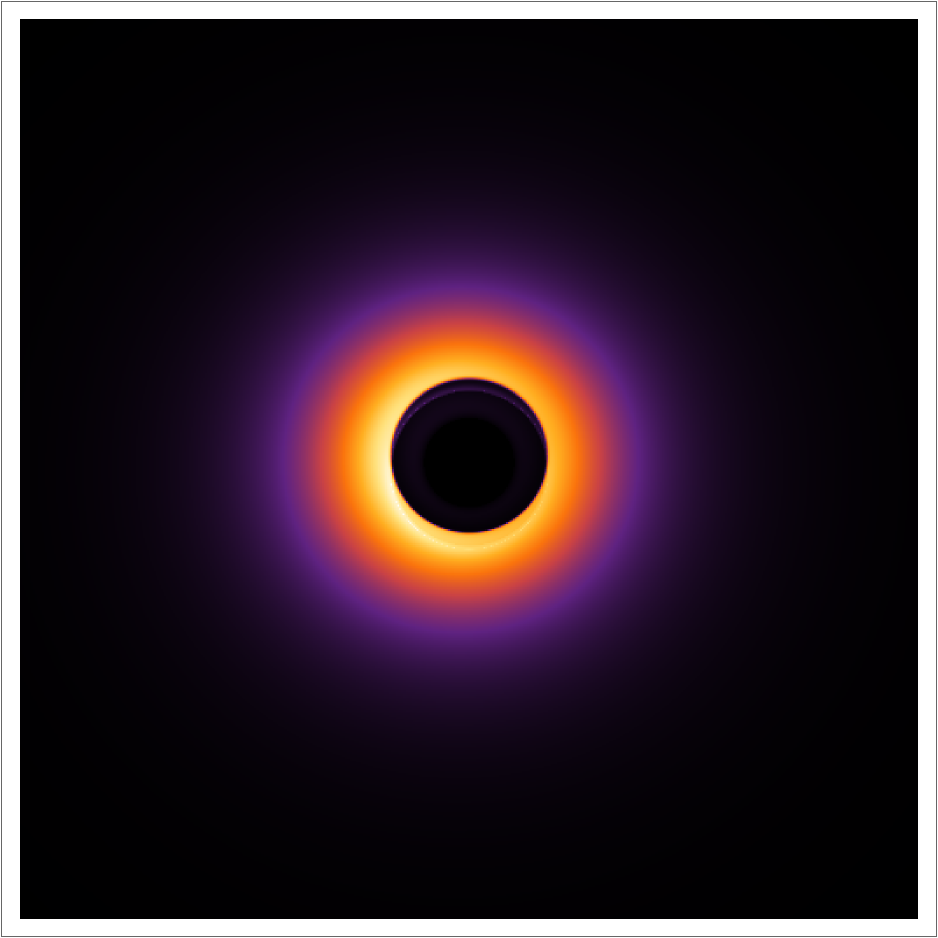}}
\subfigure[$v=2.6$]{\includegraphics[width=.2\textwidth]{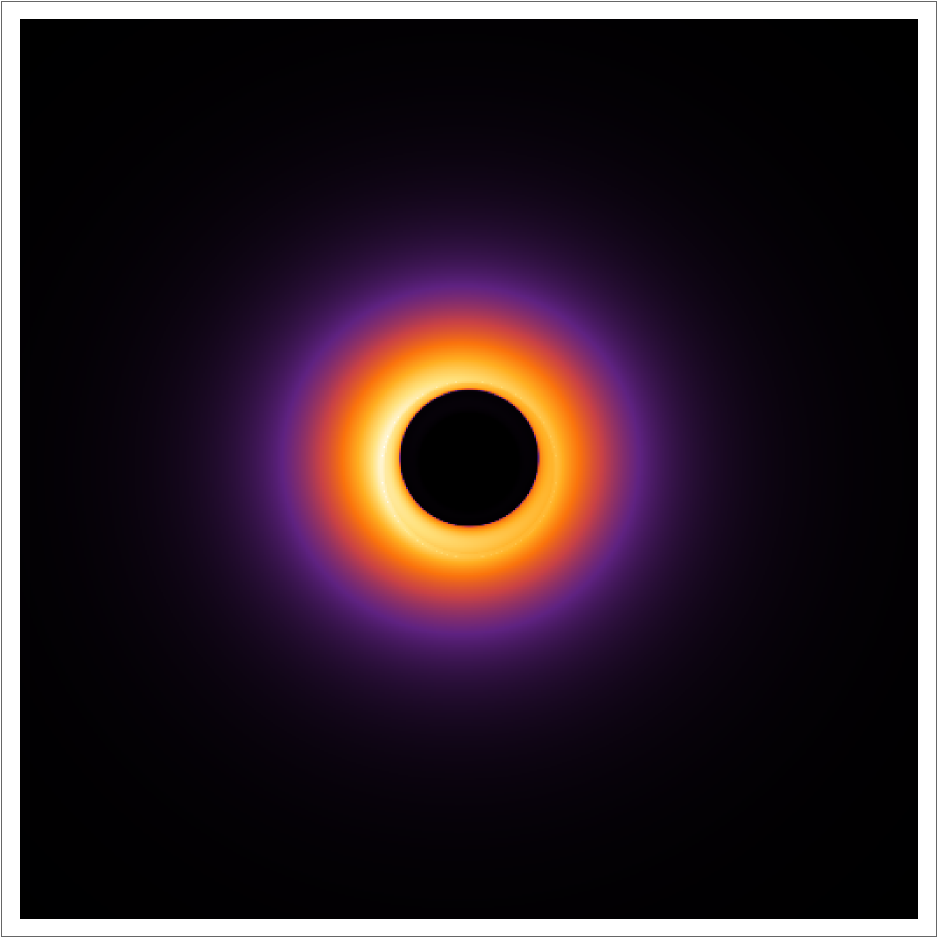}}
\subfigure[$v=3.0$]{\includegraphics[width=.2\textwidth]{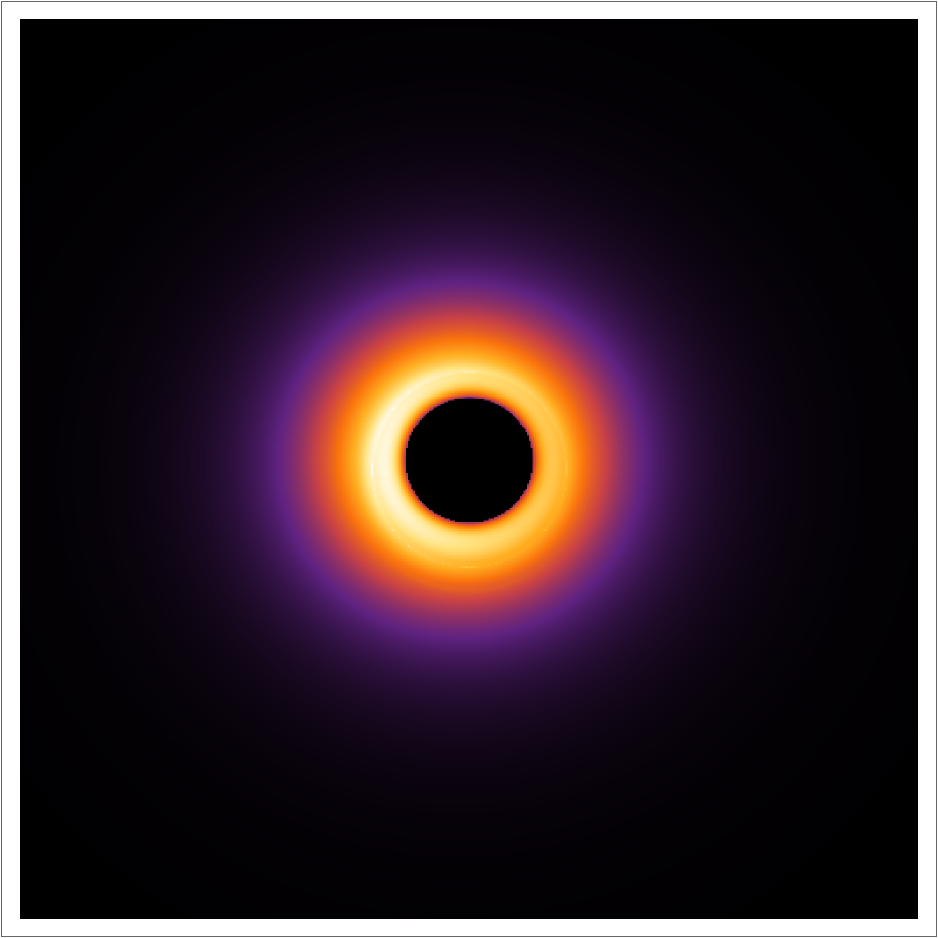}}
\subfigure[$v=3.5$]{\includegraphics[width=.2\textwidth]{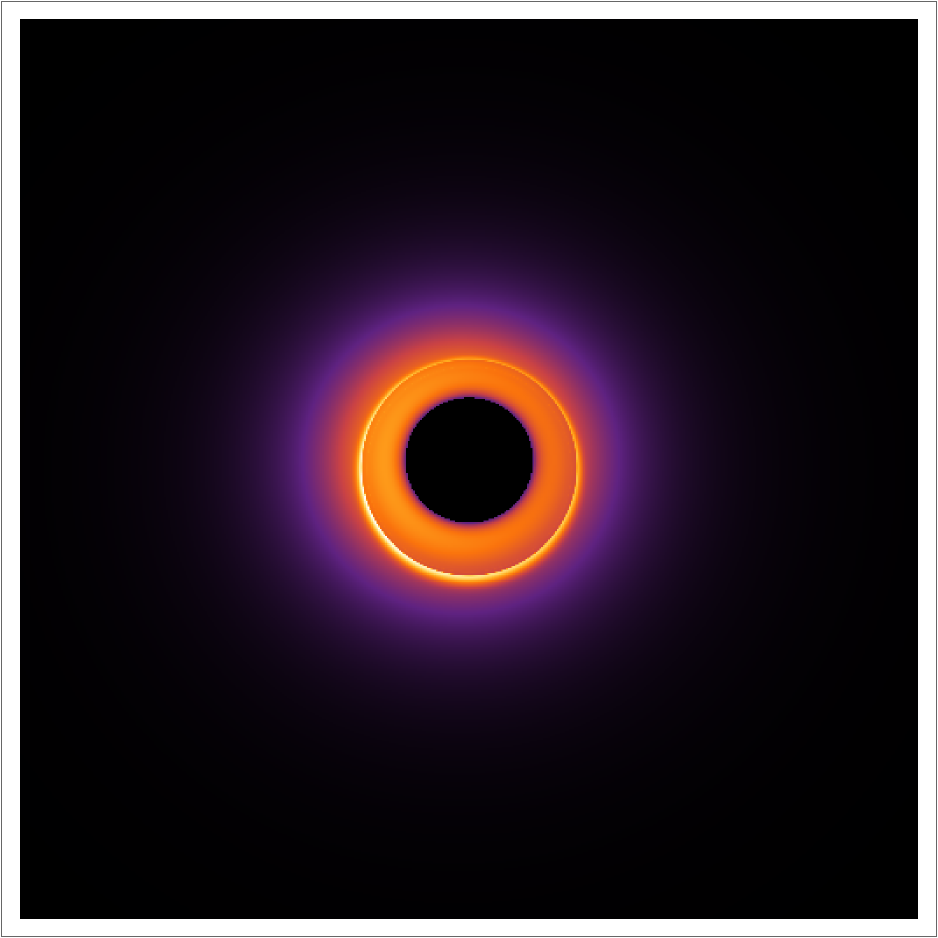}}
\caption{\label{figTD17}  In the thin accretion disk model, the evolutionary characteristics of the optical observational appearance of the Vaidya black hole, with the observational inclination angle being $\theta_{obs}= 17^{\circ}$ .}
\end{figure}

\begin{figure}[!t]
\centering 
\subfigure[$v=0$]{\includegraphics[width=.2\textwidth]{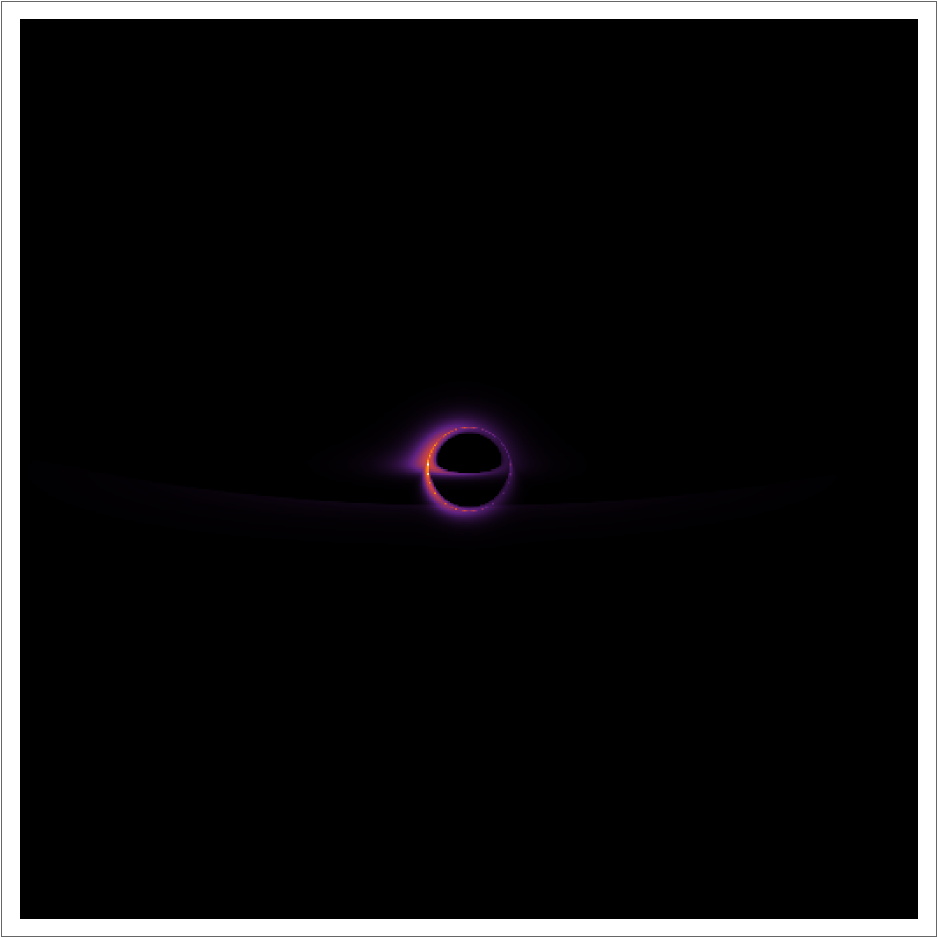}}
\subfigure[$v=0.1$]{\includegraphics[width=.2\textwidth]{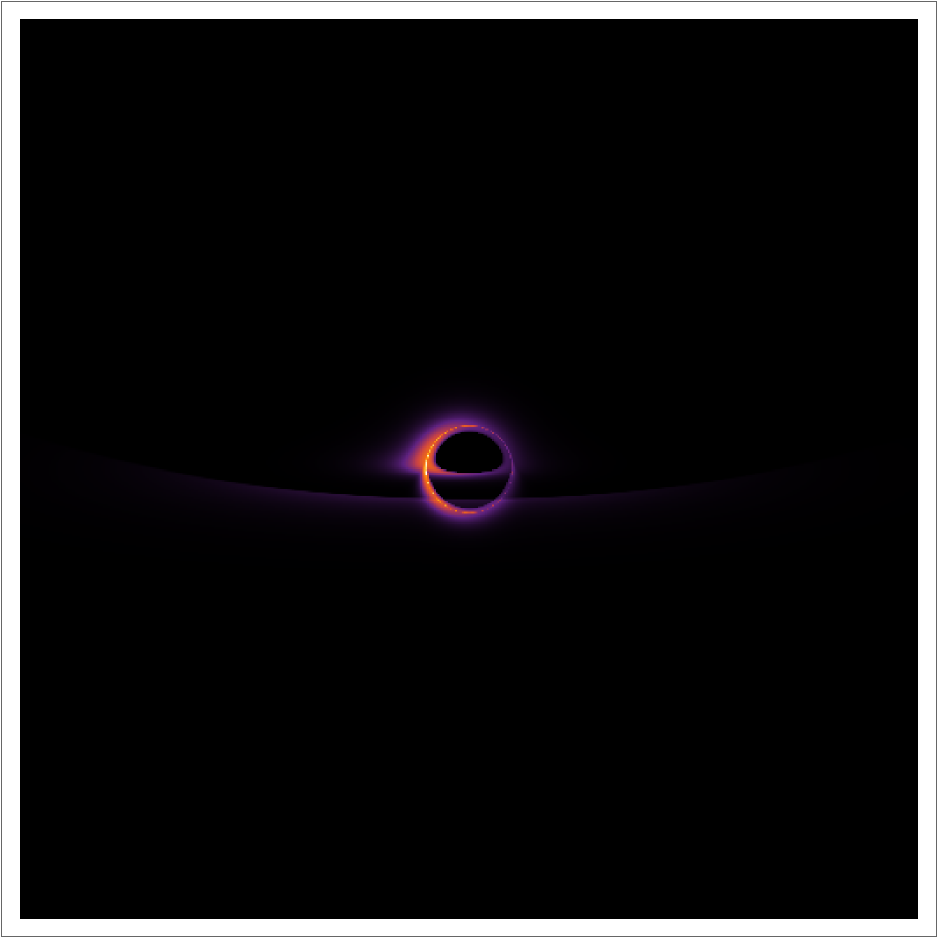}}
\subfigure[$v=0.3$]{\includegraphics[width=.2\textwidth]{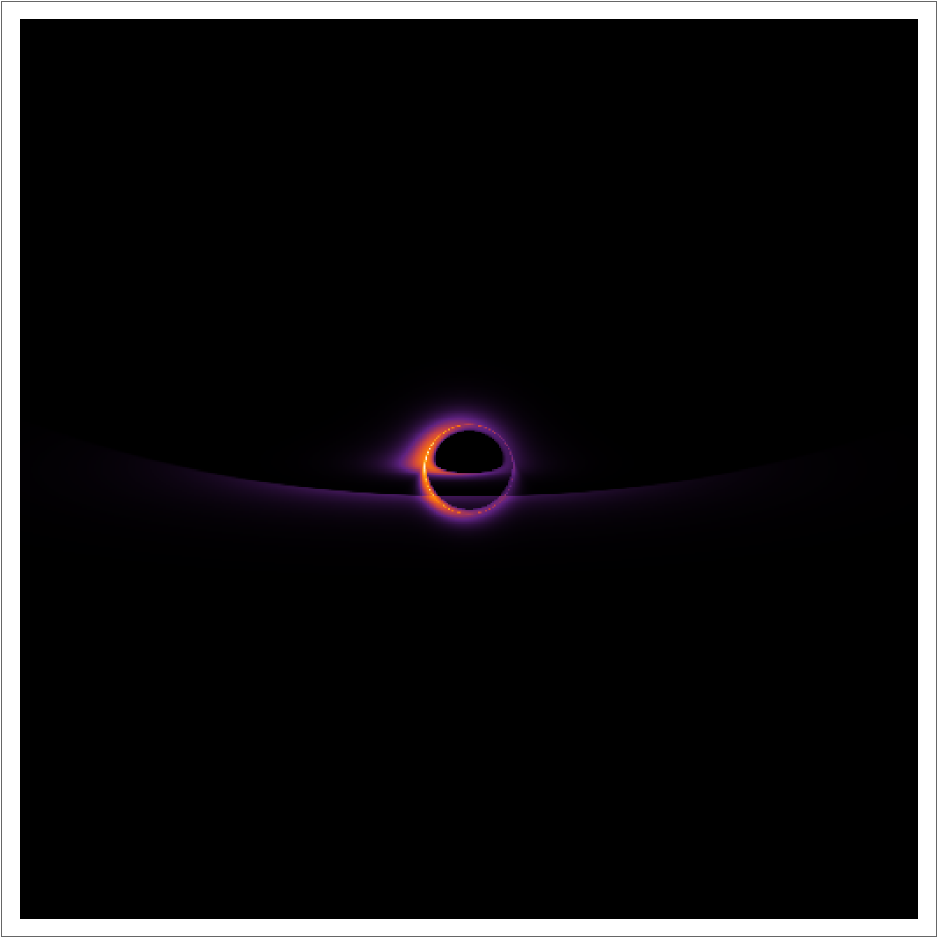}}
\subfigure[$v=0.5$]{\includegraphics[width=.2\textwidth]{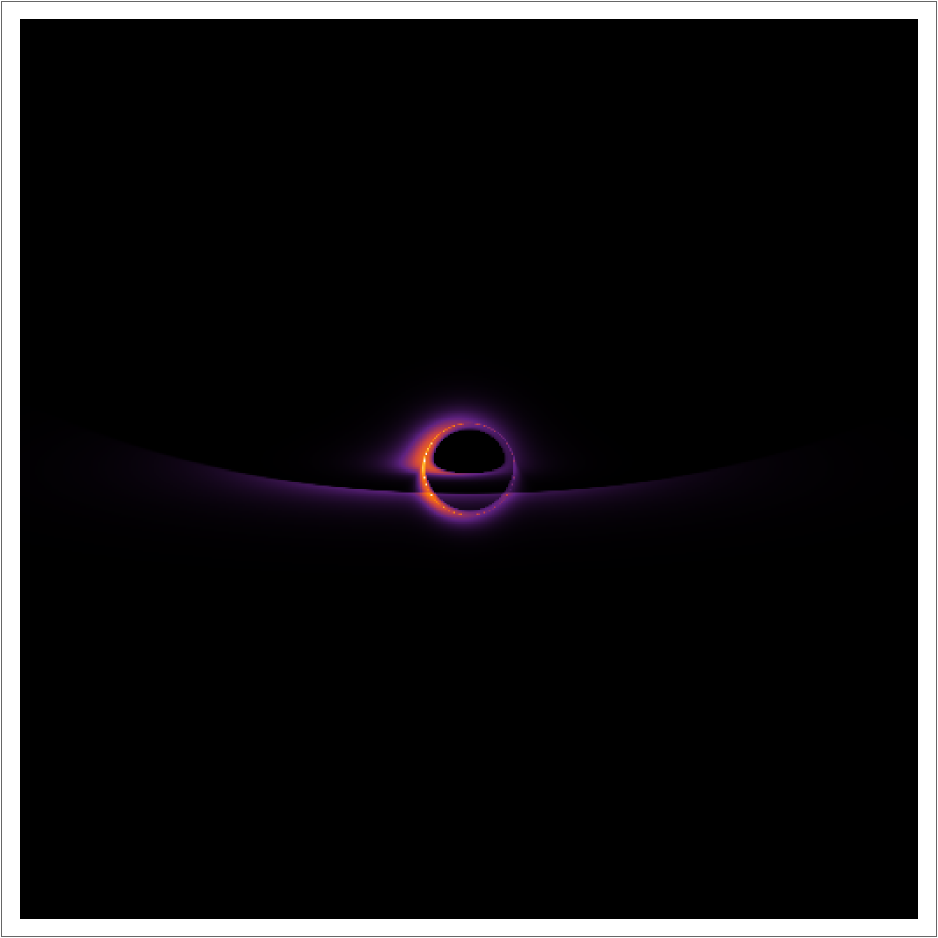}}
\subfigure[$v=0.8$]{\includegraphics[width=.2\textwidth]{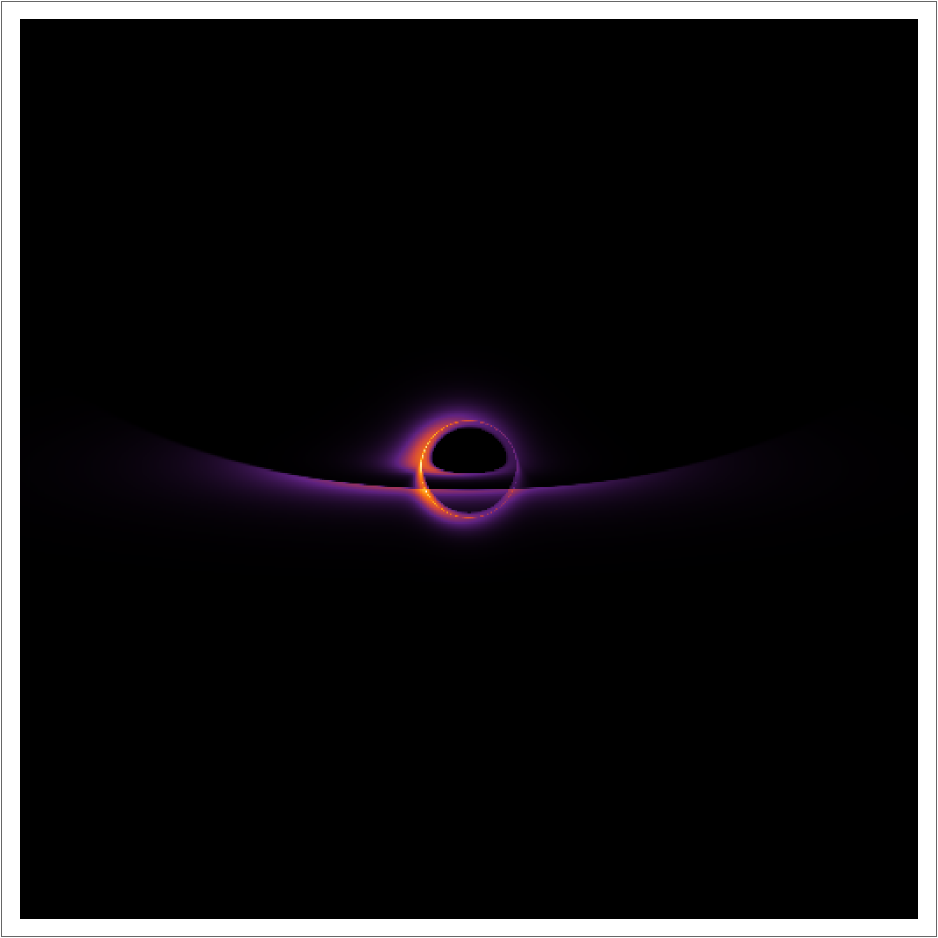}}
\subfigure[$v=1.1$]{\includegraphics[width=.2\textwidth]{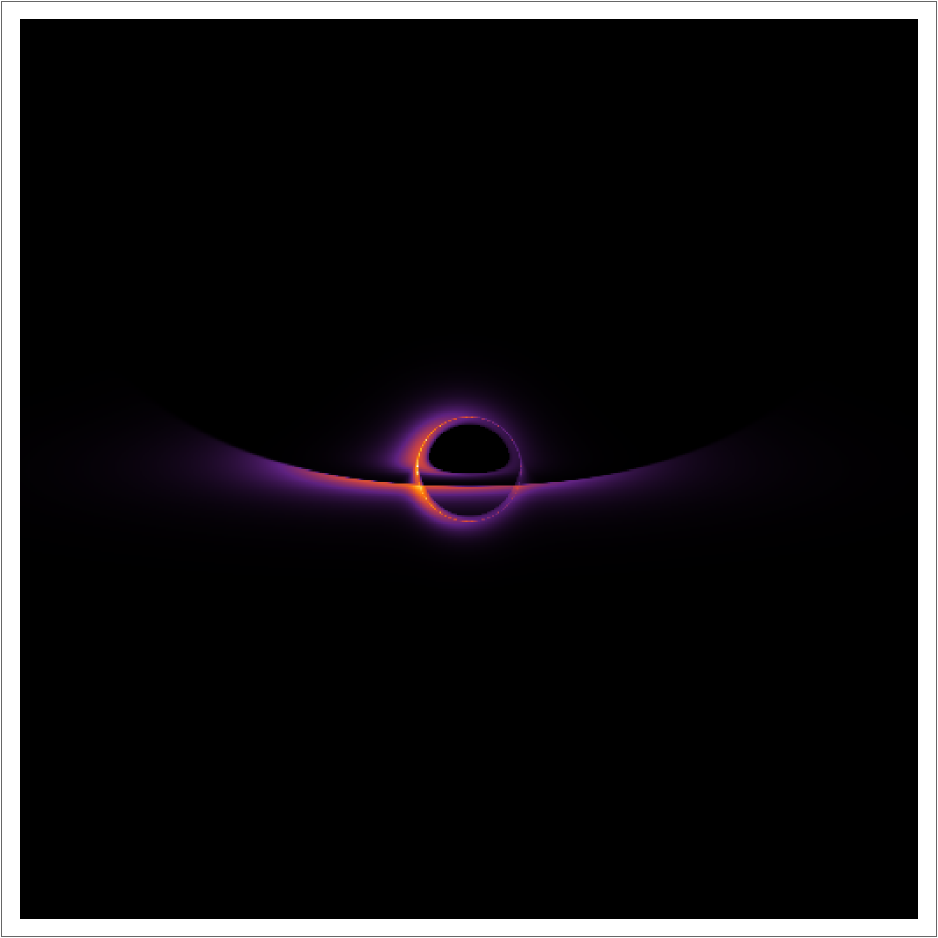}}
\subfigure[$v=1.5$]{\includegraphics[width=.2\textwidth]{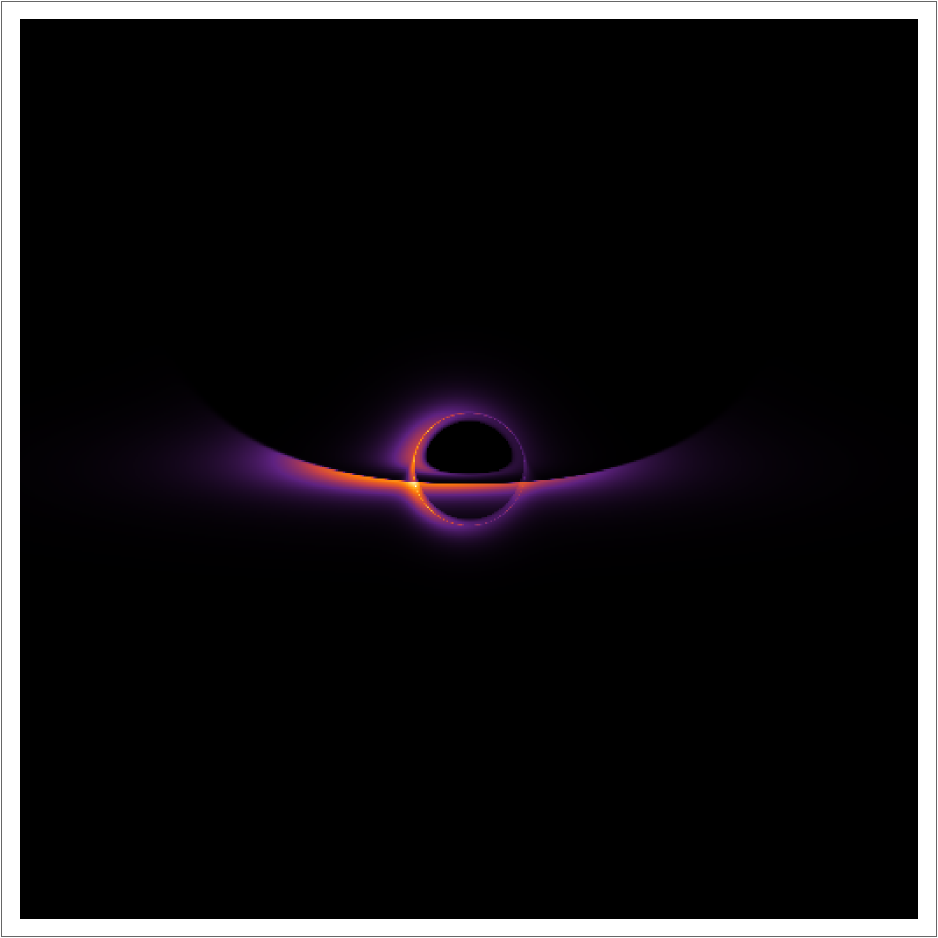}}
\subfigure[$v=2.0$]{\includegraphics[width=.2\textwidth]{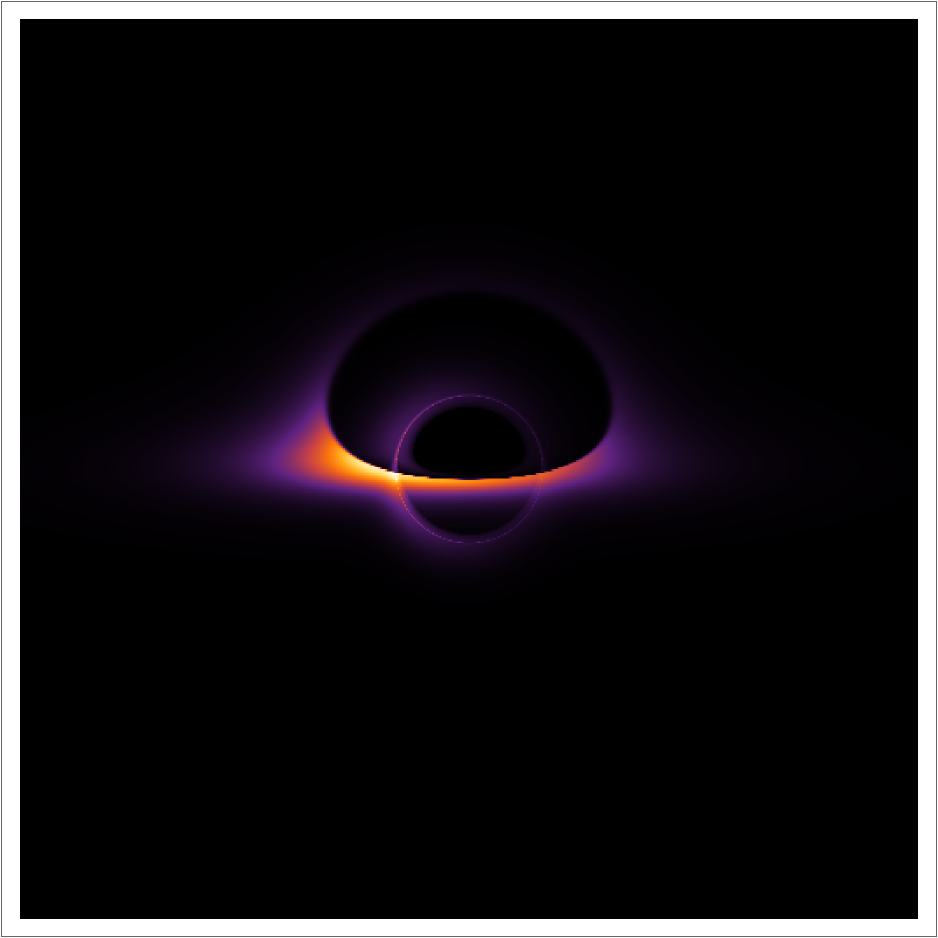}}
\subfigure[$v=2.3$]{\includegraphics[width=.2\textwidth]{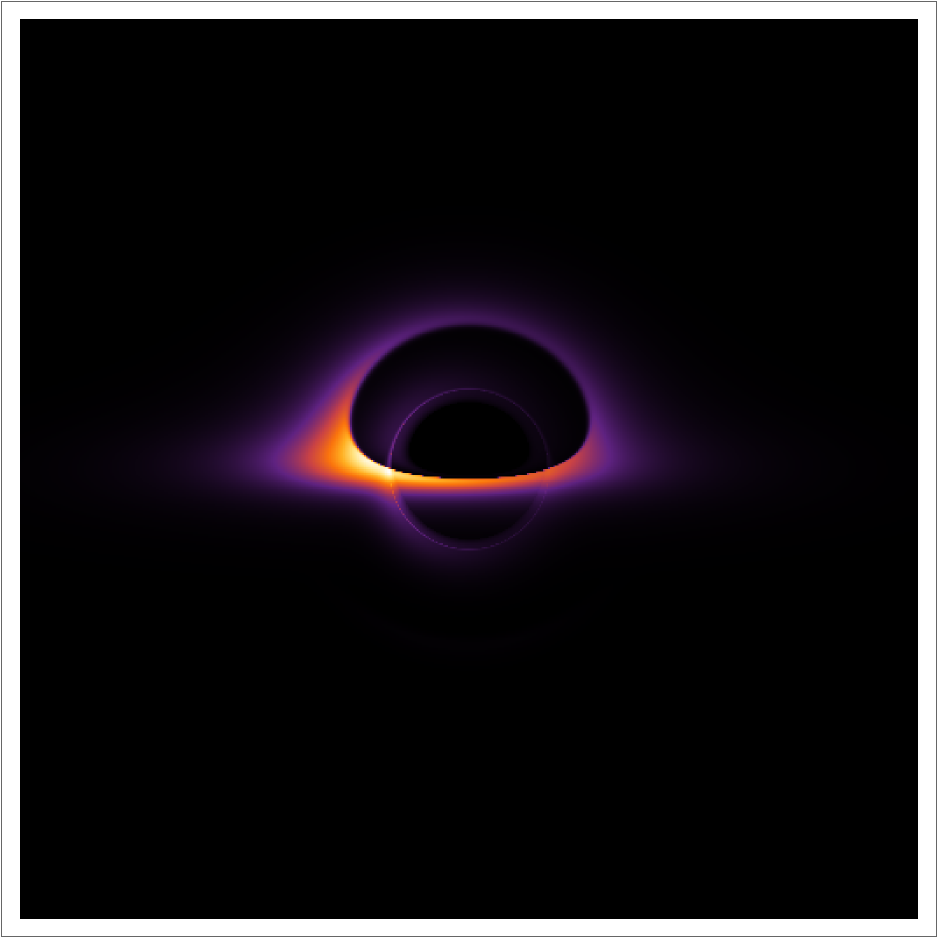}}
\subfigure[$v=2.6$]{\includegraphics[width=.2\textwidth]{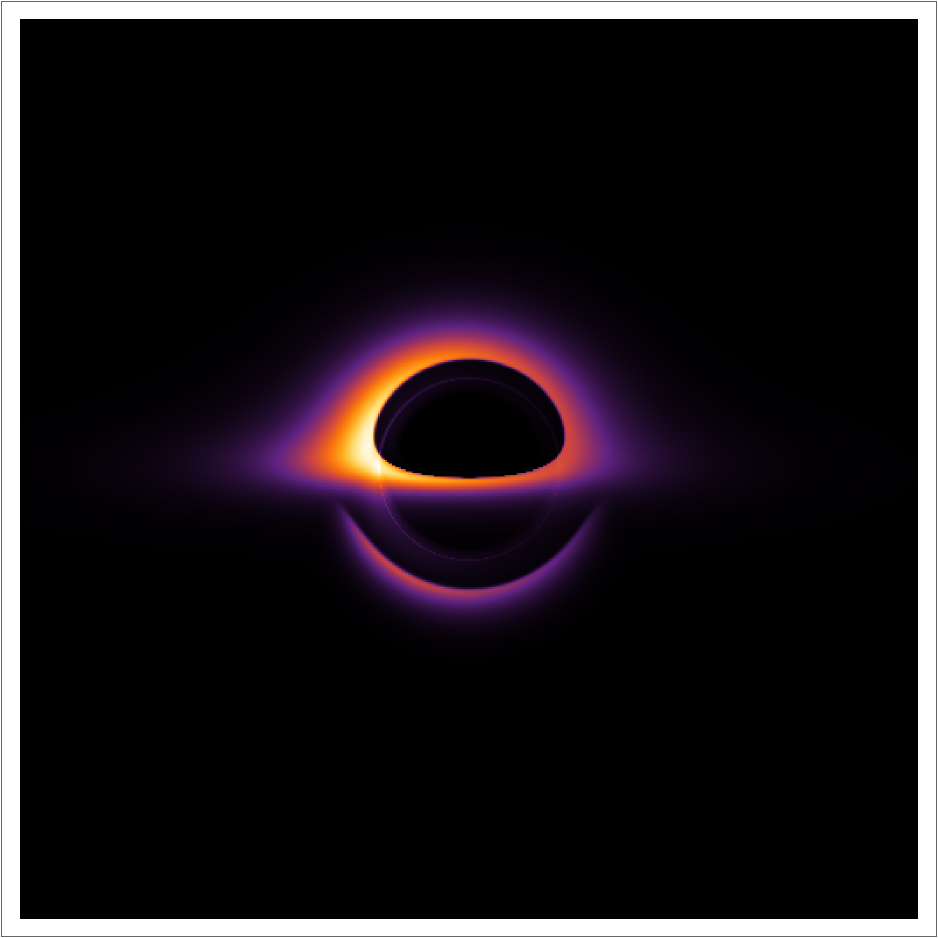}}
\subfigure[$v=3.0$]{\includegraphics[width=.2\textwidth]{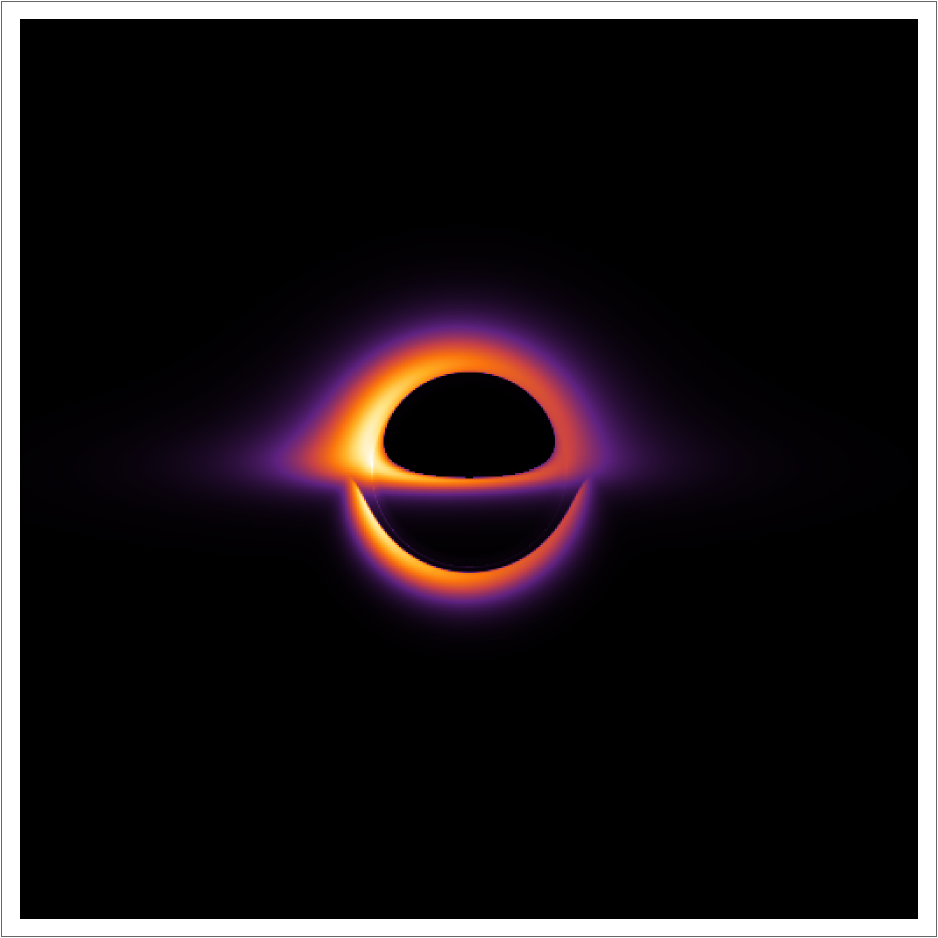}}
\subfigure[$v=4.0$]{\includegraphics[width=.2\textwidth]{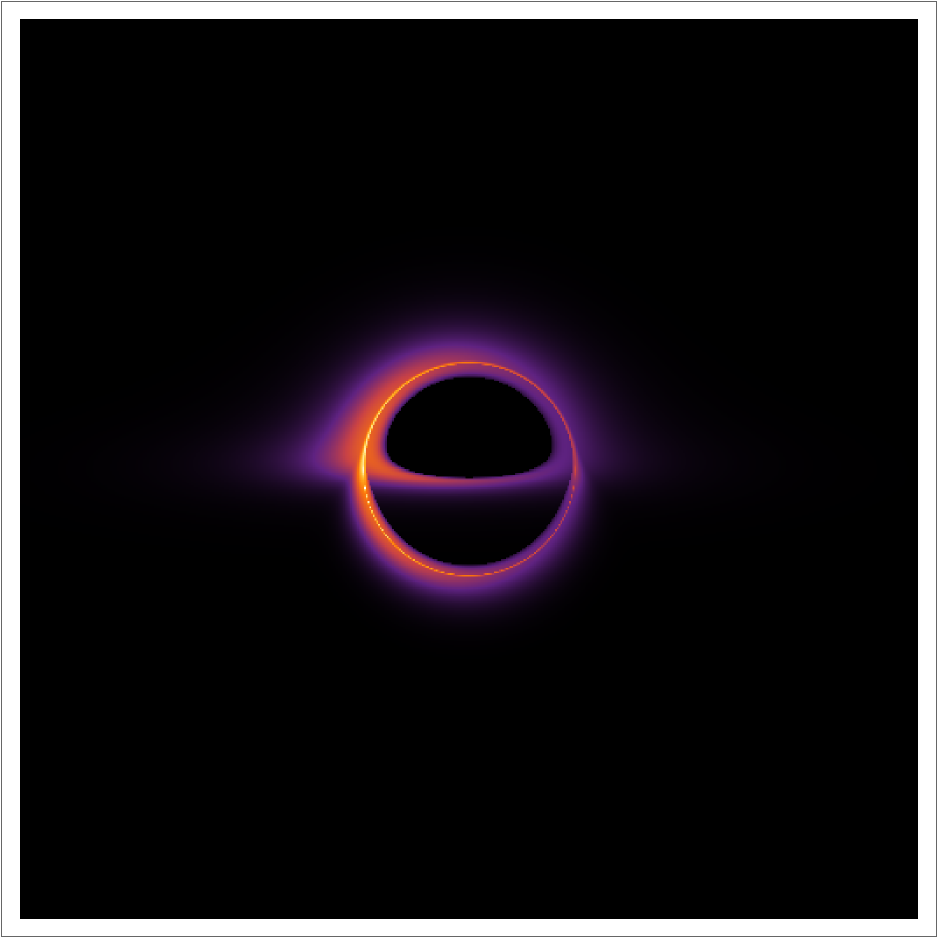}}
\caption{\label{figTD83}  In the thin accretion disk model, the evolutionary characteristics of the optical observational appearance of the Vaidya black hole, with the observational inclination angle being $\theta_{obs}= 83^{\circ}$ .}
\end{figure}

In Figures \ref{figTD17} and \ref{figTD83}, the dynamic evolution characteristics of the observational appearance of the Vaidya black hole under low observation inclination ($\theta_{obs}= 17^{\circ}$) and high observation inclination ($\theta_{obs}= 83^{\circ}$) are respectively presented.
From Figure \ref{figTD17}, it is observable that during the initial accretion phase of the black hole, a bright ring also emerges near the inner shadow region on the screen, accompanied by an additional ring structure in the distance. However, the inner shadowed area no longer maintains a perfect concentric circular symmetry with the bright ring and the additional ring, and the structures of the two rings no longer display axial symmetry in a circular form. Similarly,  the bright ring demonstrates a dynamic evolution process from existence to disappearance and then back to existence throughout the entire process. Additionally, the brightness within the image has a tendency to accumulate in the lower half of the screen, and the brightness of the two rings is asymmetrical. This phenomenon is attributed to the circular motion of the accretion flow, which results in the generation of the Doppler effect.

As can be observed from Figure \ref{figTD83}, the observation perspective exerts a substantial influence on the direct image, the lens image, and the additional ring structure. At high observation inclination angles, the direct image assumes a hat shape and gradually separates from the lens image. Owing to the influence of the Doppler effect, the brightness in the image is predominantly concentrated on the left side of the screen. Meanwhile, an arc-shaped structure resulting from dynamical redshift emerged on the equatorial plane of the black hole and gradually approached the inner shadow region as the accretion process progressed.
During the active phase of accretion, the brightness of this arc  shaped structure gradually increases and approaches the direct image region,  and its shape gradually evolves into a hat like structure, as depicted in Figure \ref{figTD83}(g)-(j). Interestingly, arc shaped structures resulting from dynamic redshift have started to emerge in the vicinity of the gravitational lensing imaging area, and the brightness also exhibits an upward trend, see Figure \ref{figTD83}(j). It can be observed that during the relatively intense accretion process of the black hole, no distinct photon ring structure emerges, which is consistent with the anticipated result. When the black hole is in an asymptotically static configuration(the extremely slow accretion in the later stage), the overall brightness of the image is once again predominantly determined by the direct image, and a distinct photon ring structure also appears, see Figure \ref{figTD83} (l).

\begin{figure}[!t]
\centering 
\subfigure[$v=0.5$, $\theta_{obs}= 0^{\circ}$ ]{\includegraphics[width=.2\textwidth]{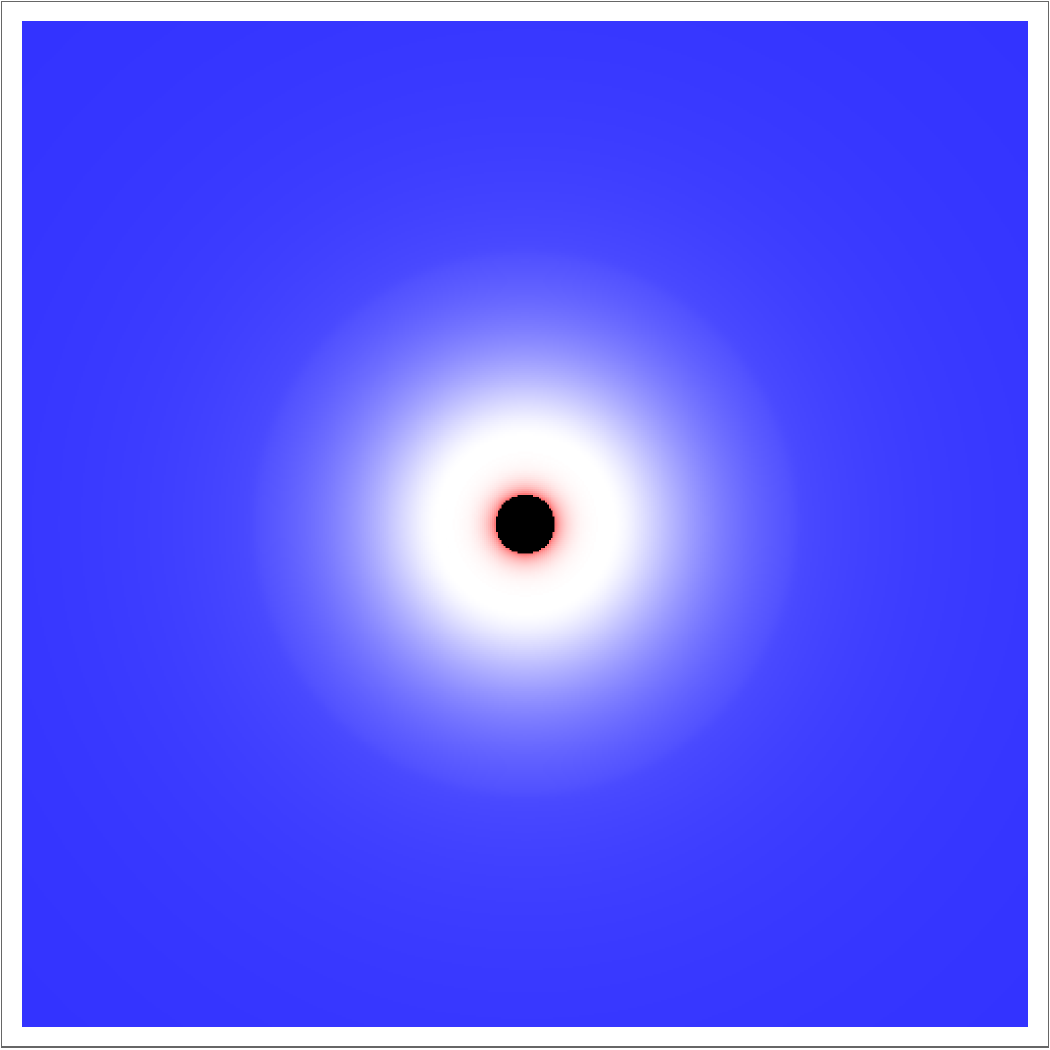}}
\subfigure[$v=1.1$, $\theta_{obs}= 0^{\circ}$]{\includegraphics[width=.2\textwidth]{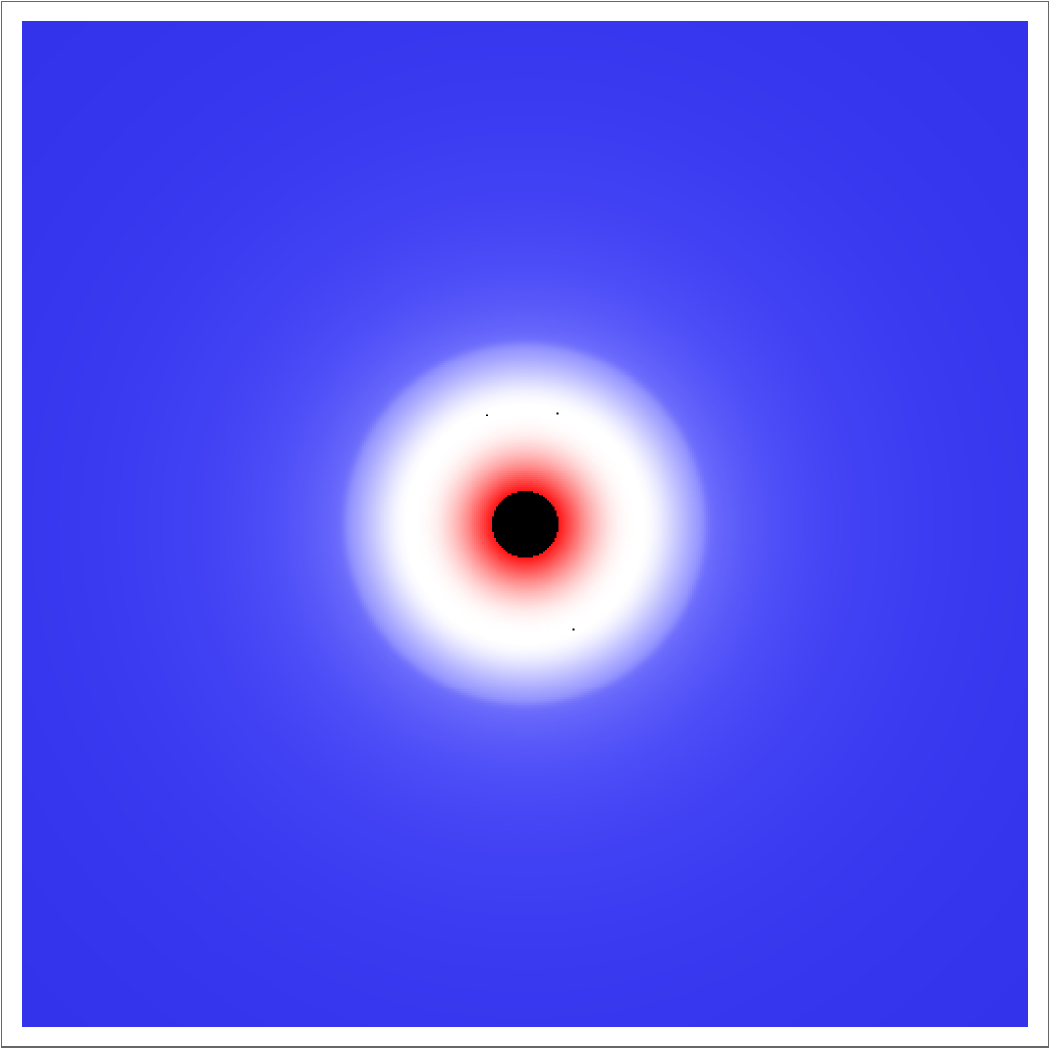}}
\subfigure[$v=2.5$, $\theta_{obs}= 0^{\circ}$]{\includegraphics[width=.2\textwidth]{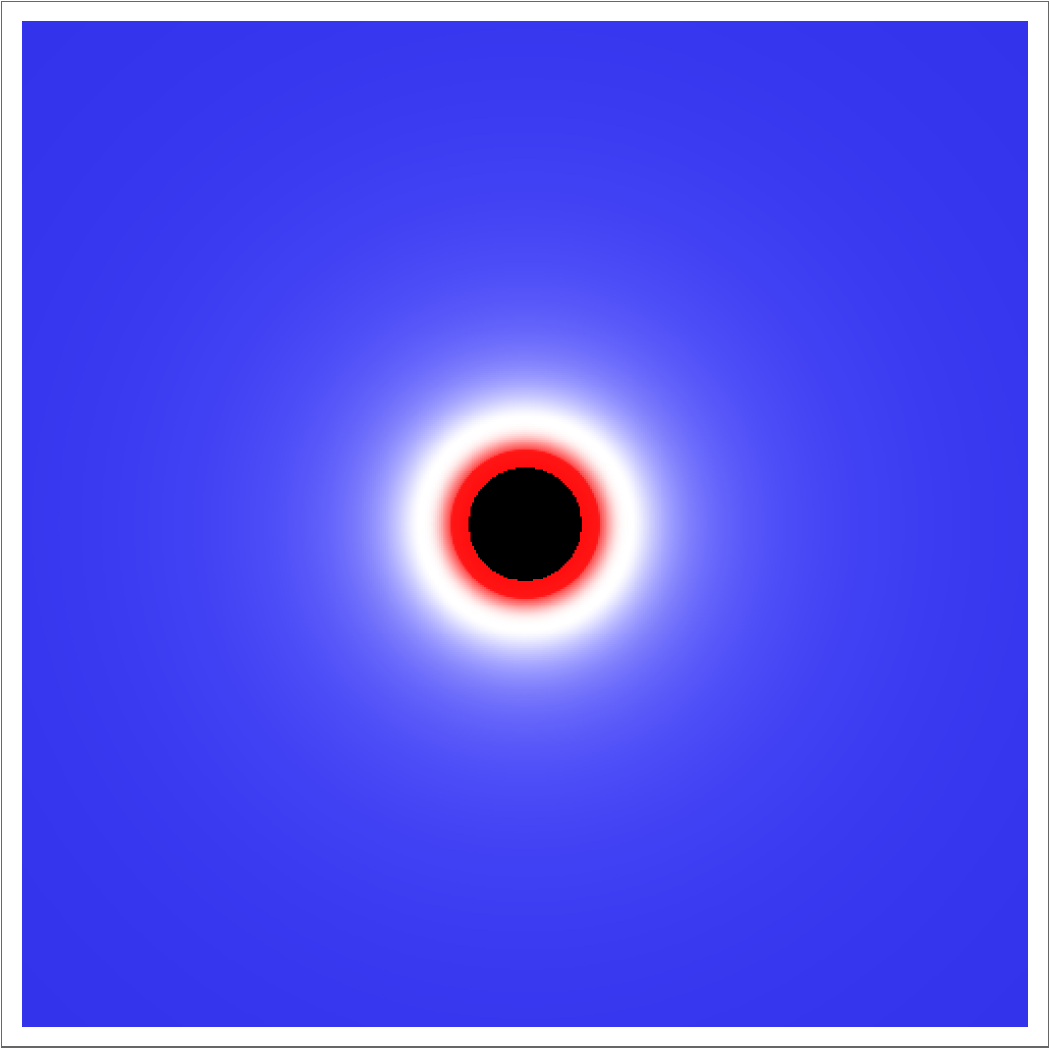}}
\subfigure[$v=4.0$, $\theta_{obs}= 0^{\circ}$]{\includegraphics[width=.2\textwidth]{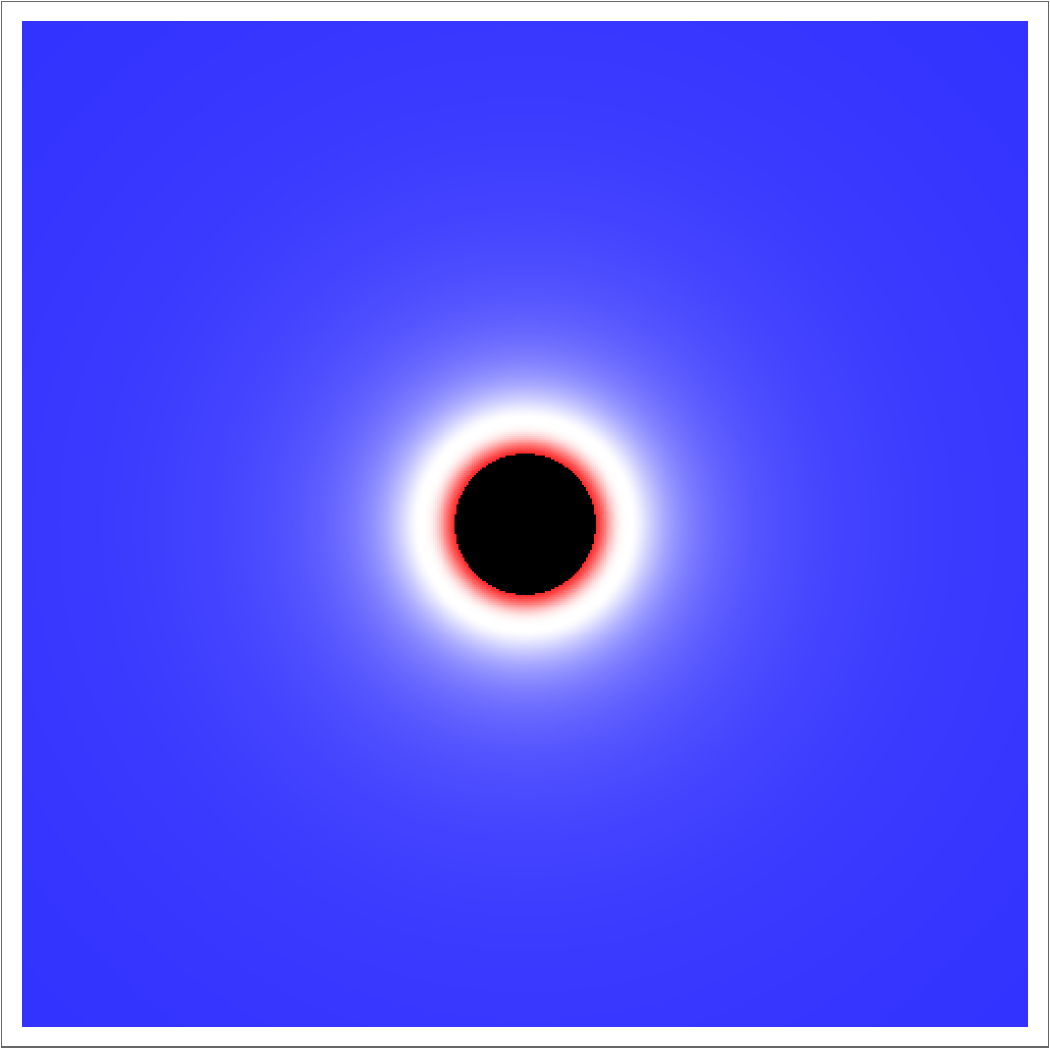}}
\subfigure[$v=0.5$, $\theta_{obs}= 17^{\circ}$]{\includegraphics[width=.2\textwidth]{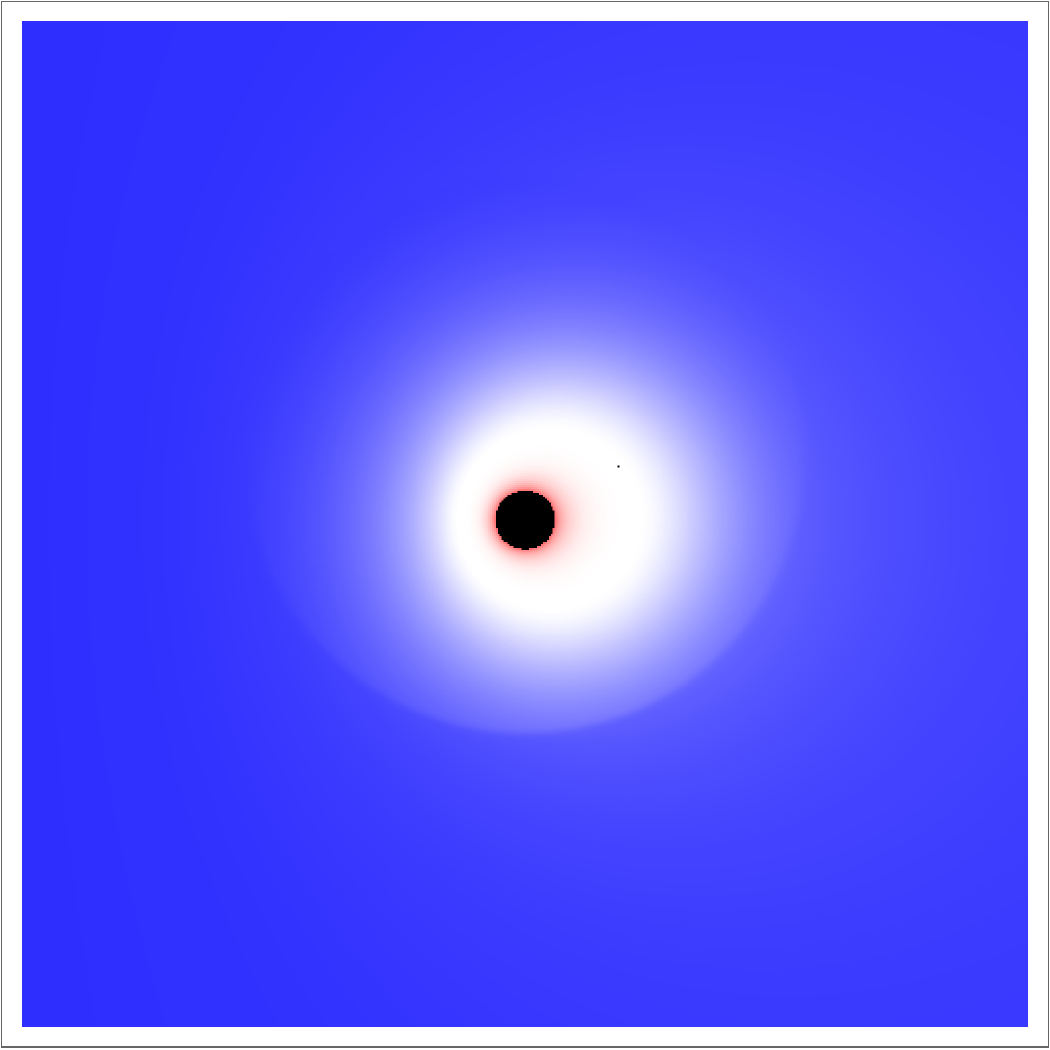}}
\subfigure[$v=1.1$, $\theta_{obs}= 17^{\circ}$]{\includegraphics[width=.2\textwidth]{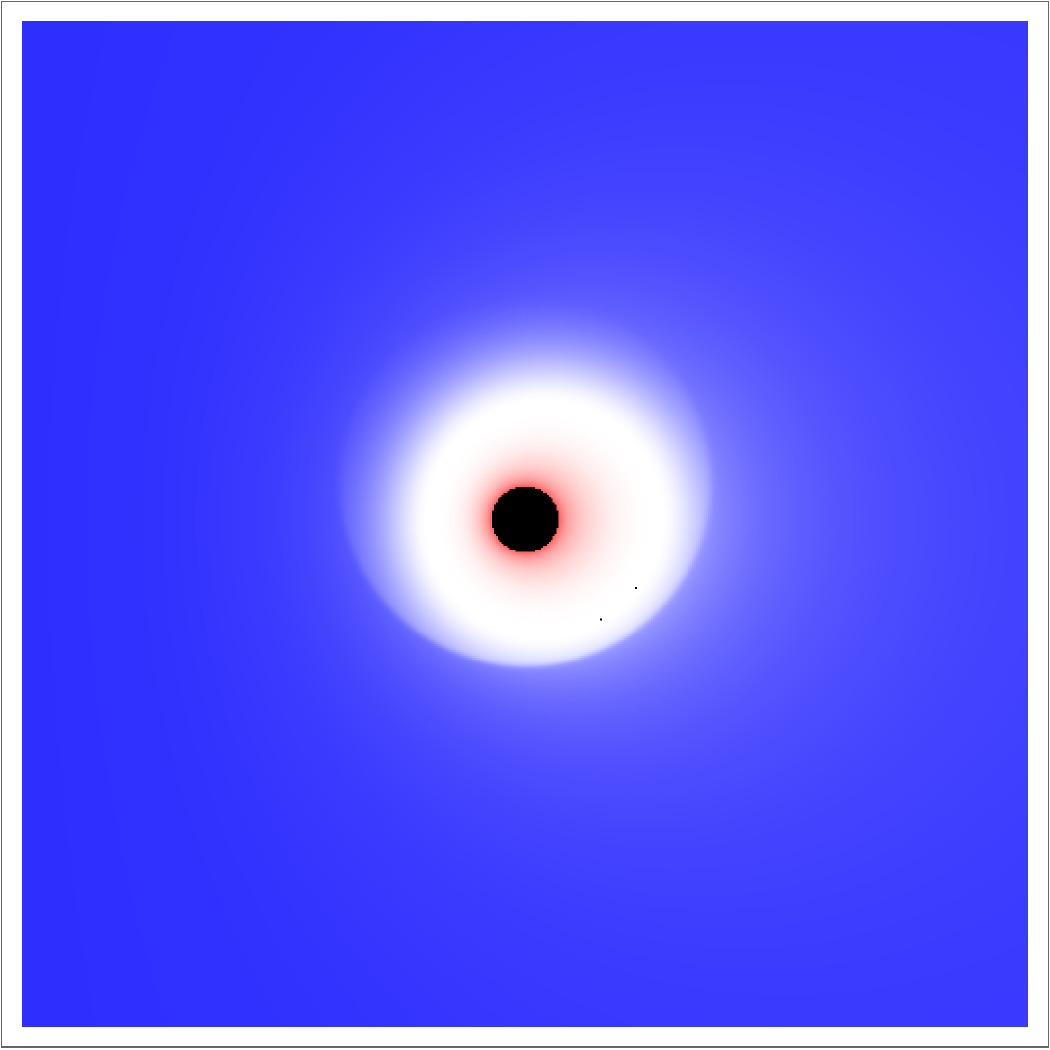}}
\subfigure[$v=2.5$, $\theta_{obs}= 17^{\circ}$]{\includegraphics[width=.2\textwidth]{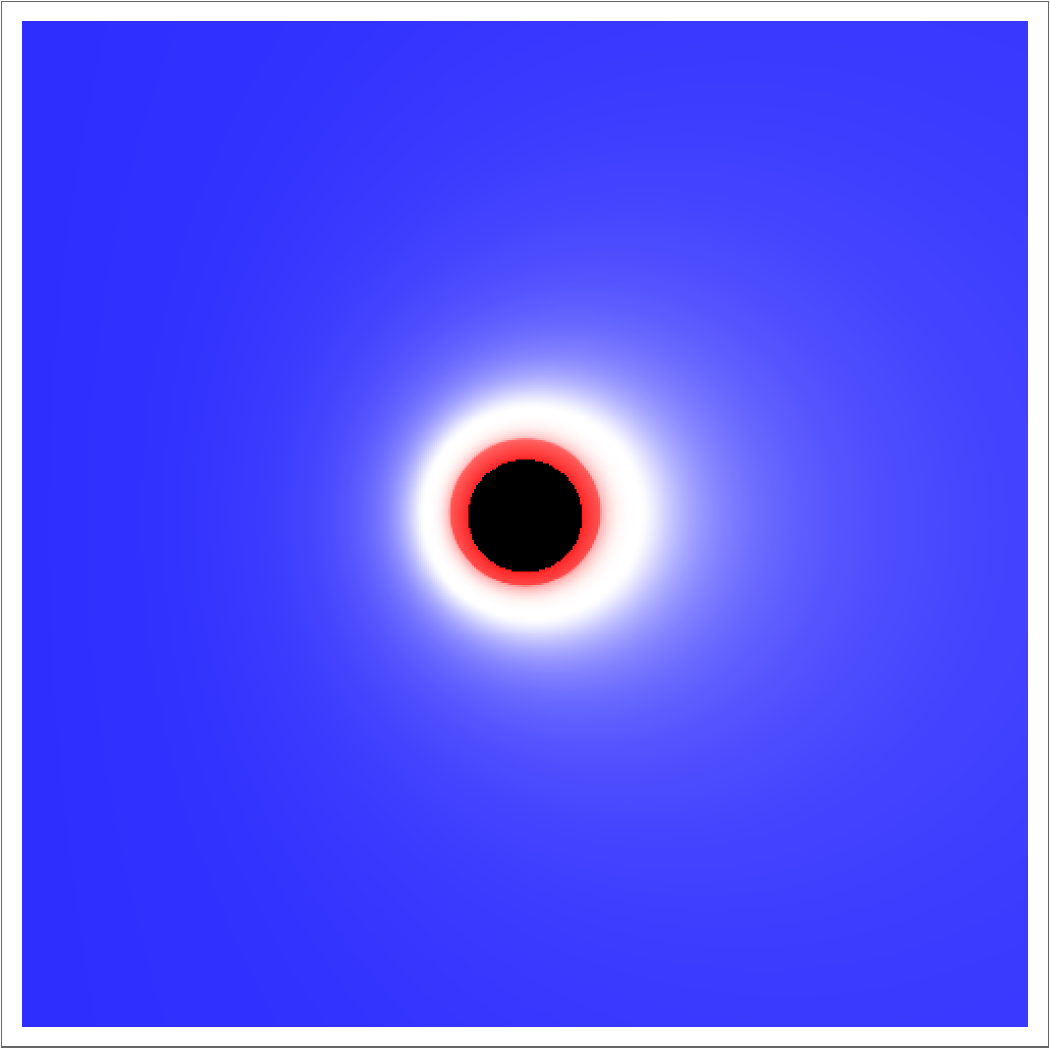}}
\subfigure[$v=4.0$, $\theta_{obs}= 17^{\circ}$]{\includegraphics[width=.2\textwidth]{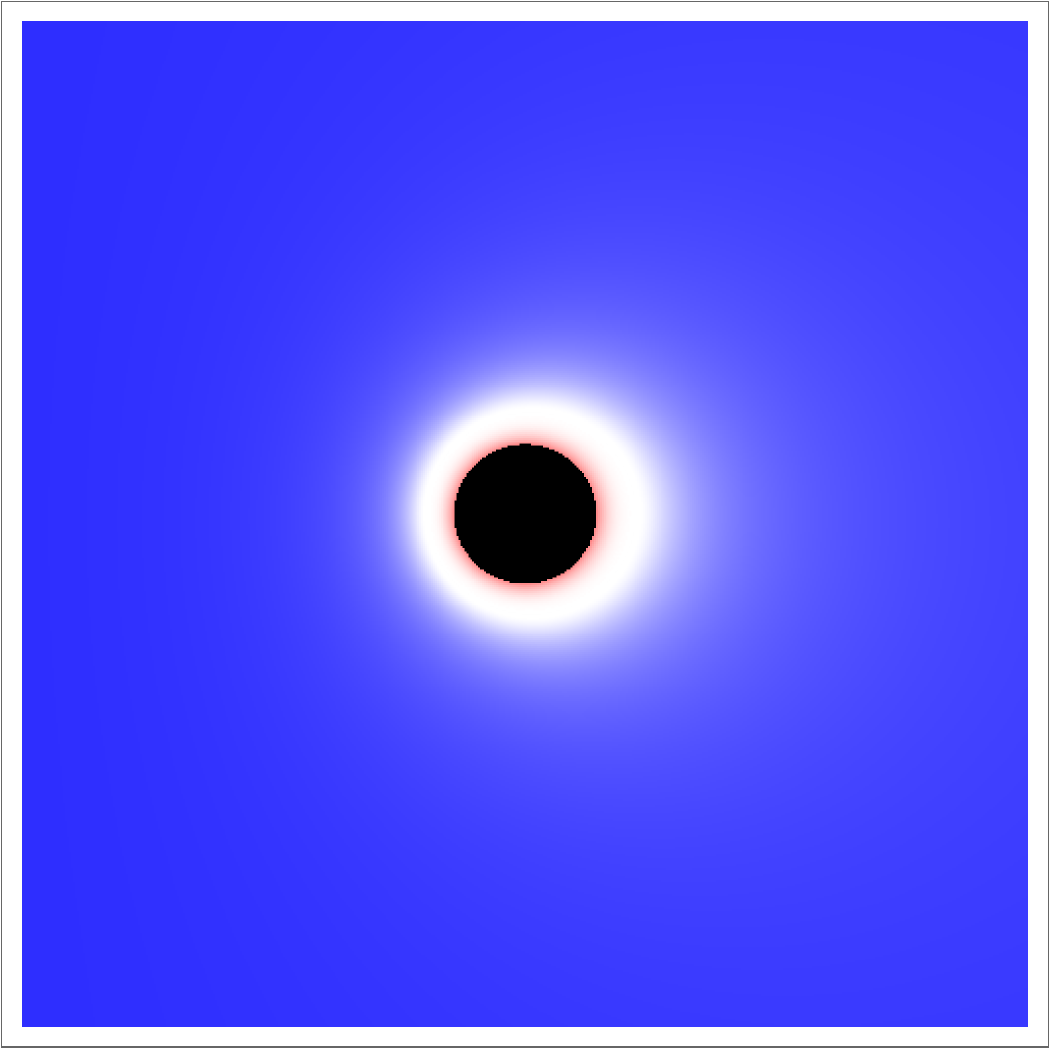}}
\subfigure[$v=0.5$, $\theta_{obs}= 83^{\circ}$]{\includegraphics[width=.2\textwidth]{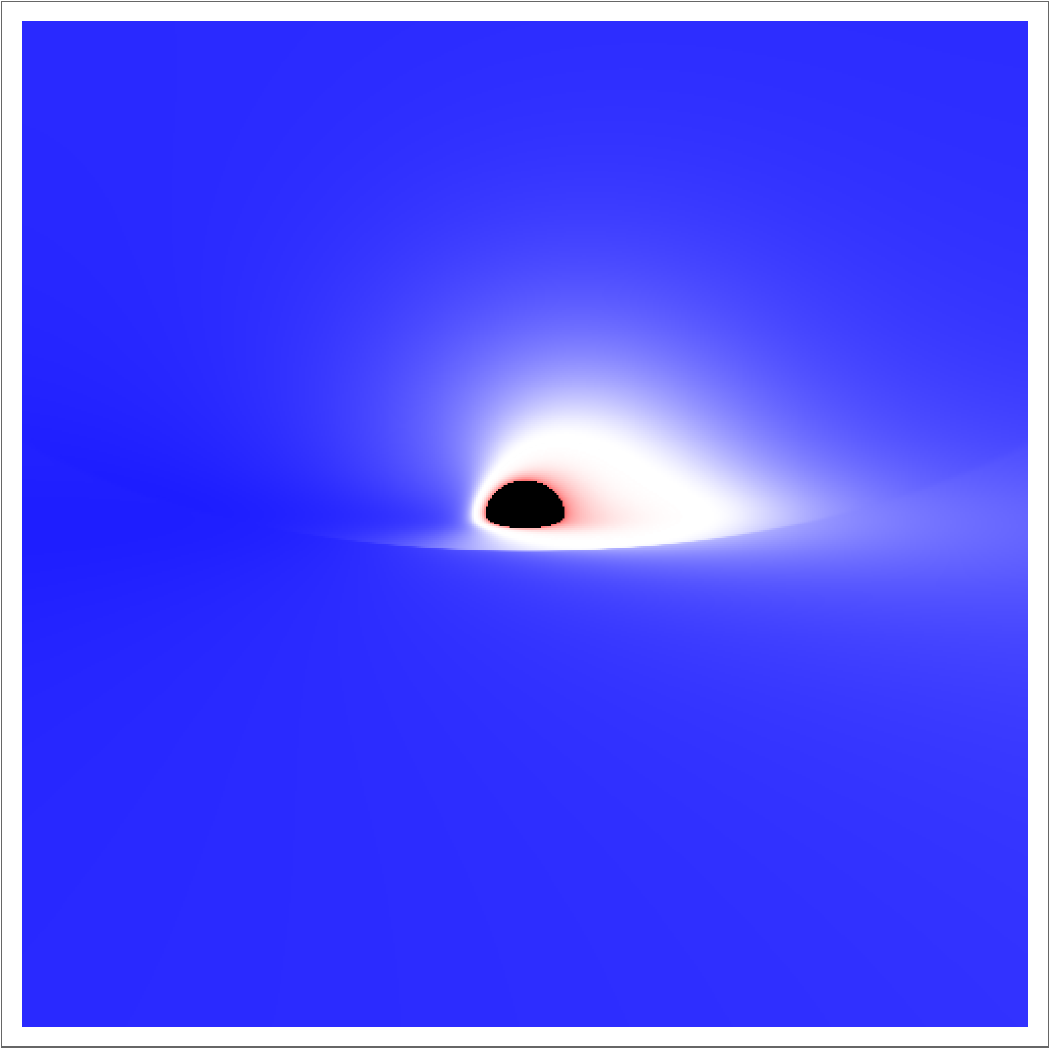}}
\subfigure[$v=1.1$, $\theta_{obs}= 83^{\circ}$]{\includegraphics[width=.2\textwidth]{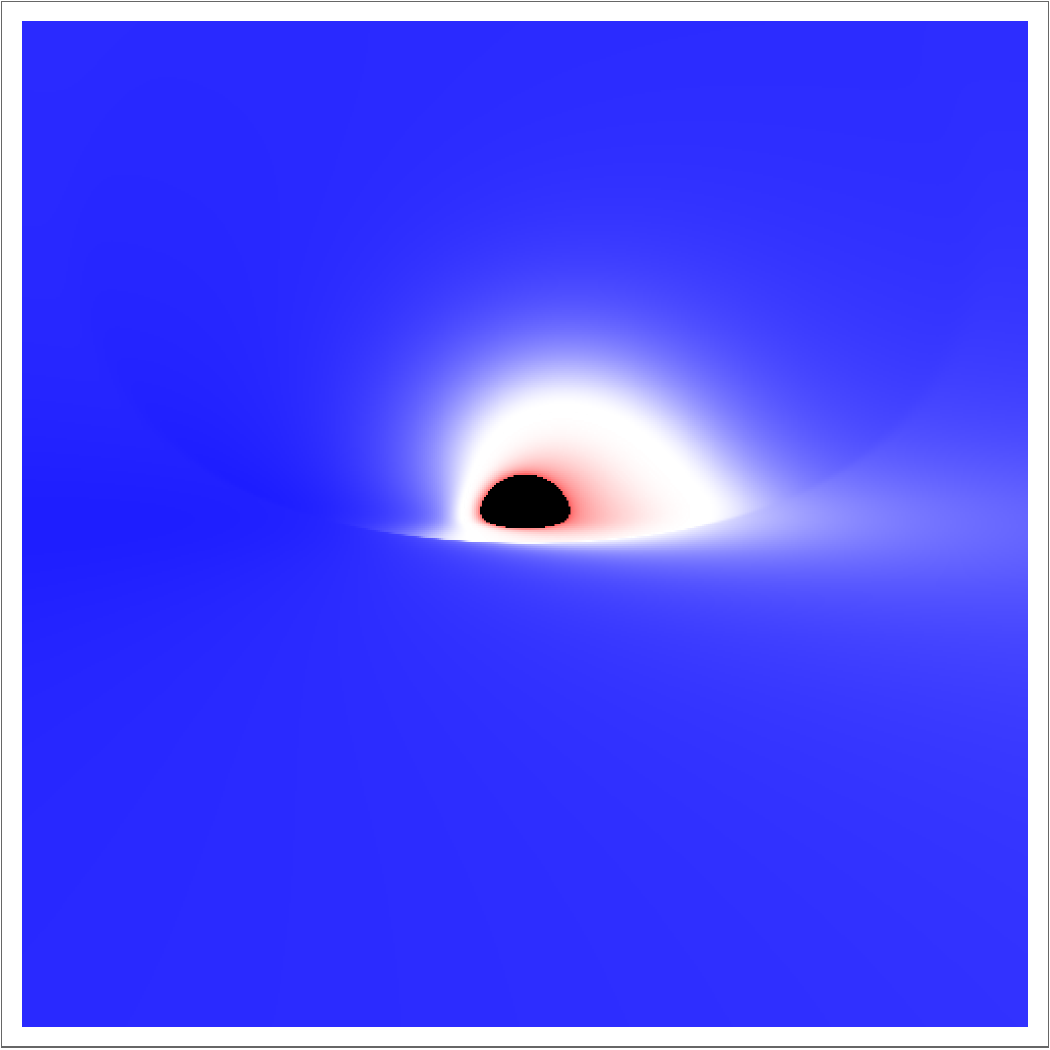}}
\subfigure[$v=2.5$, $\theta_{obs}= 83^{\circ}$]{\includegraphics[width=.2\textwidth]{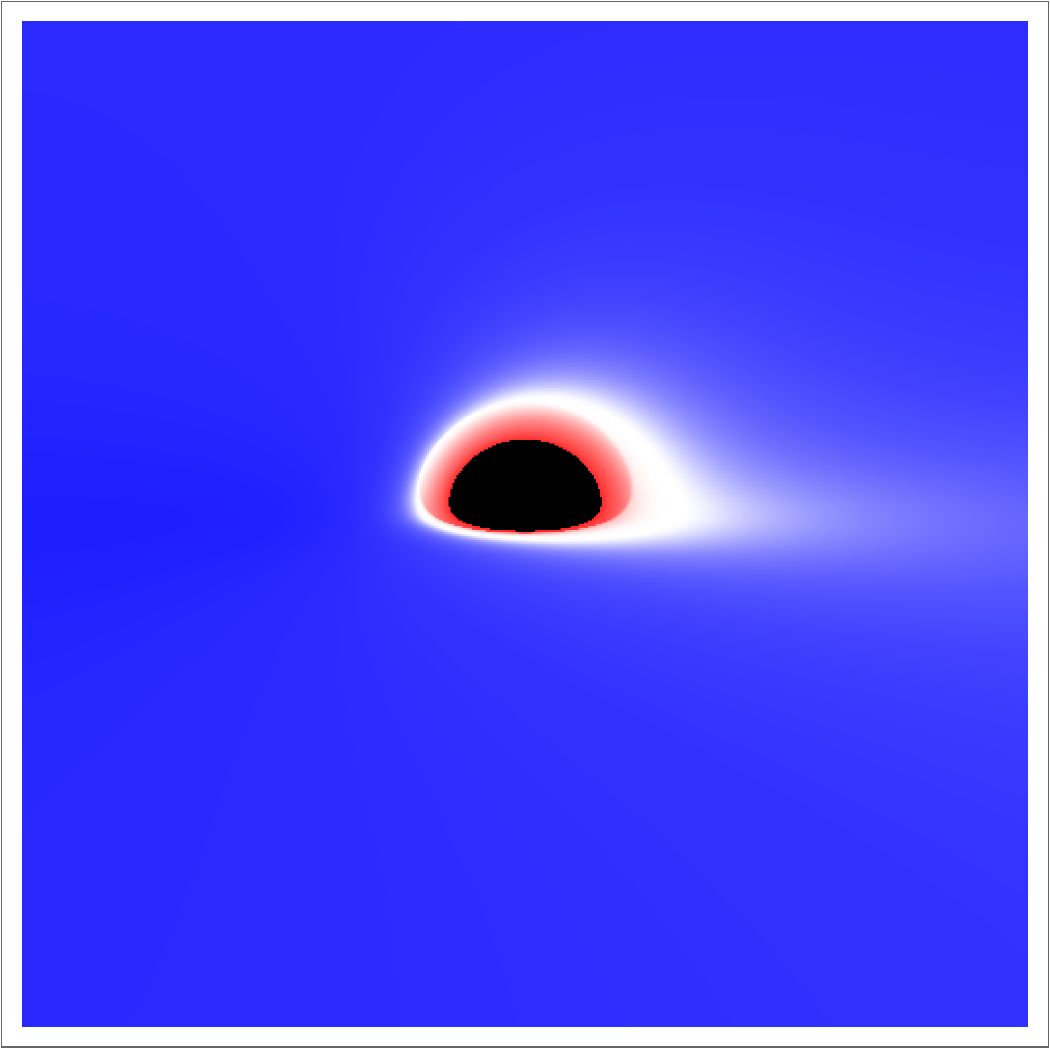}}
\subfigure[$v=4.0$, $\theta_{obs}= 83^{\circ}$]{\includegraphics[width=.2\textwidth]{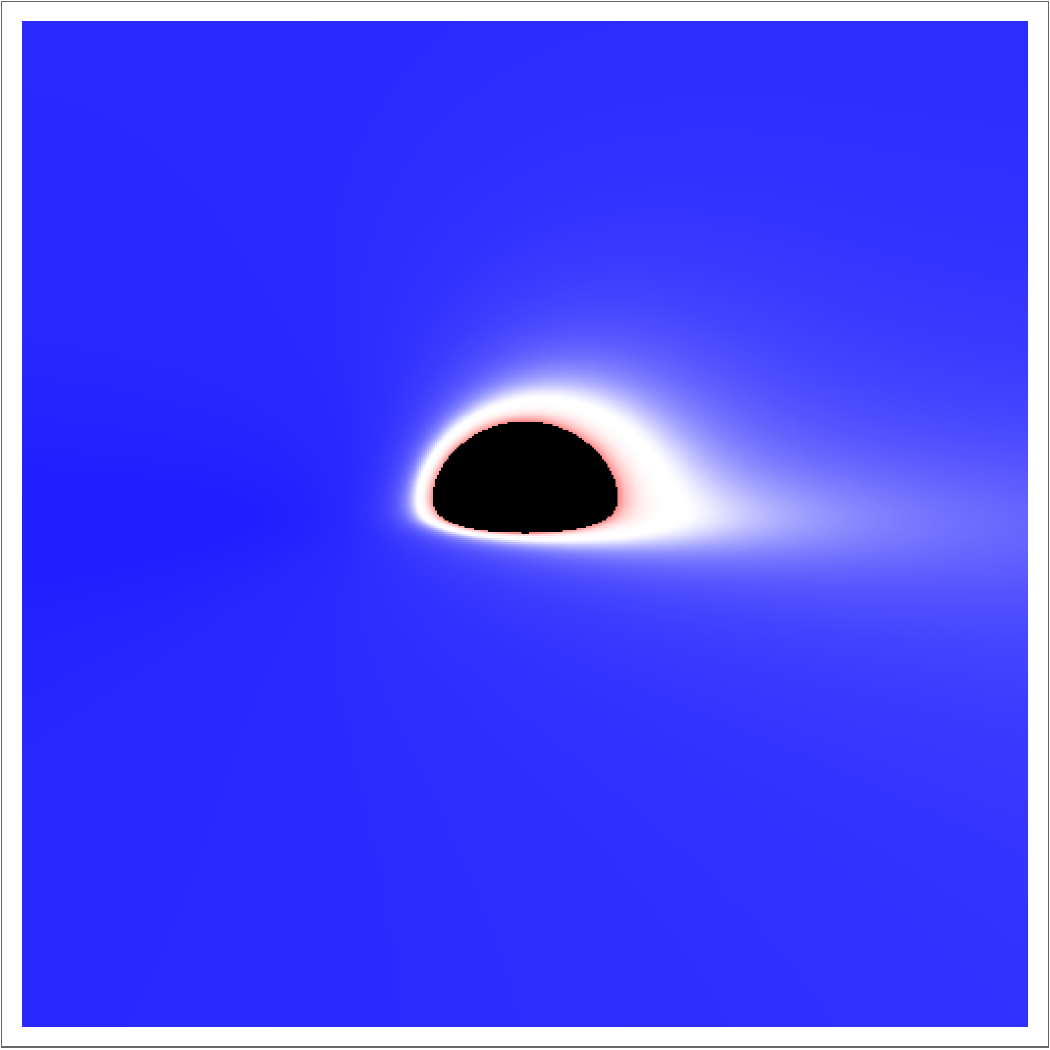}}
\caption{\label{figTDR}  The distribution of redshift and blueshift in direct imaging, where the observation inclination angles of the top row, middle row, and bottom row are $\theta_{obs}= 0^{\circ}$, $17^{\circ}$, and $83^{\circ}$, respectively. }
\end{figure}
To verify that this additional ring is caused by the redshift effect, we present in Figure \ref{figTDR} the corresponding redshift and blueshift images at different viewing angles. Among them, the red and blue pixels represent redshift and blueshift respectively, and the black circles denote the shadow areas. It is evident that there exists a red area encircling the shadow region. This phenomenon is attributable to the gravitational redshift induced by the strong gravitational field in the vicinity. Outside the red area is located the blue area, which corresponds to the dynamical redshift. As the observation inclination angle varies, the overall profiles of both redshift and blueshift undergo deformation. Particularly at high observation inclination angles, the redshifted region transforms into a hat like shape and is predominantly concentrated on the right side of the image. The inner contour of the blueshifted region also bears a resemblance to a hat; however, the majority of the pixels are concentrated on the left side of the image. By comparing the results presented in Figure \ref{figTDR} with those in Figure \ref{figTD0}-Figure \ref{figTD0}, it is evident that the position of the additional ring precisely corresponds to the inner contour position of the blueshifted region, and the evolution trend is consistent.

\section{Conclusions and discussions}
\label{conclusion}
This work conducts a systematic investigation of the gravitational lensing and dynamic evolution of shadows associated with Vaidya black holes. We employ the Vaidya metric to depict the temporal evolution of accreting black hole spacetimes. As an exact solution to general relativity that characterizes non-static, spherically symmetric gravitational fields, this metric presents the significant advantage of integrating a time-dependent mass function to account for black hole mass variations, thus offering a more realistic portrayal of astrophysical accretion environments. For the mass function, we consider a smoothly increasing functional form that enables a seamless transition from an initial low mass state to an asymptotically static final configuration. In the context of the celestial sphere model and the thin accretion disk model, the dynamic evolutionary characteristics of the gravitational lensing effect and the shadow in the Vaidya spacetime can be explored via the reverse ray-tracing technique.

Under the framework of the celestial sphere model, the shadow of the Vaidya black hole presents a complete evolutionary picture from the initial stable state, through expansion, to the final stable state.
It is noteworthy that, in contrast to the scenario where the shadow in a collapsing spacetime gradually expands from a minuscule point at the center, the shadow of a Vaidya black hole manifests as a small dark region at the center of the image from the early phase of accretion (even prior to the onset of accretion). This difference originates from the fundamental distinction in the physical scenarios of the two. Specifically, our model features a compact central object (a low  mass black hole) right from the start, whereas the black hole event horizon is formed only at the end of the collapse process. Therefore, during the entire accretion process, a shadow is consistently present, and its size steadily expands as the mass increases. This contrasting outcome has the potential to offer an observational criterion for differentiating whether the spacetime is undergoing a collapse or an accretion process. One can find that although the size of the shadow keeps changing, the spatial distribution of the Einstein ring in the distance and the external grid structure remains almost unchanged. This is because the black hole smoothly evolves from the initial low-mass state to the final static configuration, and the spacetime structure at infinity always approaches the Schwarzschild metric; for distant observers, the mass increase and horizon expansion effects caused by the accretion process are extremely weak, thus the external lensing structure remains stable In addition,  the active accretion stage and the later stage of accretion, a new lensing ring structure appears in the vicinity of the shadow, whose width first increases and then decreases, ultimately stabilizing as the black hole approaches a static state.

For the thin accretion disk model, we propose positioning it on the equatorial plane of the black hole, with its inner edge reaching the apparent horizon of the black hole. When the observation inclination angle is $\theta_{\mathrm{obs}} = 0^\circ$, the brightness distribution in the image always maintains an axisymmetric distribution. In the initial stage of accretion or prior to the commencement of accretion, a luminous ring encircles the inner shadow of the black hole. This ring is formed through the superposition of lensed images and the photon ring. As the accretion process persisted, an additional ring-like structure appeared in the image. Subsequently, this feature gradually brightens and contracts towards the inner shadow region. Its formation is attributed to the photon energy shift caused by the non - steady state of spacetime, namely, dynamic redshift. When accretion enters an active phase, the stable photon ring fails to form and subsequently disappears. The direct image then becomes the dominant feature, while the additional ring contracts towards its outer edge. It is only when the black hole approaches a static state that the photon ring structure re-emerge. For $\theta_{\mathrm{obs}} = 17^\circ$ , the inner shadow and the bright ring no longer exhibit concentric circular symmetry. The brightness of the image is concentrated in the lower half plane, and both the bright ring and the additional ring display asymmetry in terms of brightness. This asymmetry originates from the Doppler effect generated by the circular motion of the accretion flow. When superimposed with the dynamical redshift effect, these two effects jointly modulate the observed intensity distribution. When the inclination angle increases to $\theta_{\mathrm{obs}} = 83^\circ$, the direct image transforms into a cap  like configuration and gradually detaches from the lensed image. The Doppler effect leads to a significant concentration of brightness in the left half plane, whereas the ring structure formed by the dynamical redshift distorts into a distinct arc, emerging from the equatorial region and gradually contracting towards the inner shadow. Particularly during the active accretion stage, the brightness of this dynamical redshift arc even exceeds that of the direct image, emerging as the dominant component of the image, and its shape gradually transforms from an arc to a cap like structure. Meanwhile, secondary arc like structures induced by dynamical redshift also emerged near the lensed image region and gradually approached and brightened the lensed image as the accretion process progressed. Furthermore, throughout the entire accretion process, the structure of the photon ring is only clearly observable in the initial and final static stages, while it completely vanishes during the active accretion phase. This outcome aligns with theoretical anticipations, as the structure of the photon ring cannot be stably formed during the dynamic evolution stage of spacetime.

Using an idealized thin-disk model, we elucidate the observable imprints of spacetime dynamics on black hole images, with particular emphasis on dynamical redshift—a novel effect analogous to cosmological redshift yet arising from local mass evolution. This finding offers a potential observational discriminant for identifying accreting black holes in future astronomical observations. It should be noted that while the thin-disk approximation employed herein effectively isolates gravitational effects, realistic accretion flows typically possess geometric thickness and involve complex radiative processes. Consequently, subsequent efforts will extend this analysis to geometrically thick disks and incorporate radiation magnetohydrodynamic simulations to assess the detectability of dynamical redshift signatures under more astrophysically realistic conditions.

\vspace{10pt}

\noindent {\bf Acknowledgments}

\noindent

We are grateful to  Min-Yong Guo and  Jie-Wei Huang for their helpful discussions.
This work is supported by the National Natural Science Foundation of China (Grants Nos. 12505059, 12375043, 12575069), and the China Postdoctoral Science Foundation (Grants No.2025MD784184), and Chongqing Normal University Fund Project (Grants No. 26XLB001).



\end{document}